\documentclass[aps,amsfonts,amsmath,superscriptaddress,twocolumn,showpacs,floatfix,nofootinbib,showkeys]{revtex4-1}

\usepackage{marginnote}

\usepackage{epsfig}
\usepackage{graphicx}
\usepackage{mathptmx}      % use Times fonts if available on your TeX system
\usepackage{latexsym}
\usepackage{bm}

\usepackage{stackrel}

\def\slashchar#1{\setbox0=\hbox{$#1$}
   \dimen0=\wd0 \setbox1=\hbox{/} \dimen1=\wd1
   \ifdim\dimen0>\dimen1 \rlap{\hbox to \dimen0{\hfil/\hfil}} #1
   \else  \rlap{\hbox to \dimen1{\hfil$#1$\hfil}} / \fi}

\def\tr{{\rm tr}}
\def\Tr{{\rm Tr}}
\newcommand{\MeV}{\,{\mathrm{MeV}}}
\newcommand{\GeV}{\,{\mathrm{GeV}}}
\newcommand{\fm}{\,{\mathrm{fm}}}
\newcommand{\SU}{{\mathrm{SU}}}

\newcommand{\Eq}[1]{Eq.~(\ref{eq:#1})}

\newcommand{\vx}{{\bm{x}}}

\newcommand{\vp}{{\bm{p}}}

\newcommand{\ignore}[1]{}

\newcommand{\QCD}{{\textrm{\scriptsize QCD}}}
\newcommand{\HRG}{{\textrm{\scriptsize HRG}}}
\newcommand{\cons}{{\textrm{\scriptsize cons}}}
\newcommand{\lat}{{\textrm{\scriptsize lat}}}
\newcommand{\param}{{\textrm{\scriptsize param}}}

\begin{document}

\title{Baryonic susceptibilities, quark-diquark models, and
  quark-hadron duality at finite temperature}
%Quark-diquark model: baryon spectrum and fluctuations of conserved charges}

\author{E. Meg\'{\i}as}
\email{emegias@ugr.es}

\author{E. \surname{Ruiz Arriola}}
\email{earriola@ugr.es}

\author{L.L. Salcedo}
\email{salcedo@ugr.es}

\affiliation{Departamento de F{\'\i}sica At\'omica, Molecular y Nuclear and
  Instituto Carlos I de F{\'\i}sica Te\'orica y Computacional \\ Universidad
  de Granada, E-18071 Granada, Spain.}

\date{\today}

\begin{abstract}
  \medskip
  Fluctuations of conserved charges such as baryon number, electric
  charge and strangeness may provide a test for completeness of states
  in lattice QCD for three light flavors. We elaborate on the idea
  that the corresponding susceptibilities can be saturated with
  excited baryonic states with an underlying quark-diquark structure
  with a linearly confining interaction. Using Polyakov-loop
  correlators we show that in the static limit, the quark-diquark
  potential coincides with the quark-antiquark potential in marked
  agreement with recent lattice studies. We thus study in a
  quark-diquark model the baryonic fluctuations of electric charge,
  baryon number and strangeness: $\chi_{BQ}$, $\chi_{BB}$ and
  $\chi_{BS}$; by considering a realization of the hadron resonance
  gas model in the light flavor sector of QCD. These results have been
  obtained by using the baryon spectrum computed within a relativistic
  quark-diquark model, leading to an overall good agreement with the
  spectrum obtained with other quark models and with lattice data for
  the fluctuations.
\end{abstract}



\maketitle

\tableofcontents

\section{Introduction}
\label{sec:introduction}

Quantum chromodynamics (QCD) is the fundamental non-Abelian gauge
theory of strong interactions in terms of $ 2 N_c N_f $ quarks and
antiquarks and $2(N_c^2-1)$ gluons with $N_c=3$ the number of colors
and $N_f=6 $ the number of flavour species $u,d,s,c,b,t$. Quark and
gluon confinement requires all physical states to be color singlet,
but are the hadronic states a complete set of eigenstates of QCD
spanning the Hilbert space ${\cal H}_{\rm QCD}$? This question is
related to the validity and meaning of quark-hadron duality.

In the case of $N_f=3$ flavors, which will be assumed throughout the
paper, the stable and low-lying bound states, such as the baryon octet
and the pseudoscalar nonet are unambiguously part of the discrete
spectrum.\footnote{Finite stable nuclei and anti-nuclei such as ${\rm
    d}$, $\overline{\rm d}$, $^3 {\rm H}$, $^3 \overline{\rm H}$, $^3
  {\rm He}$,$^3 \overline{\rm He}$, $^4 {\rm He}$,$^4 \overline{\rm
    He}$, etc. are also a part of the spectrum, despite being weakly
  bound states on a hadronic scale.} The remaining states belong to
the continuum spectrum and are experimentally spotted in strong
hadronic reactions, interpreted as unstable resonances and
characterized by a mass and a width. They have been reported over the
years by the Particle Data Group (PDG)
booklet~\cite{Patrignani:2016xqp} as single states rated with
*,**,***,**** (for baryons) depending on the increasing confidence on
their existence. However, are the PDG states complete and if so in
what sense would they be complete?

For such resonance states, the verification of completeness for QCD in
the continuum is subtle since within a Hamiltonian perspective they are not
proper (normalizable) eigenstates of the QCD Hamiltonian, and besides,
they correspond to unconventional representations of the Poincar\'e
group~\cite{Bohm:2004zi}. Lattice QCD presents the clear advantage
that in a finite box all states are discretized and hence become
countable; below a certain maximal mass the number of states is finite
as long as the volume $V$ remains finite (typically $ V \sim (2-3\,\mathrm{fm})^3$).  Resonances are extracted from those particular energy
levels which become insensitive to the box size.  The extent to which
these states play a key role in the completeness issue is uncertain
since, although there is a larger concentration of states around the
resonance, in the bulk, the mass separation is $\Delta M \sim
V^{-1/3}$.

\begin{figure*}[htp!]
\centering
 \begin{tabular}{c@{\hspace{2.5em}}c}
 \includegraphics[width=0.43\textwidth]{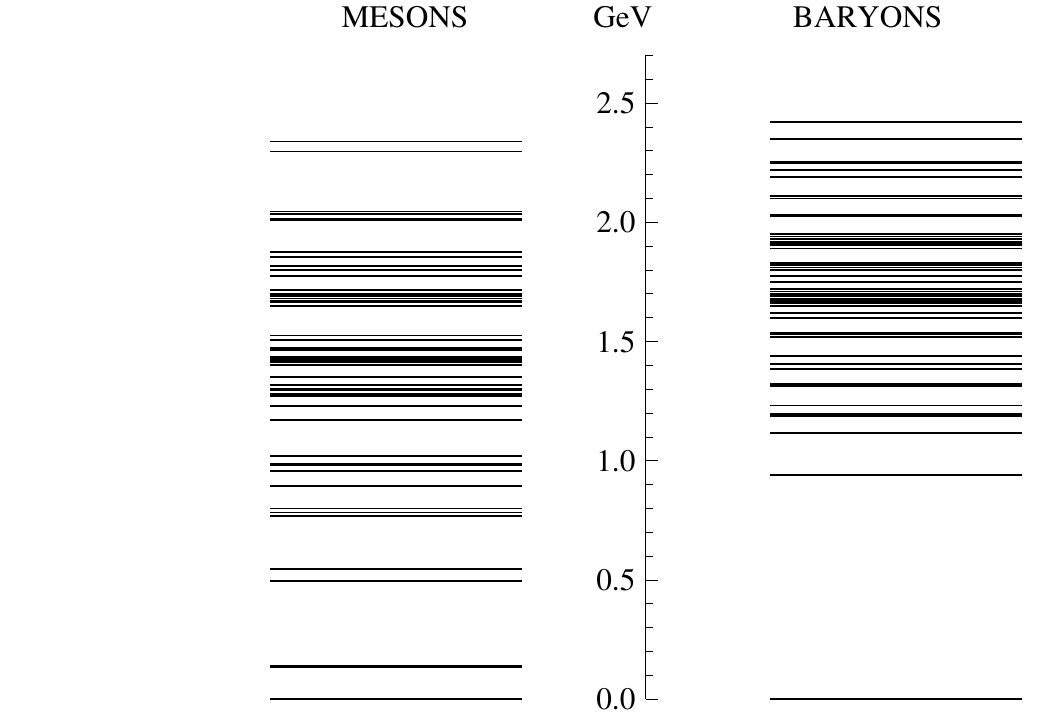} &
 \includegraphics[width=0.43\textwidth]{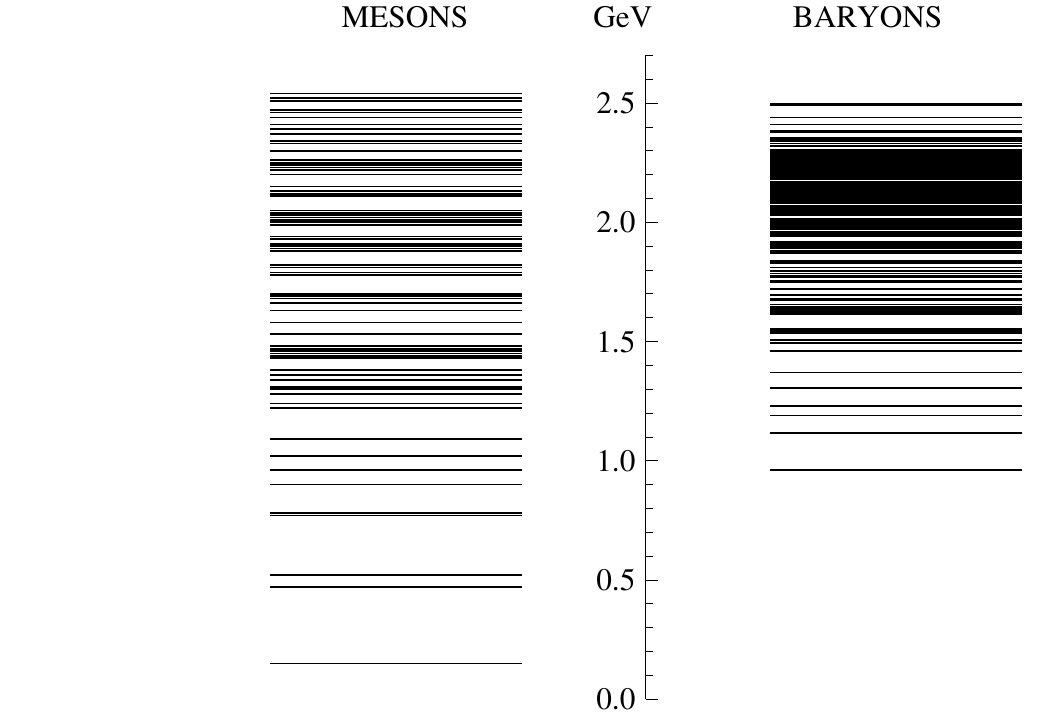} 
\end{tabular}
%\vspace{-0.4cm}
 \caption{Left panel: Mesons and baryons spectrum made of $u$, $d$ and
   $s$ quarks from the PDG~\cite{Tanabashi:2018oca} (left panel), and
   from the Relativized Quark
   Model~\cite{Godfrey:1985xj,Capstick:1986bm} (right panel).}
\label{fig:spectrum}
\end{figure*}

\begin{figure*}[htb!]
\centering
\begin{tabular}{c@{\hspace{4.5em}}c}
\includegraphics[width=0.43\textwidth]{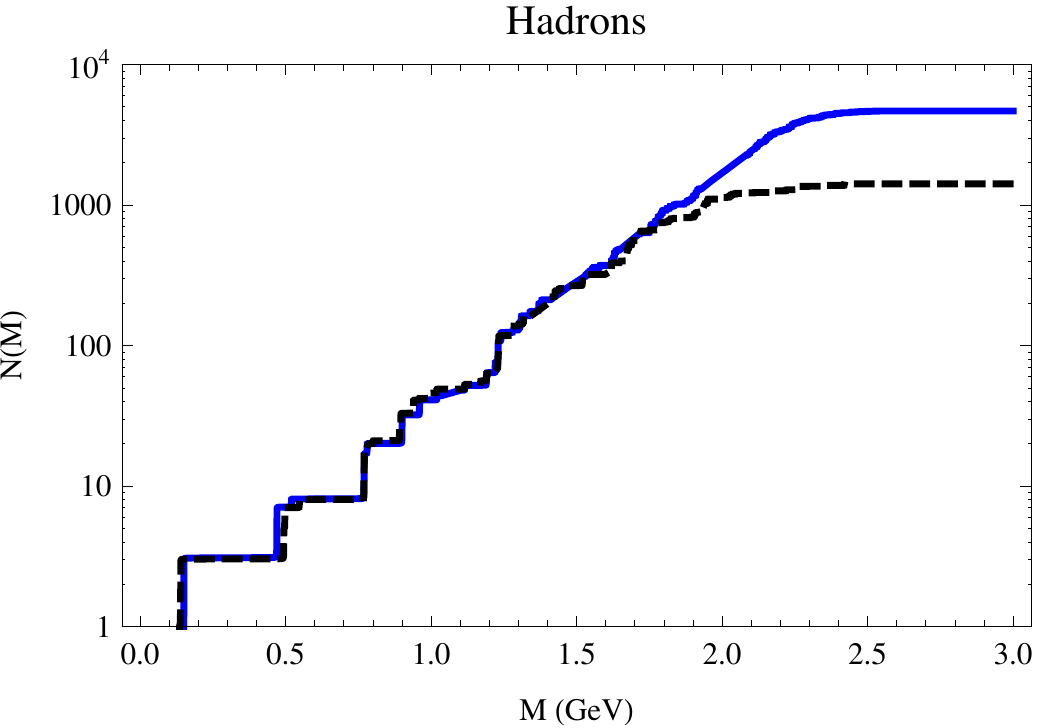} &
\includegraphics[width=0.43\textwidth]{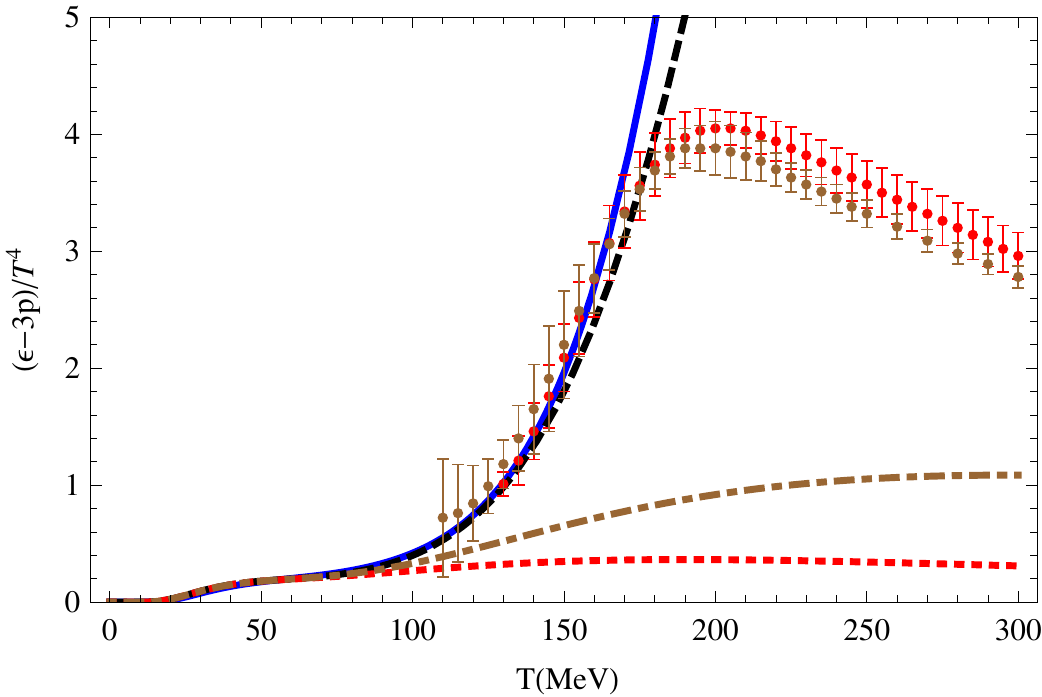} 
\end{tabular}
\vspace{-0.4cm}
 \caption{Left panel: Cumulative number for the
   PDG~\cite{Tanabashi:2018oca} (dashed line) and the RQM
   (solid)~\cite{Godfrey:1985xj,Capstick:1986bm}. Right panel: Trace
   anomaly as a function of temperature in lattice
   QCD~\cite{Borsanyi:2013bia,Bazavov:2014pvz} vs HRG using PDG
   (dashed) and RQM (solid) spectra. We also plot just the
   contribution of states with $M < 0.6 \GeV$ (dotted) and $M < 0.8
   \GeV$ (dotted-dashed).  }
\label{fig:trace_anomaly}
\end{figure*}

In recent years, the thermodynamic approach to strong interactions
pioneered by Hagedorn~\cite{Hagedorn:1965st,Hagedorn:1984hz} where the
vacuum is represented by a non-interacting Hadron Resonance Gas (HRG),
has emerged as a practical and viable path to establish completeness
of hadronic states on a quantitative level in the hadronic phase. With
all the provisos regarding the nature of resonance and bound states,
the most impressive and vivid verification has been the recent study
of the trace anomaly, $(\epsilon-3P)/T^4$ with $\epsilon$ energy
density and $P$ the pressure. It was computed directly in lattice QCD
by the Wuppertal-Budapest (WB) and the HotQCD
collaborations~\cite{Borsanyi:2013bia,Bazavov:2014pvz} and the HRG
model using the most recent compilation of the PDG states taking just
their masses, i.e. assuming zero widths~\cite{Patrignani:2016xqp} and
for temperatures below $T \sim 170 \MeV$. Remarkably, a similar degree
of success is achieved within uncertainties~\cite{Arriola:2014bfa}
(see Figs.~\ref{fig:spectrum} and \ref{fig:trace_anomaly}) using the
hadronic spectrum obtained in the Relativized Quark Model (RQM) of
Capstick, Godfrey and Isgur~\cite{Godfrey:1985xj,Capstick:1986bm}
which predated the lattice QCD calculations by about 30 years. Finite
width effects on the PDG (PDG-$\Gamma$) naturally provide a shift
towards lower masses~\cite{Arriola:2012vk,Broniowski:2016hvt} as a
consequence of the mass spectrum spread weighted by the exponentially
decreasing Boltzmann factor.

This remarkable agreement becomes significantly spoiled when
susceptibilities involving conserved charges such as the baryon
number, the electric charge and the strangeness are considered
\cite{Borsanyi:2011sw,Bazavov:2012jq}. In particular, the differences
between PDG, PDG-$\Gamma$ and the RQM become more visible and an excess of
baryonic states as compared to the lattice QCD results is
observed~\cite{RuizArriola:2016qpb} illustrating the so-called {\it
  missing resonance problem}.

In fact, since the early days of the quark model the extreme abundance
of predicted and experimentally missing baryonic resonances has been a
major cause of concern both at the theoretical as well as at the
experimental level~\cite{Capstick:1992uc} (see also \cite{Hey:1982aj}
and \cite{Capstick:2000qj} for reviews and references therein).
Possible ways out of the difficulties have traditionally been
attributed either to a weak coupling of the predicted states to the
particular production process (photo-production, $\pi N$ scattering
etc.) or to a dynamical reduction of degrees of freedom due to diquark
clustering.  

The suspicion that the baryonic spectrum can be understood in terms of
quark-diquark degrees of freedom is rather old (see
e.g. Ref.~\cite{Anselmino:1992vg} for a review, but also
Ref.~\cite{Klempt:2017lwq} for evidence against it). This includes
diquark clustering studies~\cite{Fleck:1988vm},
non-relativistic~\cite{Santopinto:2004hw} and
relativistic~\cite{Ferretti:2011zz,Gutierrez:2014qpa} analyses where
scalar and axial-vector diquarks have a mass of about $600\MeV$ and
$800\MeV$ respectively (the diquark mass difference seems quite model
independent and about $200\MeV$). As expected, in diquark models many
states predicted by the quark model do not
appear~\cite{Capstick:1986bm}. More specifically, while not strictly
forbidden, the relativistic diquark
model~\cite{Ferretti:2011zz,Gutierrez:2014qpa} does not predict any
missing states below $2\GeV$, whereas Isgur and Capstick have
predicted five unobserved states~\cite{Capstick:1986bm}.  Lattice QCD has
also provided insights, as some evidence on diquarks correlations in
the nucleon~\cite{Alexandrou:2006cq} and the dominance of the scalar
diquark channel~\cite{DeGrand:2007vu} have been reported. Moreover,
the diquark-approximation has been found to work well in
Dyson-Schwinger and Faddeev equations studies~\cite{Eichmann:2016hgl}.
The radial Regge behavior found in the relativistic quark-diquark
picture from a numerical study of the spectrum of the relativistic
two-body problem~\cite{Ebert:2011kk} has been confirmed from a direct
PDG analysis when the widths of the resonances are implemented in the
analysis~\cite{Masjuan:2017fzu,RuizArriola:2017ggc}.

While the missing resonance problem has attracted a lot of interest
both experimentally as well as theoretically, much of the discussion
is focused on the {\it individual} one-to-one mapping of resonance
states which have a mass spectrum and which are produced with
different backgrounds. In contrast, the thermodynamic approach offers
the possibility to perform a more global analysis where many of the
fine details will hopefully be washed out due to the presence of the
heat bath. In the present paper we profit from the new perspective
provided by lattice QCD at finite temperature, based on separation of
quantum numbers with the study of susceptibilities of conserved
charges, where a combination of degeneracy and level density is
involved. We try to answer the question whether or not quark-diquark
baryonic states saturate the baryonic susceptibilities below the
deconfinement crossover as compared to the available lattice QCD
calculations~\cite{Borsanyi:2011sw,Bazavov:2012jq}. Aspects of
quark-hadron duality at finite temperature have been discussed in
pedagogical way in Ref.~\cite{Arriola:2014bfa}. Actually, in a recent
work it has been discussed how quark-hadron duality in deep inelastic
scattering for baryons suggests an asymptotic quark-diquark spectrum
with a linearly rising potential for scaling to hold in the structure
functions~\cite{Masjuan:2017fzu,RuizArriola:2017ggc}. Motivated by
this we want to establish if a similar pattern holds also at finite
temperature.

The paper is organized as follows. In Section~\ref{sec:HRG} we review
the relevant aspects of the HRG model from the point of view of the
Equation of State and the trace anomaly as compared to lattice QCD.
In Section~\ref{sec:Fluctuations} we analyze the implications for
fluctuations of conserved charges at finite temperature in the vacuum.
then analyze in Section~\ref{sec:VqD_potential} the quark-diquark
potential obtained as a correlation function involving Polyakov
loops. This allows to define our model and compute the spectrum by
diagonalization in Section~\ref{sec:Model} where the
susceptibilities are analyzed in terms of the free parameters of the
theory. Finally in Section~\ref{sec:Conclusions} we come to the
conclusions. In the appendices we provide details on the semiclassical
determination of the spectrum and also prove a theorem on the sign of
susceptibilities which is verified by lattice calculations.

\begin{figure*}
  \centering
\begin{tabular}{c@{\hspace{4.5em}}c}
  \includegraphics[width=0.43\textwidth]{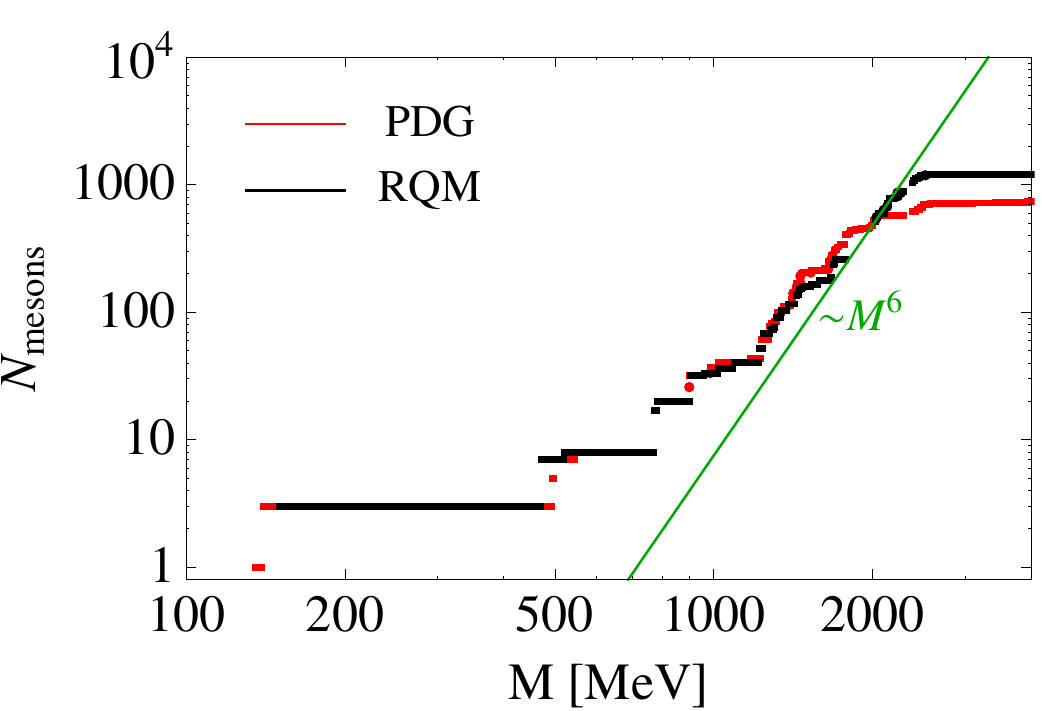} &
\includegraphics[width=0.43\textwidth]{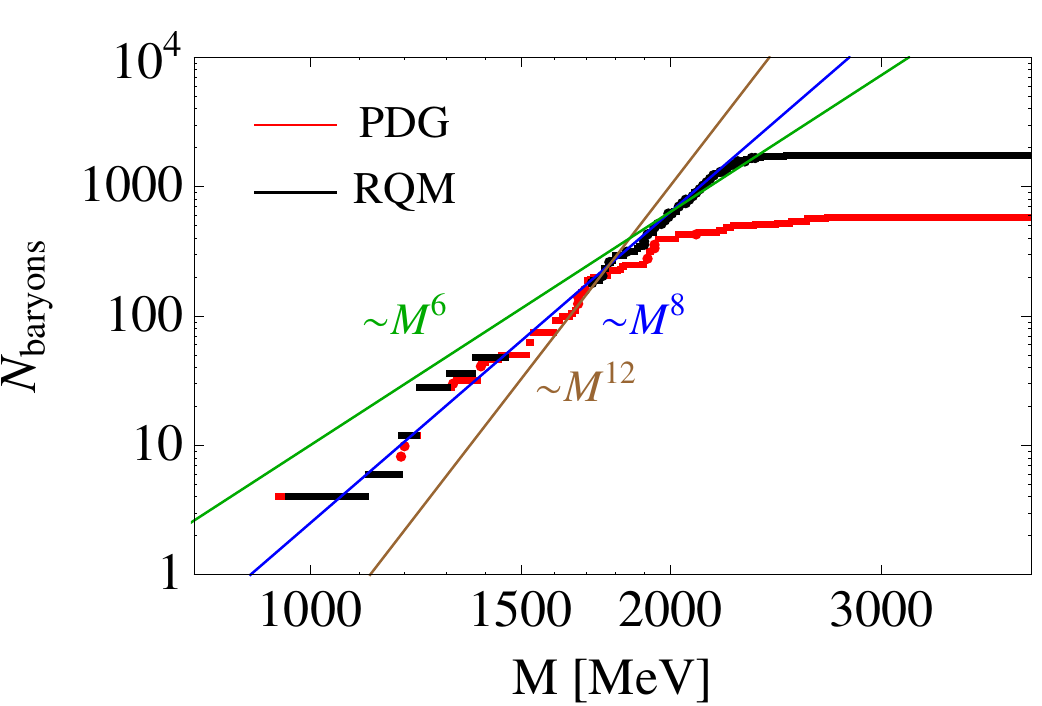} 
\end{tabular}
\vspace{-0.4cm}
\caption{Cumulative numbers for the PDG (red line) and the RQM
  (black line). Left panel: mesonic states (log-log
  scale). Right panel: Baryonic states (log-log scale).}
\label{fig:nucum-meson-baryon}
\end{figure*}

\section{Completeness and thermodynamic equivalence}
\label{sec:HRG}

\subsection{QCD spectrum and thermodynamics}

The cumulative number of states may be used as a
characterization of the QCD
spectrum~\cite{Broniowski:2000bj,Broniowski:2004yh}. It is defined as
the number of bound states below some mass $M$, i.e.
\begin{equation}
N(M) = \sum_i g_i \, \Theta(M-M_i) \,,
\label{eq:ncum}
\end{equation}
where $g_i$ is the degeneracy factor, $M_i$ is the mass of the $i$-th
hadron and $\Theta(x)$ is the step function, so that the density of
states is given by~$\rho(M) = dN(M)/dM$.  A way to provide a practical
meaning of the previous equation is to use finite box periodic
(antiperiodic) boundary conditions for gluon/quark fields. In such a
case {\it all states} contribute on equal footing. In such setting,
stable bound states correspond to eigenvalues which do not depend on
the volume of the box for sufficiently large boxes, whereas unstable
resonance states correspond to eigenvalues which are volume
independent within a given volume interval~~\cite{Briceno:2015rlt}.
Colour neutral eigenstates of the QCD Hamiltonian and in fact excited
states have been determined on the lattice in this
way~\cite{Edwards:2012fx}. Despite its importance, the cumulative
number of states, Eq.~(\ref{eq:ncum}) has never been evaluated
directly in QCD.

Within a thermodynamic setup, the issue of completeness acquires a
precise meaning where the global aspects of the spectrum rather than
individual features are highlighted.  The partition function of QCD is
given by the standard relation
\begin{equation}
Z_{\QCD} = \Tr \, e^{-H_{\QCD}/T}= \sum_n e^{-E_n/T} \,, \label{eq:Zqcd}
\end{equation}
and is the fundamental quantity to study the thermodynamic properties of the
theory. Written in terms of the eigenvalues of the QCD Hamiltonian,
i.e.~$H_{\QCD} \psi_n = E_n \psi_n$, Eq.~(\ref{eq:Zqcd}) illustrates the
relation between the thermodynamics of the confined phase and the spectrum of
QCD. Although $Z_{\rm QCD}$ has been determined independently as an Euclidean
path integral on the lattice~\cite{Borsanyi:2011sw,Bazavov:2012jq}, to our
knowledge the energy levels contribution to Eq.~(\ref{eq:Zqcd}) has not been
tested explicitly against the partition function.\footnote{As is well known
  reflection positivity of the Euclidean action on the lattice guarantees the
  existence of a transfer matrix and hence of a Hamiltonian
  ~\cite{Osterwalder:1977pc}; the issue is to list the pertinent set of
  eigenvalues.} Actually, the primary quantity is the trace
anomaly~\cite{Borsanyi:2011sw,Bazavov:2012jq}
\begin{eqnarray}
\Delta \equiv \frac{\epsilon - 3 P}{T^4} 
= T^5 \partial_T \left( \frac{P}{T^4} \right) 
\end{eqnarray}
whence the equation of state (EoS) can be obtained by integration of
the rhs of this equation with suitable boundary conditions and
physical quantities such as the entropy density $s = \partial_T P$ or
the sound velocity $c_s^2 =\partial P /\partial \epsilon \equiv
\partial_T P /\partial_T \epsilon$ may be determined.

From this thermodynamic point of view it is fair to say that the
completeness of the QCD-spectrum remains to be checked in practice.

%Before presenting our analysis, it is instructive to review
%some relevant issues of the HRG model motivating the present study.  

\subsection{Hadron Resonance Gas model: the PDG and RQM spectra}

The completeness of the listed PDG states~\cite{Patrignani:2016xqp} is
equally a subtle issue.  On the one hand they are mapped into the $q
\bar q$ and $qqq$ quark model states. On the other hand, most reported
states by the PDG are not stable particles but resonances which are
produced as intermediate steps in a variety of scattering processes.

The HRG model was originally proposed by
Hagedorn~\cite{Hagedorn:1984hz}. In spirit, this is valid under the
assumption that physical quantities in the confined phase of QCD admit
a representation in terms of hadronic states, which are considered as
stable, non-interacting and point-like particles. Within this approach
the EoS of QCD is described in terms of a gas of non-interacting
hadrons~\cite{Hagedorn:1984hz,Tawfik:2004sw}, and the grand-canonical
partition function turns out to be
\begin{equation}\begin{split}
\log Z_\HRG   &=
 -V \int \frac{d^3p}{(2\pi)^3}  \sum_{i \in {\rm Hadrons}}  \zeta_i g_i 
\\ & \quad\times
\log\left( 1 - \zeta_i e^{-(\sqrt{p^2 + M_i^2} - \sum_a  \mu_a q_a^i )/T} \right) , 
\end{split}
 \label{eq:Zhrgm}
\end{equation}
with~$\zeta_i = \pm 1$ for bosons and fermions respectively. We
consider several conserved charges labeled by the index~$a$, with
$q_a^i$ the charge of the $i$-th hadron for symmetry $a$, and $\mu_a$
the chemical potential associated with this symmetry. The obvious
consequence is that a good understanding of the spectrum of QCD turns
out to be crucial for a precise determination of the thermodynamic
properties of this theory.

In QCD, the quantized energy levels are the masses of low-lying and
stable colour singlet states, which are commonly identified with
mesons $[q\bar{q}]$ and baryons $[qqq]$ in the quark model.  As
already mentioned, unstable hadronic resonances are not proper
eigenstates which in the HRG model are regarded as bound states.  We
reiterate that while there is currently a large scale on-going effort
to determine hadronic resonances from lattice QCD by means of the
L\"uscher formula~\cite{Edwards:2012fx} (see
e.g. \cite{Briceno:2015rlt} for a review) these states are not
numerous enough to verify the HRG model.

So far, the states listed by PDG echo the standard quark model
classification for mesons~$[q{\bar q}]$ and baryons~$[qqq]$. Then, it
would be pertinent to consider also the Relativized Quark Model (RQM)
for mesons~\cite{Godfrey:1985xj} and
baryons~\cite{Capstick:1986bm}.\footnote{The consideration of this
  model is motivated not only by its success concerning the
  thermodynamic equivalence with the PDG (see below), but also on the
  fact that this is done with a comparable number of parameters as in
  QCD itself.} We show in Fig.~\ref{fig:spectrum} the hadron spectrum
with the PDG compilation (left) and the RQM spectrum (right). The
comparison clearly shows that there are {\it further states} in the
RQM spectrum above some scale $M > M_{\min}$ that may or may not be
confirmed in the future as mesons or hadrons. In addition there could
be exotic, glueballs or hybrids states, predicted by other hadronic
models.

The contribution to $N(M)$ obtained {\it just} adding the $q \bar q$ (mesons),
$qqq$ (baryons) and $\bar{q}\bar{q}\bar{q}$ (antibaryons) components
as\,\footnote{Antibaryons $\bar{q}\bar{q}\bar{q}$ contribute to the cumulative
  number in the same amount as baryons $qqq$, i.e. $
  N_{[\bar{q}\bar{q}\bar{q}]}(M) = N_{[qqq]}(M)$. We display Eq.~(\ref{eq:NM})
  by distinguishing between them for clarity.}
\begin{equation}
N (M) = N_{[q{\bar q]}}(M) + N_{[qqq]}(M) + N_{[\bar{q}\bar{q}\bar{q}]}(M)  \,, \label{eq:NM}
\end{equation}
can be evaluated either from the PDG booklet~\cite{Patrignani:2016xqp}
or alternatively from the RQM of Capstick, Godfrey and
Isgur~\cite{Godfrey:1985xj,Capstick:1986bm}.
The resemblance of both schemes below $1.7\GeV$, see
Fig.~\ref{fig:trace_anomaly} (left), is noteworthy, particularly if we
take into account the 30 years elapsed between the RQM and the current
PDG. Finite width effects on the PDG (PDG-$\Gamma$) have been
estimated ~\cite{Arriola:2012vk,Broniowski:2016hvt} and naturally
provide a shift towards lower masses as a consequence of the mass
spectrum spread. The flattening of the curves indicates lack of
reported states on the PDG side, as well as a the higher computational
cut-off $M_{\rm high}=2.3\GeV$ in the numerical calculation on the RQM
side.

This resemblance is also realized at the
level of the trace anomaly which in the HRG model is given by
\begin{equation}
  \Delta_{\rm HRG} (T) =
\frac{1}{T^4}\sum_{i \in {\rm Hadrons}} \!\! g_i \int \frac{d^3 p}{(2\pi)^3} \frac{E_i(p)- \vp
  \cdot \nabla_p E_i (p)}{e^{E_i(p)/T} - \zeta_i} \, ,
\label{eq:tran-HRG}
\end{equation}
and compared with lattice QCD by the
Wuppertal-Budapest (WB) and the HotQCD
collaborations~\cite{Borsanyi:2013bia,Bazavov:2014pvz} in
Fig.~\ref{fig:trace_anomaly} (right), suggesting the thermodynamic equivalence
of QCD, PDG and RQM below the crossover.

\subsection{Missing states}

Given the resemblance of the PDG to the RQM we may speculate on the
nature of mesonic and baryonic states {\it separately}. While the PDG
is a consented compilation of numerous analyses, the RQM corresponds
{\it by construction} to a solution of the quantum mechanical problem
for both $q \bar q$-mesons and $qqq$-baryons. For colour-singlet
states the $n$-parton Hamiltonian takes the schematic form
\begin{eqnarray}
H_n = \sum_{i=1}^n \sqrt{p_i^2+m^2} + \sum_{i< j}^n v_{ij}(r_{ij}) \, .
\end{eqnarray}
Neglecting spin dependent contributions, not essential in the
following argument, and assuming Casimir scaling, the two-body interactions
take the form
\begin{eqnarray}
v_{q \bar q} = \sigma r - \frac{4 \alpha_S}{3 r}  = (N_c-1) v_{qq}
\,.
\label{eq:8}
\end{eqnarray}  

Asymptotic estimates may be undertaken to sidestep the actual numerical
evaluation of the eigenvalues. Specifically a semiclassical expansion
describes the high mass spectrum for systems where interactions are dominated
by linearly rising potentials with a string tension $\sigma$, in the range $M
\gg \sqrt{\sigma}$.  Such an approach relies on a derivative expansion of the
cumulative number of states for a given
Hamiltonian~\cite{Caro:1994ht}.\footnote{Unfortunately the derivative
  expansion involves smoothness assumptions for the Hamiltonian which are not
  met in detail by the Coulomb potential, the $\sigma r$ potential or the
  relativistic kinetic energy in the massless case, therefore a distortion in
  the semiclassical counting could arise beyond the leading order.}  At
leading order the number of states below a certain mass $M$ takes the form
\begin{eqnarray}
N_n (M) &\sim& g_n \int \prod_{i=1}^n \frac{d^3 x_i d^3 p_i}{(2\pi)^3}
\, \delta\!\!\left({\textstyle \sum_{i=1}^n \vx_i}\right) 
\delta\!\!\left({\textstyle \sum_{i=1}^n \vp_i}\right) \nonumber \\
&\times& \theta (M-H_n(p,x)) \,,
\label{eq:ncum-n}
\end{eqnarray}
where $g_n$ takes into account the degeneracy.

For the sake of the argument, let us neglect the Coulomb term in (\ref{eq:8}),
thus $v(r)=\sigma\,r$, as well as the current quark masses.  In this case, a
dimensional argument, $p\to M p$, $r\to Mr/\sigma$, gives
\begin{eqnarray}
N_n (M) 
\sim \left(\frac{M^2}{\sigma}\right)^{3 n-3}
.
\end{eqnarray}
Using these techniques, one can predict that the large mass expansion
of these contributions is~$N_{[q{\bar q}]} \sim M^6$, $N_{[qqq]} \sim
M^{12}$, $N_{[q{\bar q}q{\bar q}]} \sim M^{18} \,,$
etc~\cite{Arriola:2014bfa}.
\ignore{ with corrections down by two relative
powers in $M$. Besides being asymptotic estimates based on linearly
confining interactions}

The separate contributions of mesons and baryons are presented in
Fig.~\ref{fig:nucum-meson-baryon} on a log-log scale where we see that
again PDG and RQM largely agree and present an approximate linear
behaviour on this scale, indicating as expected a power behaviour.
However, while in the meson case the $ N_{[q{\bar q}]} \sim M^6 $
seems to conform with the asymptotic estimate, in the baryon case much
lower powers, $M^{6}-M^8$, than the expected $ N_{[qqq]} \sim M^{12} $
are identified. We note that $M^6$ suggests a similar two-body
behavior as in the case of mesons (with a linearly rising
potential). We take this feature as a hint that the $qqq$ excited
spectrum effectively conforms to a two body system of particles
interacting with a linearly growing potential, as we will analyze
below in detail in Section~\ref{subsec:Spectrum}.

\section{Fluctuations of conserved charges in a thermal medium}
\label{sec:Fluctuations}

While mesonic and baryonic contributions can be explicitly
distinguished the thermodynamic separation of mesons and baryons in
the EoS cannot be done at the QCD level. In order to achieve directly
such a separation for baryons in particular we analyze fluctuations
containing at least one baryonic charge as well as charge and
strangeness, the three of them being conserved charges in strong
interactions.

Conserved charges $[Q_a,H]=0$ play a fundamental role in the
thermodynamics of QCD. In the ({\it uds}) flavor sector of QCD the
conserved charges are the electric charge~$Q$, the baryon number~$B$,
and the strangeness~$S$. While their thermal expectation values in the
hot vacuum are vanishing, i.e.  in the absence of chemical potentials
~$\langle Q_a\rangle_T = 0$, where $Q_a \in \{ Q, B, S \}$, they
present statistical fluctuations characterized by
susceptibilities~\cite{Bazavov:2012jq,Bellwied:2015lba,Asakawa:2015ybt,RuizArriola:2016qpb}
\,\footnote{One can also work in the quark-flavor basis, $Q_a \in \{
  u, d, s\}$, where $u$, $d$ and $s$ is the number of up, down and
  strange quarks. In this basis~$B = \frac{1}{3}(u+d+s)$, $Q =
  \frac{1}{3}(2u-d-s)$ and $S= -s$, }
\begin{equation}
\chi_{ab}(T) \equiv \frac{1}{V T^3} \langle \Delta Q_a \Delta
Q_b \rangle_T \,, \quad \Delta Q_a = Q_a - \langle Q_a \rangle_T \,.
\end{equation}
The susceptibilities can be computed from the grand-canonical partition
function by differentiation with respect to the chemical potentials, i.e.
\begin{equation}
\langle Q_a \rangle_T = 
-\frac{\partial \Omega}{\partial\mu_a} %\Bigg|_{\mu_a = 0} 
 \,, 
\qquad  
\langle \Delta Q_a \Delta Q_b \rangle_T = 
-T \frac{\partial^2 \Omega}{\partial\mu_a \partial\mu_b}  %\Bigg|_{\mu_a = 0 = \mu_b} 
\,,  
\label{eq:susc}
\end{equation}
where~$\Omega = -T \log Z$ is the thermodynamical potential.

\begin{figure*}[htb]
\centering
 \begin{tabular}{c@{\hspace{4.5em}}c@{\hspace{4.5em}}c}
 \includegraphics[width=0.26\textwidth]{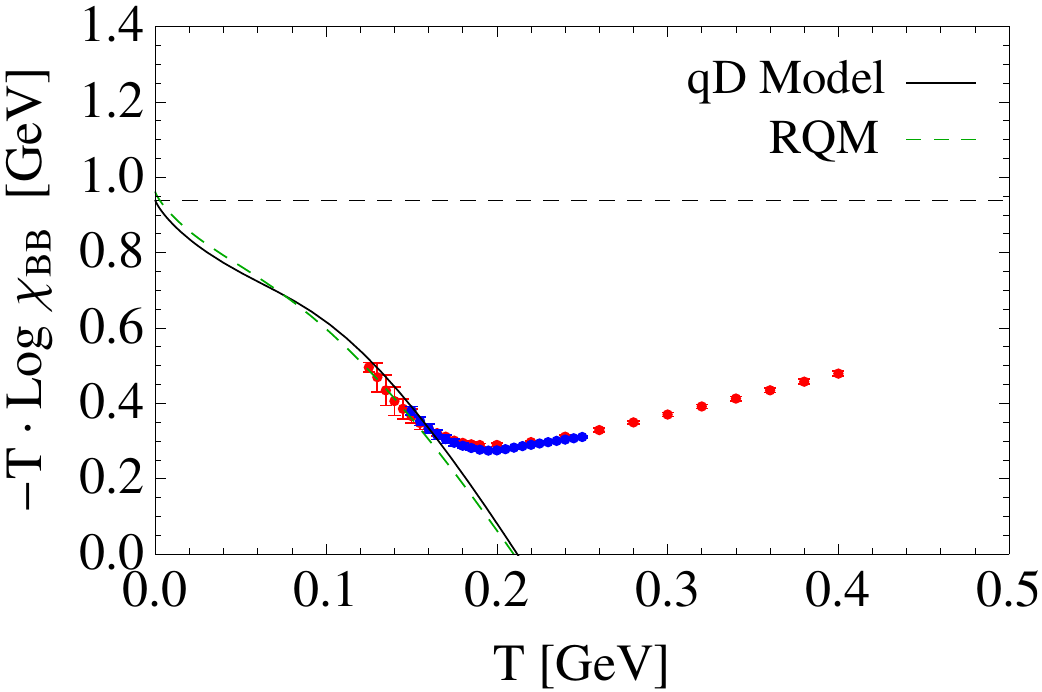} &
 \includegraphics[width=0.26\textwidth]{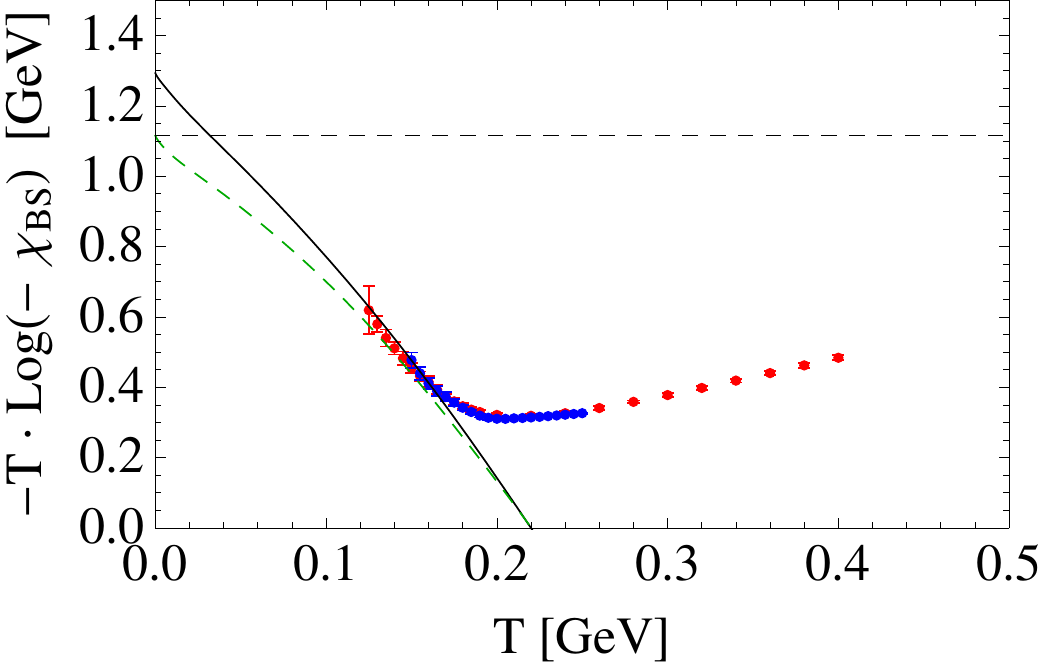} &
 \includegraphics[width=0.26\textwidth]{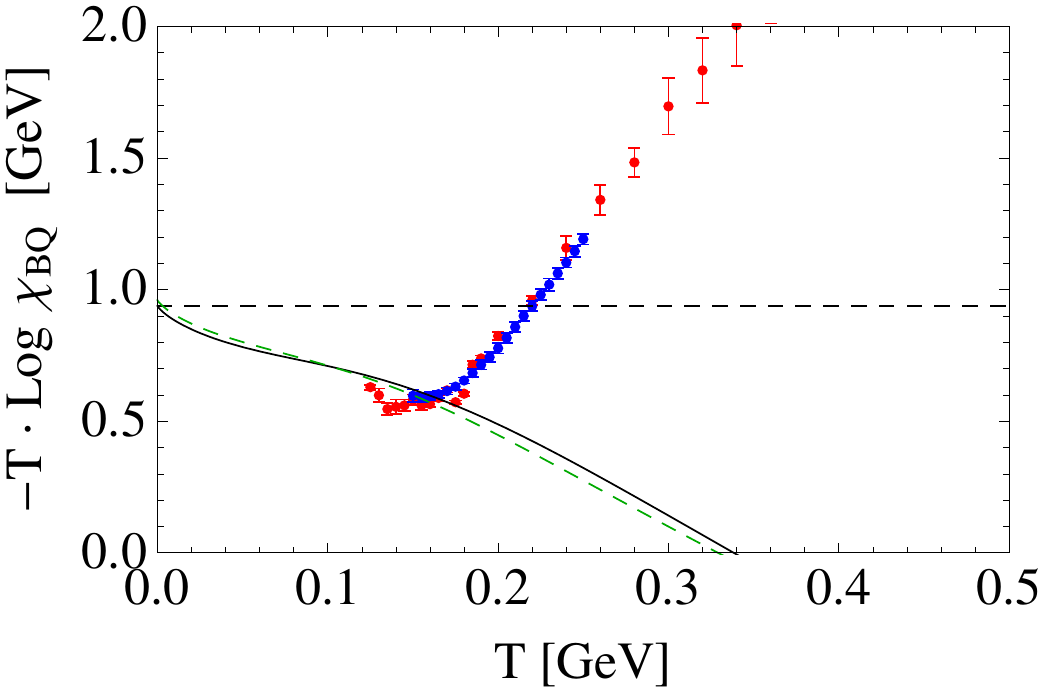} \\
 \includegraphics[width=0.26\textwidth]{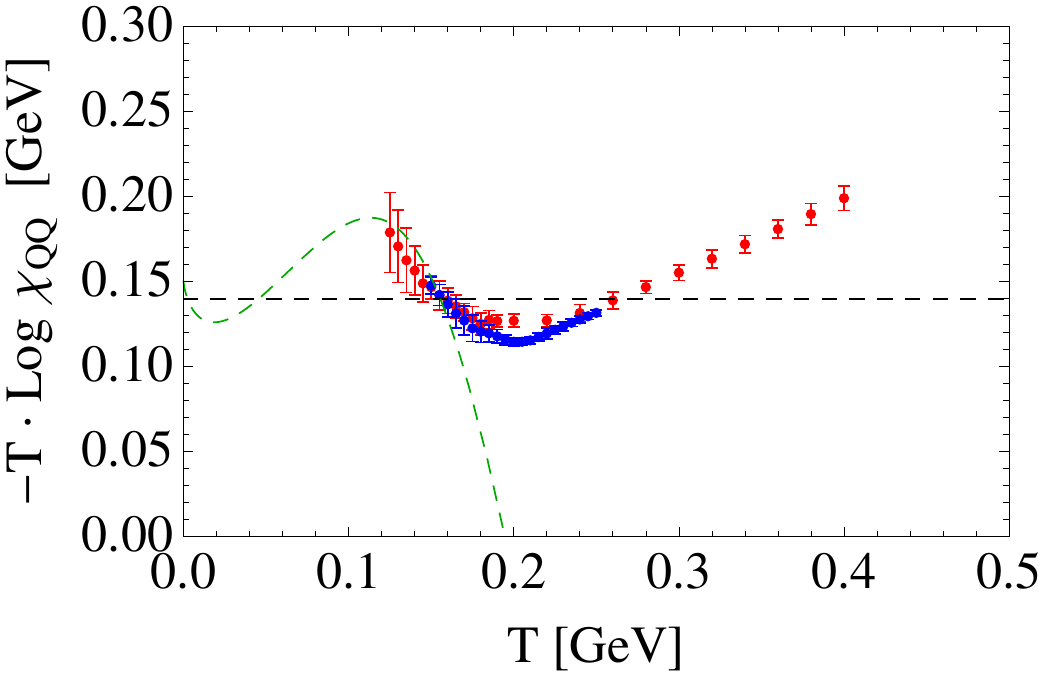} &
 \includegraphics[width=0.26\textwidth]{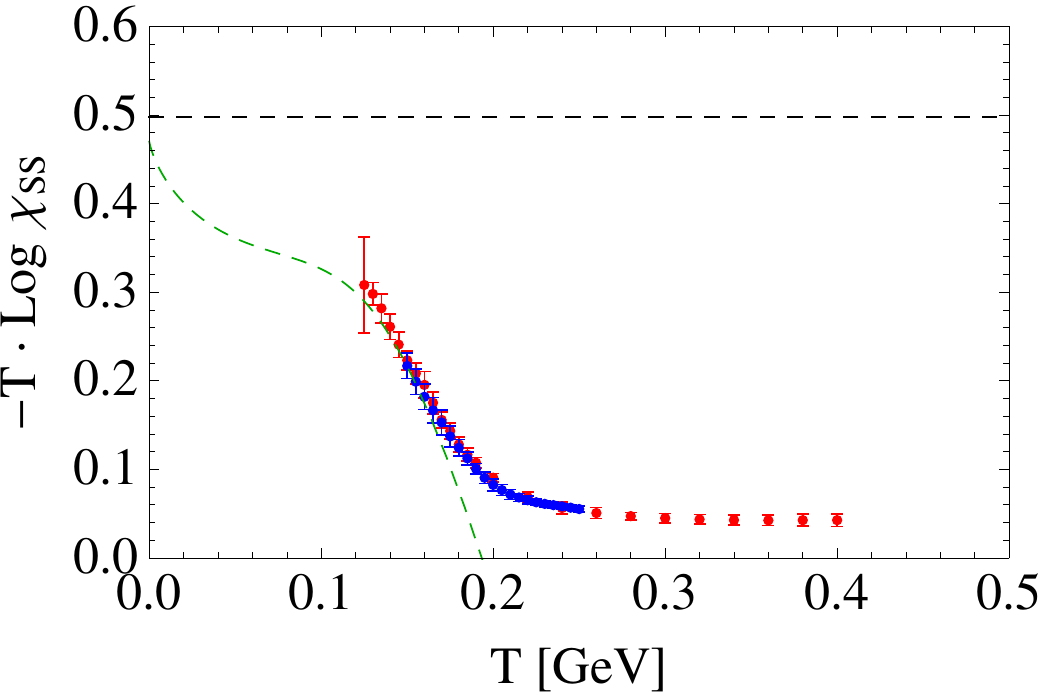} &
 \includegraphics[width=0.26\textwidth]{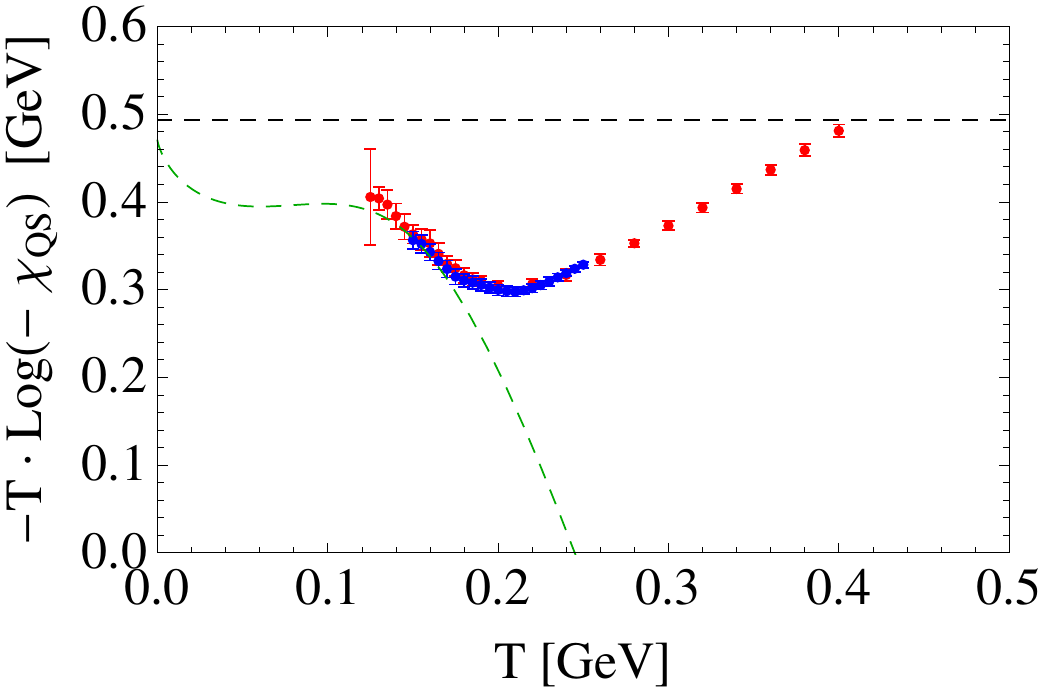}
\end{tabular}
%\vspace{-0.4cm}
 \caption{Plot of $-T \log |\chi_{ab}|$ as a function of temperature. We
   display as dots the lattice data from Refs.~\cite{Bazavov:2012jq} (blue)
   and \cite{Borsanyi:2011sw} (red). We also display the HRG model results
   including the spectrum of the RQM~\cite{Godfrey:1985xj,Capstick:1986bm}
   (dashed green) and the baryon spectrum from the quark-diquark model
   computed in Sec.~\ref{subsec:Spectrum} (solid black). Horizontal dashed
   lines represent the values of the lowest-lying states contributing to the
   fluctuations, as it is shown in Eq.~(\ref{eq:Chi_M0}).  }
\label{fig:LogChi1}
\end{figure*}

QCD at high temperature behaves as an ideal gas of quarks and gluons. In this
limit the susceptibilities approach to
\begin{equation}
\chi_{ab} = \frac{N_c}{3}\sum_{i=1}^{N_f} q^i_a q^j_b  \,;
\end{equation}
hence for $N_c=3$ and flavors $u,d,s$
\begin{equation}\begin{split}
\chi_{BB}(T)& \to 1/3 \,, \quad\!\!\!\!\!\!
\chi_{BQ}(T)  \to 0\,,  \qquad
\chi_{BS}(T)  \to -1/3  \,, \\
\chi_{SS}(T)& \to 1 \,, \quad
\;\, \chi_{QS}(T)  \to 1/3\,,  \quad
 \chi_{QQ}(T)  \to 2/3  \,.
\end{split}\end{equation}
Within the HRG approach, the charges are carried by various species of
hadrons, $Q_a = \sum_i q^i_a N_i$, where $q_a^i \in \{ Q_i , B_i , S_i \}$,
and $N_i$ is the operator number of hadrons of type~$i$. By using in
Eq.~(\ref{eq:susc}) the thermodynamic potential of this model,
cf. Eq.~(\ref{eq:Zhrgm}), one gets
\begin{equation}\begin{split}
\chi_{ab}(T) =   
 \frac{1}{2\pi^2}  \sum_{i \in {\rm Hadrons}}
\!\! g_i \, q_i^a \, q_i^b \sum_{n=1}^\infty  \zeta_i^{n+1} \frac{M_i^2}{T^2} 
 \, K_2\!\!\left( \frac{nM_i}{T} \right) 
, 
\end{split}
\label{eq:chi_HRGM}
\end{equation}
where $K_2(z)$ refers to the Bessel function of the second
kind.\footnote{While these formulas display explicitly the Dirac-Fermi
  or Bose-Einstein nature of the hadronic states, in practice the
  first term, $n=1$, in the thermal sum suffices for the considered
  temperatures, so that quantum statistical effects are
  marginal.} This formula will be used to compute the baryonic
susceptibilities, namely $\chi_{BB}$, $\chi_{BQ}$ and
$\chi_{BS}$. Eq.~(\ref{eq:chi_HRGM}) predicts the asymptotic behavior
\begin{equation}
\chi_{ab}(T) \stackbin[T \to 0]{}{\sim} e^{-M_0/T}  \,,
\end{equation}
where $M_0$ is the mass of the lowest-lying state in the spectrum with quantum
numbers $a$ and $b$. Therefore
\begin{equation}
\begin{aligned}
&\chi_{BB}(T) \stackbin[T \to 0]{}{\sim} e^{-M_p/T}  \,, 
\quad\;\;\, \chi_{BQ}(T) \stackbin[T \to 0]{}{\sim} e^{-M_p/T}  \,,  \\
&\chi_{BS}(T) \stackbin[T \to 0]{}{\sim} e^{-M_{\Lambda^0}/T}  \,, 
\quad\,\, \chi_{SS}(T) \stackbin[T \to 0]{}{\sim} e^{-M_{K^\pm}/T}  \,, \\
&\chi_{QS}(T) \stackbin[T \to 0]{}{\sim} e^{-M_{K^\pm}/T}  \,, 
\quad \chi_{QQ}(T) \stackbin[T \to 0]{}{\sim} e^{-M_{\pi^\pm}/T}  \,,
\end{aligned} \label{eq:Chi_M0}
\end{equation}
where $M_p = 938\MeV$ is the proton mass, $M_{\Lambda^0}$ the mass of the
$\Lambda^0$ baryon, etc. These relations go beyond the HRG model, since for
each sector the lightest hadron must certainly saturate the QCD partition
function at low enough temperature. This observation makes it appealing to
plot the lattice data for the fluctuations in a logarithmic scale. These plots
are shown in Fig.~\ref{fig:LogChi1}.

Of the six susceptibilities only four are independent, on account of isospin
symmetry which requires $\chi_{uu}=\chi_{dd}$ and $\chi_{us}=\chi_{ds}$. On
the other hand, the following inequality holds in QCD for {\em degenerate}
$u,d$ flavors
\begin{equation}
\chi_{ud}(T) \le 0   %\quad \textrm{with} \quad  a \ne b 
\,,  
\label{eq:chiab_neg}
\end{equation}
at any temperature, even in the deconfined phase. This is proven in
App. \ref{app:B}. Because strange and light-quark flavors are not exactly
degenerate, the relation $\chi_{us}(T) \le 0$ does not follow as a QCD theorem
but it is nevertheless supported by the lattice results.

\begin{figure*}[htb]
\centering
 \begin{tabular}{c@{\hspace{4.5em}}c@{\hspace{4.5em}}c}
 \includegraphics[width=0.26\textwidth]{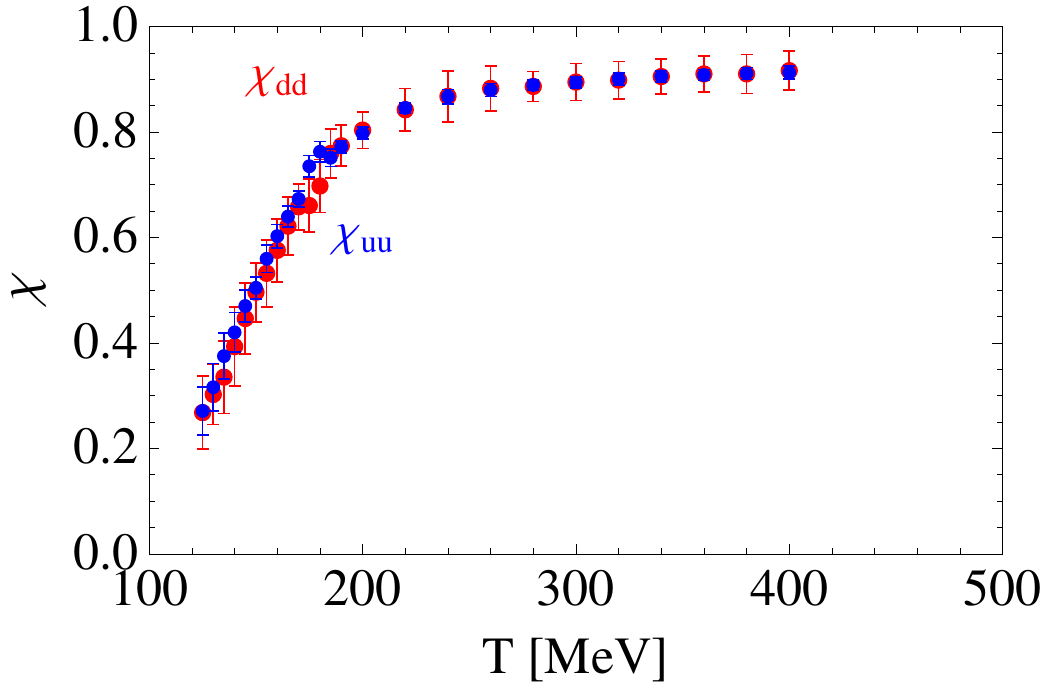} &
 \includegraphics[width=0.27\textwidth]{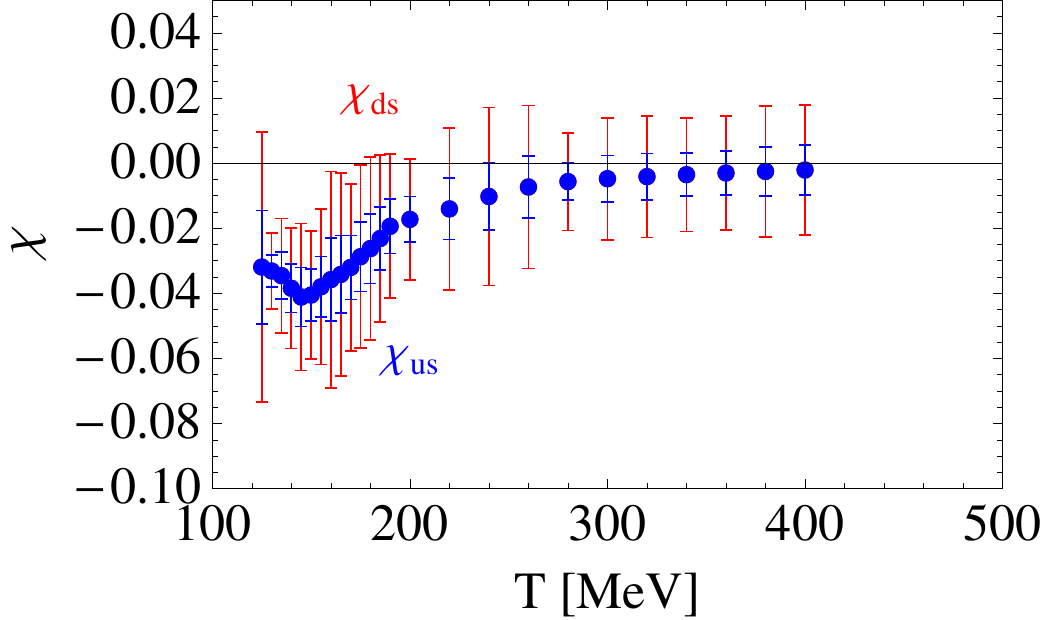} &
 \includegraphics[width=0.27\textwidth]{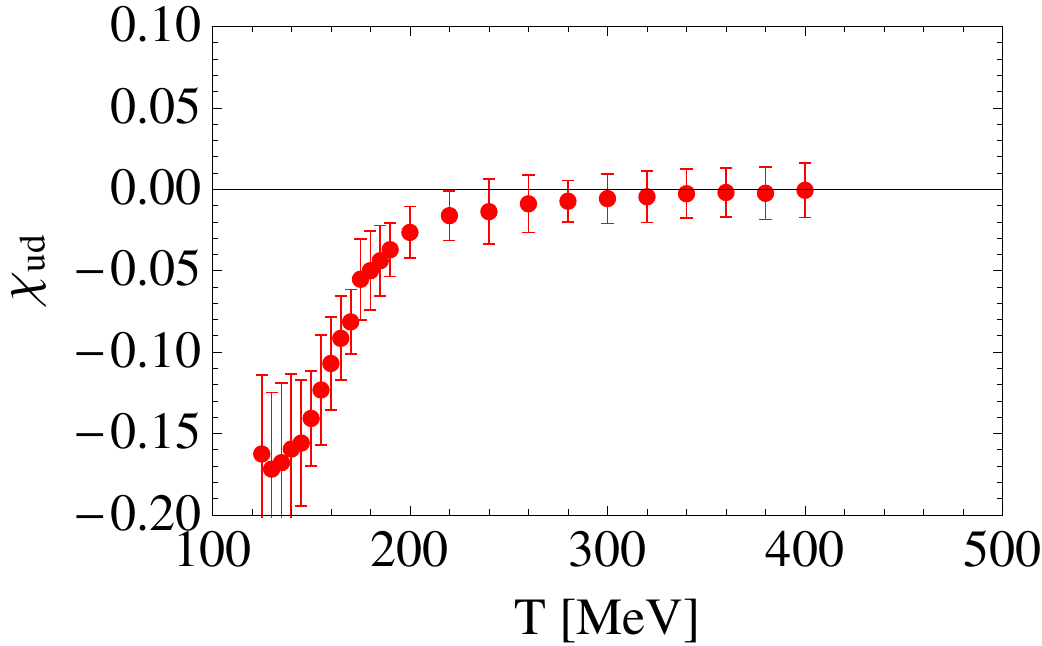} \\
 \includegraphics[width=0.27\textwidth]{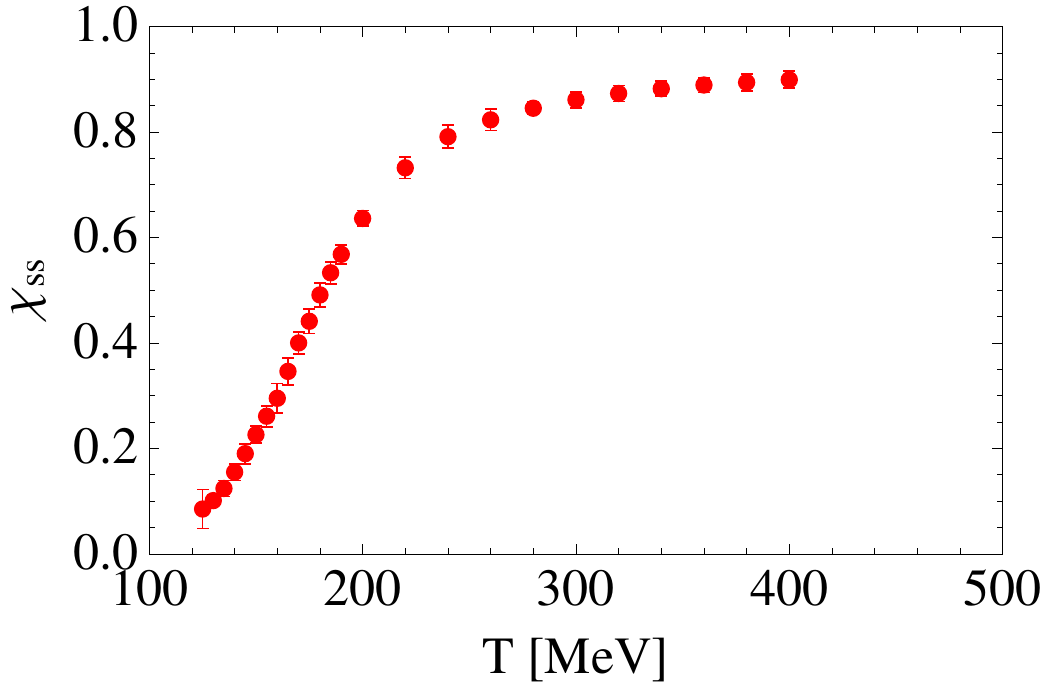} &
 \includegraphics[width=0.27\textwidth]{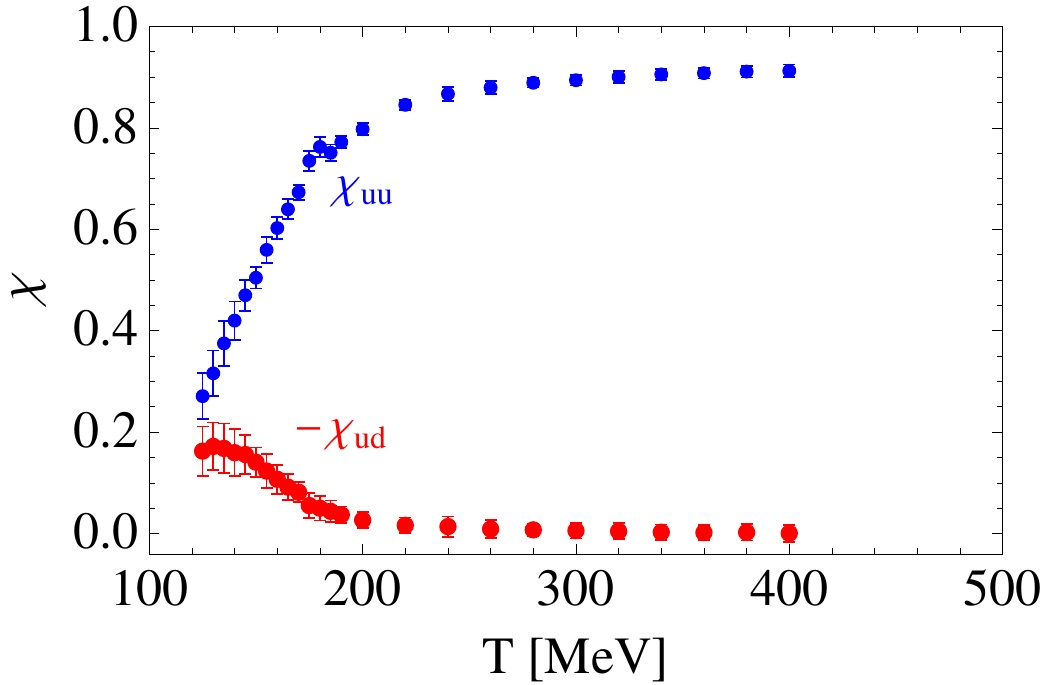} &
 \includegraphics[width=0.27\textwidth]{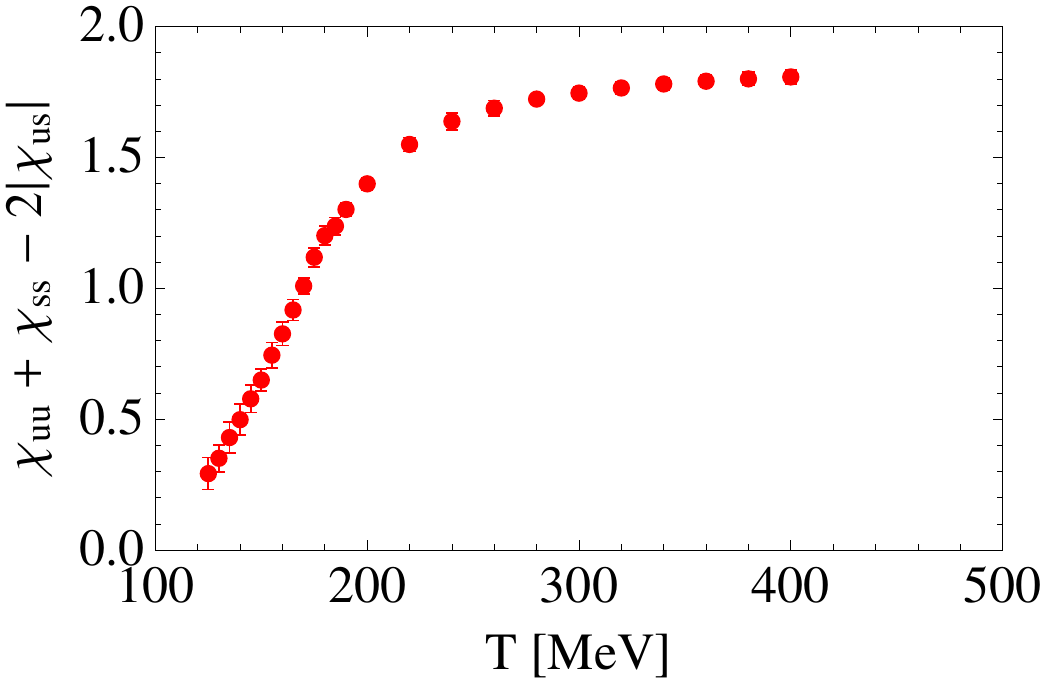}
\end{tabular}
%\vspace{-0.4cm}
 \caption{ Plot of the susceptibilities $\chi_{ab}$ in the
   quark-flavor basis as a function of temperature. The dots are the
   lattice data from Ref.~\cite{Borsanyi:2011sw}. We are checking
   numerically in these plots the relations
   Eqs.~(\ref{eq:chiab_neg})-(\ref{eq:chiab_ine2}). Similar results
   are obtained with the lattice data of
   Ref.~\cite{Bazavov:2012jq}.}
\label{fig:Chiuds}
\end{figure*}

In addition, the following relation holds for any pair of charges,
\begin{equation}
\chi_{aa} + \chi_{bb}  \ge  2 | \chi_{ab} | \,,  \label{eq:chiab_ine1}
\end{equation}
since $\langle (\Delta Q_a \pm \Delta Q_b)^2 \rangle$ is non-negative. In
particular it follows
\begin{equation}
 \chi_{uu} \ge -\chi_{ud} \ge 0 \,.  \label{eq:chiab_ine2}  
\end{equation}
The meson contribution to $\chi_{ud}$ is always negative while the baryon
contribution is always positive. The negative global result indicates that
mesons dominate over baryons in $\chi_{ud}$. Going to low temperatures, this
provides yet another proof that in QCD with two degenerate light flavors the
lightest meson is lighter than the lightest baryon. We display in
Fig.~\ref{fig:Chiuds} some of these relations for the lattice data of the
susceptibilities, and check that they are fulfilled.

Finally, let us mention that higher order fluctuations can be obtained within
the HRG model by taking higher derivatives of the thermodynamics
potential. This leads to
\begin{equation}\begin{split}
 \chi^{BQS}_{pqr}(T) &= 
  \frac{1}{2\pi^2}  \sum_{i \in {\rm Hadrons}} 
g_i \, B_i^p  Q_i^q  S_i^r   
\\ & \quad \times 
\sum_{n=1}^\infty \zeta_i^{n+1} 
n^{p+q+r-2} \,\frac{M_i^2}{T^2} K_2\!\left( \frac{nM_i}{T} \right) 
.
\end{split}
 \label{eq:chi_HRGMhigher}
\end{equation}
Some results of fourth-order fluctuations are presented in
Section~\ref{subsec:baryonic_fluctuations}.

\section{Quark-diquark potential from Polyakov loop correlators}
\label{sec:VqD_potential}

In view of the missing resonance problem alluded to previously there is
nowadays some discussion about the most probable spatial configuration of
quarks inside baryons, and more specifically the structure of excited states.
In the quark model such as the RQM~\cite{Godfrey:1985xj,Capstick:1986bm}
baryons are $qqq$ states where the interaction is given by a combination of
$\Delta$-like pairs of $qq$ interactions and a genuinely $Y$-like $qqq$
interaction. Remarkably the $Y$-stringlike behavior of a static baryon energy at
finite temperature has been observed in Ref.~\cite{Bakry:2014gea}.
An interesting possibility would be that the quarks are distributed
according to an isosceles triangle. This is the idea behind an easily
tractable class of models, the so-called relativistic quark-diquark
($qD$) models with non-relativistic~\cite{Santopinto:2004hw} and
relativistic~\cite{Ferretti:2011zz,Gutierrez:2014qpa} variants where a
linearly rising potential is shown to work phenomenologically. In
addition to this success, there was no further reason to invoke such a
behaviour.  In the next paragraphs we elaborate on this and provide a
theoretical argument in favor of the presence of a linear potential.

It should be noted that by definition, a potential can only be
evaluated unambiguously (up to an additive constant) in the heavy
particles or static limit, since then the position operator is well
defined. The calculation of $qD$ static interactions for heavy sources
has been addressed on the lattice in several
works~\cite{Bissey:2009gw,Bakry:2017jna,Koma:2017hcm}, where it has
actually been found that up to numerical uncertainties the potential
is linearly rising, and the corresponding string tension is
numerically identical to the one of the $q \bar q$ system (see
specifically Ref.~\cite{Koma:2017hcm}).
In this section we show analytically that under very specific
assumptions this must actually be true. To show how this comes about
we will rely heavily on our previous work on free
energies~\cite{Megias:2013xaa} where a more detailed discussion can be
found. Here we just summarize the main issues relevant to the problem
at hand.

An operational way of placing static sources in a gauge theory such as
QCD, is by introducing in the Euclidean formulation a local gauge
rotation $\Omega$ realized by the Polyakov loop, i.e., the gauge
covariant operator defined as
\begin{equation}
\Omega(\vx) = P\, e^{\, i \int_0^\beta A_0(x) \, dx_0}
,
\end{equation}
where $P$ indicates path ordering and $A_0$ and $x$ are Euclidean.  The colour
source may be decomposed into irreducible representations, say $\mu$ so that
one can build the gauge invariant combinations corresponding to the character
in that representation, $\chi_\mu[ \Omega(x)]$.\footnote{$\chi_\mu(g)$ stands
  here for the character of the element $g$ in the representation~$\mu$, not
  to be confused with the susceptibilities $\chi_{ab}$ introduced earlier.}
For instance, the trace in the (anti)fundamental representation corresponds to
the character of the $\SU(3)$ colour gauge group,
\begin{eqnarray}
\tr \,\Omega(\vx) \equiv \chi_{\bf 3} [ \Omega(\vx) ] \, , \qquad 
\tr \,\Omega(\vx)^\dagger \equiv \chi_{\bf 3} [ \Omega(\vx) ]^*  \,.
\end{eqnarray}
Thus, the $q \bar q$ free energy is given by a thermal expectation value 
\begin{eqnarray}
e^{-F_{q \bar q} (r,T)/T}  = 
\langle \tr\, \Omega(\vx_1)  \tr\, \Omega(\vx_2)^\dagger \rangle_T  
\end{eqnarray}
where due to translational invariance the free energy depends only on the
separation $r= |\vx_1-\vx_2|$. In this and the following expressions, an
ambiguity of an additive constant must be allowed in the free energies coming
from the renormalization of the Polyakov loop operator. The potential is
obtained as the zero temperature limit of the free energy
\begin{eqnarray}
V_{q \bar q}(r) = F_{q \bar q}(r,0)
\,.
\end{eqnarray}
Likewise the $qqq$ free energy is given by
\begin{eqnarray}
e^{-F_{qqq} (\vx_1, \vx_2 , \vx_3 ,T)/T} = \langle \tr\, \Omega(\vx_1)
\tr\, \Omega(\vx_2) \tr\, \Omega(\vx_3) \rangle_T
\,.
\end{eqnarray}
In the limit $\vx_3 \to \vx_2 $ we get, for the quark-diquark free energy 
\begin{eqnarray}
e^{-F_{qD} (\vx_1, \vx_2 , T)/T} = \langle \tr\, \Omega(\vx_1)
\tr (\Omega(\vx_2)^2)   \rangle_T 
\,.
\end{eqnarray} 
This limit is singular since at very small distances the interaction is
dominated by one gluon exchange, $\sim 1/r $, and a self-energy must be
added. The renormalization of the new composite operator $\tr (\Omega(\vx_2)^2)$ yields a new ambiguity in the form of an additive constant from
$F_{qqq}(\vx_1, \vx_2 , \vx_2 , T)$ to $F_{qD}(\vx_1, \vx_2 ,
T)$.

Using the Clebsch-Gordan series
\begin{eqnarray}
{\bf 3}\otimes {\bf 3} = 
{\bf \bar{3}} \oplus {\bf 6} \,, 
\end{eqnarray}
yields the equivalent character relations 
\begin{eqnarray}
[ \chi_{\bf 3} ]^2  = 
\chi_{\bf \bar{3}} + \chi_{\bf 6}  \,,
\end{eqnarray}
and therefore 
\begin{eqnarray}
e^{-F_{qD} (\vx_1, \vx_2 , T)/T} \equiv e^{-F_{q \bar q} (\vx_1, \vx_2, T )/T} + 
e^{-F_{6 \otimes  3} (\vx_1, \vx_2 , T )/T}
.
\end{eqnarray}
The $qD$ potential is obtained in the zero temperature limit. 
In this limit we expect the energy ${\bf 6} \otimes  {\bf 3}$
configuration to be larger than that of the $q \bar q$ one, 
hence
\begin{eqnarray}
V_{q D} (r) = V_{q \bar q} (r) + {\rm const} \,.
\end{eqnarray}

\section{Quark-diquark model for the baryon}
\label{sec:Model}

\subsection{The model Hamiltonian}
\label{subsec:model}

In these models, the baryons are assumed to be composed of a constituent
quark, $q$, and a constituent diquark, $D \equiv
(qq)$~\cite{Santopinto:2014opa}, and the Hamiltonian writes\,\footnote{In this
  work we are concerned with the overall features of the quark-diquark
  spectrum, relevant to the thermodynamics of the system, hence we will
  consider a simplified version of the model, neglecting possible fine
  interaction terms, like contact terms in the $s$-channel and spin-dependent
  interactions, cf. Ref.~\cite{Santopinto:2014opa}.}
\begin{equation}
H_{qD} =  \sqrt{\vp^2 + m_q^2 } + \sqrt{\vp^2 + m_D^2}  + V_{qD}(r) \,.  \label{eq:H_qD}
\end{equation}
In Sec.~\ref{sec:VqD_potential} we have argued that the quark-diquark
potential is the same as the quark-antiquark potential, up to an additive
constant. For the latter we assume $V_{q{\bar q}}(r) = - \frac{\tau}{r} +
\sigma r + c$, hence
\begin{equation}
V_{qD}(r) = - \frac{\tau}{r} + \sigma r + \mu \,,  
\label{eq:VqD}
\end{equation}
with $\sigma = (0.42\GeV)^2$. In addition, we adopt the value $\tau = \pi/12$,
as expected from the L\"uscher term in the potential~\cite{Luscher:1980ac}.

In this model we can distinguish between two kinds of diquarks: scalar, $D$,
and axial-vector, $D_{AV}$. When considering the quark content notation of a
diquark, we will use $[q_1 q_2]$ to denote scalar diquarks, and $\{q_1 q_2\}$
for axial-vector diquarks. Some studies on QCD indicate that the mass
difference between these diquarks is~\cite{Jaffe:2004ph}
\begin{equation}
\Delta m_D := m_{D_{AV}} - m_{D} \simeq 0.21 \GeV \,. \label{eq:DeltamD}
\end{equation}
In what follows we adopt this value for $\Delta m_D$. We will assume that the mass parameters of the model are
controlled by a constituent quark mass, $m_\cons$, and the current quark mass
for the strange quark, $\hat{m}_s$, in the following way:
\begin{equation}
m_{u,d} = m_\cons   \,, \qquad m_s = m_\cons + \hat{m}_s \,.  \label{eq:muds}
\end{equation}
The breaking of flavor $\SU(3)$ for diquarks will be modeled as
\begin{equation}
m_D = m_{D, \textrm{{ns}}} + n_s  \hat{m}_s  \,,
\label{eq:mDsAV2}
\end{equation}
where $m_{D, \textrm{{ns}}}$ is the scalar mass of non strange
diquarks, and $n_s = 0, 1, 2,$ is the number of $s$-quarks in the
diquark.  The further choice
\begin{equation}
m_{D, \textrm{{ns}}} = 2 m_{\cons}
\label{eq:mDsAV}
\end{equation}
is rather natural, but we will not always enforce it.

With these assumptions, the only free parameters of the model are $m_\cons$, $\hat{m}_s$ and $\mu$, plus $m_{D, \textrm{{ns}}}$ if (\ref{eq:mDsAV}) is not
enforced. The goal is to reproduce the lattice results for the baryonic
fluctuations $\chi_{BB}$, $\chi_{BQ}$ and $\chi_{BS}$ for temperatures below
the crossover. For the baryonic states $B = 1$ and $S = -n_s$. The electric
charges and the degeneracies of the states are summarized in
Table~\ref{tab:degeneracy}.  The spectrum is computed as explained below. With
these ingredients the susceptibilities can be evaluated in the model using
Eq.~(\ref{eq:chi_HRGM}).

\begin{table}[htb!]
\centering
\begin{tabular}{||c|c|c|c|c|c||}
\hline\hline
Baryon & Total deg. & $Q= -1$ & $Q=0$ & $Q=1$ & $Q=2$  \\ \hline
$[nn]n$     &   4    &  -  &   2  &  2  &  -  \\   \hline
$\{nn\}n$   &  36    &  6  &  12  & 12  &  6   \\  \hline
$[nn]s$     &   2    &  -  &   2  &  -  &  -   \\  \hline
$\{nn\}s$   &  18    &  6  &   6  &  6  &  -   \\  \hline
$[ns]n$     &   8    &  2  &   4  &  2  &  -   \\  \hline
$\{ns\}n$   &  24    &  6  &  12  &  6  &  -   \\  \hline
$[ns]s$     &   4    &  2  &   2  &  -  &  -   \\  \hline
$\{ns\}s$   &  12    &  6  &   6  &  -  &  -   \\  \hline
$\{ss\}n$   &  12    &  6  &   6  &  -  &  -   \\  \hline
$\{ss\}s$   &   6    &  6  &   -  &  -  &  -   \\ 
\hline\hline
\end{tabular}
\caption{ Spin-isospin degeneracies of the baryonic states within the
  quark-diquark model of Sec.~\ref{sec:Model}. The second column contains the
  total degeneracy of each state, while the columns from $Q=-1$ to $Q=2$
  contain the degeneracies by distinguishing between the electric charges of
  the states. $n$ represents the light flavors $u,d$.}
\label{tab:degeneracy}
\end{table}

\subsection{Baryon spectrum}
\label{subsec:Spectrum}

In order to obtain the baryon spectrum with the quark-diquark model we have to
diagonalize the Hamiltonian of Eq.~(\ref{eq:H_qD}). Since the problem does not
admit a closed analytic solution we will obtain the spectrum numerically after
truncation to a model space and diagonalization of the corresponding finite
dimensional matrix. This is a variational procedure. Because of the form of
the Hamiltonian $H=f_1(p) + f_2(r)$ a convenient basis is that of the
isotropic harmonic oscillator, with normalized wave functions of the form
\ignore{
\begin{equation}\begin{split}
R_{nl}(r) &=  N_{nl} \! \left(\frac{r}{b} \right)^l 
L_{n-1}^{l+\frac{1}{2}} \! \left( \frac{r^2}{b^2} \right) 
e^{-\frac{r^2}{2b^2}}
\\
N_{nl} &= 
\sqrt{ \frac{(n-1)! \, 2^{l+n+1}}{ \sqrt{\pi} b^3 (2l + 2(n-1) + 1)!!}  } 
\end{split}
\label{eq:Rnl}
\end{equation}
\begin{equation}
R_{nl}(r) =  
\sqrt{ \frac{(n-1)! \, 2^{l+n+1}}{ \sqrt{\pi} b^3 (2l + 2(n-1) + 1)!!}  } 
\! \left(\frac{r}{b} \right)^l 
L_{n-1}^{l+\frac{1}{2}} \! \left( \frac{r^2}{b^2} \right) 
e^{-\frac{r^2}{2b^2}}
\,.
\label{eq:Rnl}
\end{equation}
}
\begin{equation}
R_{nl}(r) =  
\frac{e^{-\frac{r^2}{2b^2}}}{\sqrt[4]{\pi}} 
\! \left(\frac{r}{b} \right)^l 
\! \sqrt{ \frac{(n-1)! \, 2^{l+n+1}}{ b^3 (2l + 2(n-1) + 1)!!}  } 
L_{n-1}^{l+\frac{1}{2}} \! \left( \frac{r^2}{b^2} \right) 
\,.
\label{eq:Rnl}
\end{equation}
\ignore{
\begin{equation}
R_{nl}(r) =  
\frac{e^{-\frac{r^2}{2b^2}}}{\sqrt[4]{\pi}} 
\! \left(\frac{r}{b} \right)^l 
\! \sqrt{ \frac{(n-1)! \, 2^{l+n+1}}{ b^3 (2l + 2(n-1) + 1)!!}  } 
L_{n-1}^{l+\frac{1}{2}} (r^2/b^2)
\,.
\label{eq:Rnl}
\end{equation}
}
$L_{n-1}^{l+\frac{1}{2}}(x)$ are the generalized Laguerre polynomials and the
positive parameter $b$ is related to the oscillator mass and frequency. $b$
can be optimized as a variational parameter. The reduced wave
functions~$u_{nl}(r)$ are normalized to unity
\begin{equation}
u_{nl}(r) =  r \, R_{nl}(r) ,
\qquad
\int_0^\infty dr \, u_{nl}(r)^2 = 1 \,.
\end{equation}
The corresponding wave functions in momentum space, $\hat{R}_{nl}(p)$, have
the same form as those in Eq.~(\ref{eq:Rnl}) up to a phase and $b\to 1/b$,
namely
\begin{equation}
\hat{u}_{nl}(p;b)=(-i)^{l+2n-2}u_{nl}(p;1/b)
.
\end{equation}
Then the matrix elements $\langle nl | H_{qD} | n^\prime l \rangle$ are
obtained from
\begin{eqnarray}
 \langle nl | H_{qD} | n^\prime l \rangle &&= 
 \int_0^\infty dp \, \hat{u}_{nl}^\ast(p) \hat{u}_{n^\prime l}(p) 
\left[ \sqrt{p^2 + m_q^2 } + \sqrt{p^2 + m_D^2}  \right]   \nonumber \\
&&\quad 
+  \int_0^\infty dr \, u_{nl}^\ast(r) u_{n^\prime l}(r) V_{qD}(r) \,, 
\label{eq:matrix_element_H}
\end{eqnarray}
with $n = 1, \dots , n_{\max}$, and $l = 0, \dots , l_{\max}$. The
value of the parameter $b$ is fixed to minimize the averaged value of
the energy levels in the spectrum for each of the multiplets in
Table~\ref{tab:degeneracy}. Typical values of this parameter are in
the range $0.55 \fm \lesssim b \lesssim 0.65 \fm$.

\begin{figure}[t]
\centering
 \includegraphics[width=0.43\textwidth]{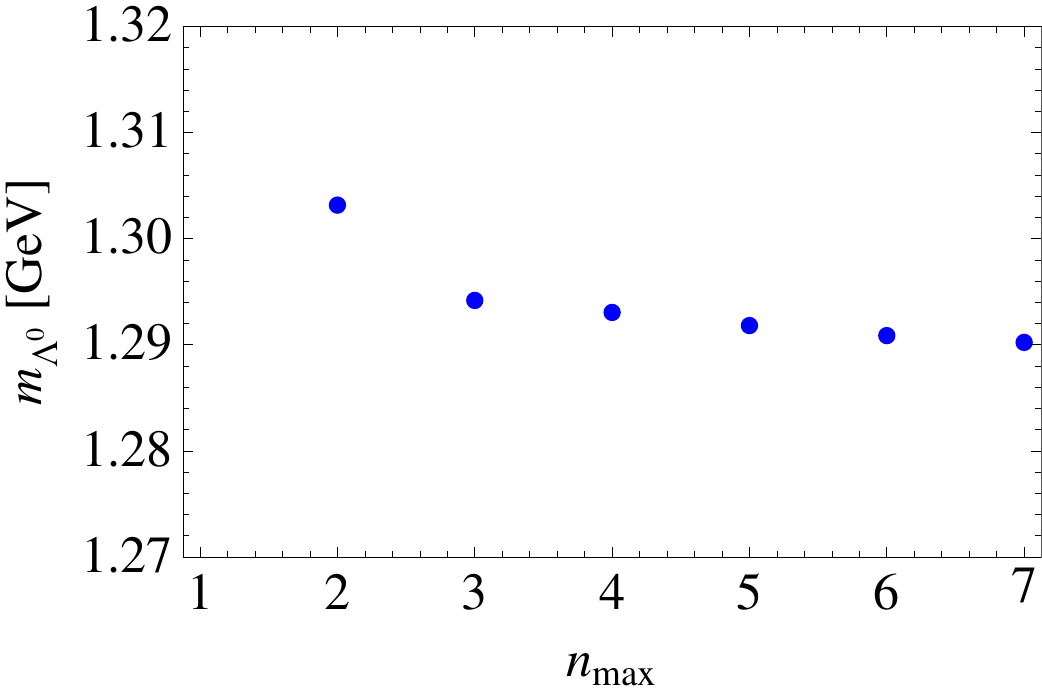}
%\vspace{-0.4cm}
 \caption{Mass of the baryonic state, $\Lambda^0$, as a function of $n_{\max}$ from diagonalization of Eq.~(\ref{eq:matrix_element_H}) with the parameters in (\ref{eq:param}).}
\label{fig:mn}
\end{figure}

We display in Fig.~\ref{fig:mn} the dependence of the mass of the
baryonic state, $\Lambda^0$, as a function of $n_{\max}$, taking
$l_{\max} = n_{\max}-1$. We can see that the dependence on $n_{\max}$ is
very weak already for $n_{\max} = 4$. We have cross-checked many of
the results presented subsequently by including larger values of
$n_{\max}$ and $l_{\max}$, and we find that they do not change
appreciably.

%%%%%%%%%%%%
%%%%%%%%%%%%%

\begin{figure}[t]
\centering
\includegraphics[width=0.43\textwidth]{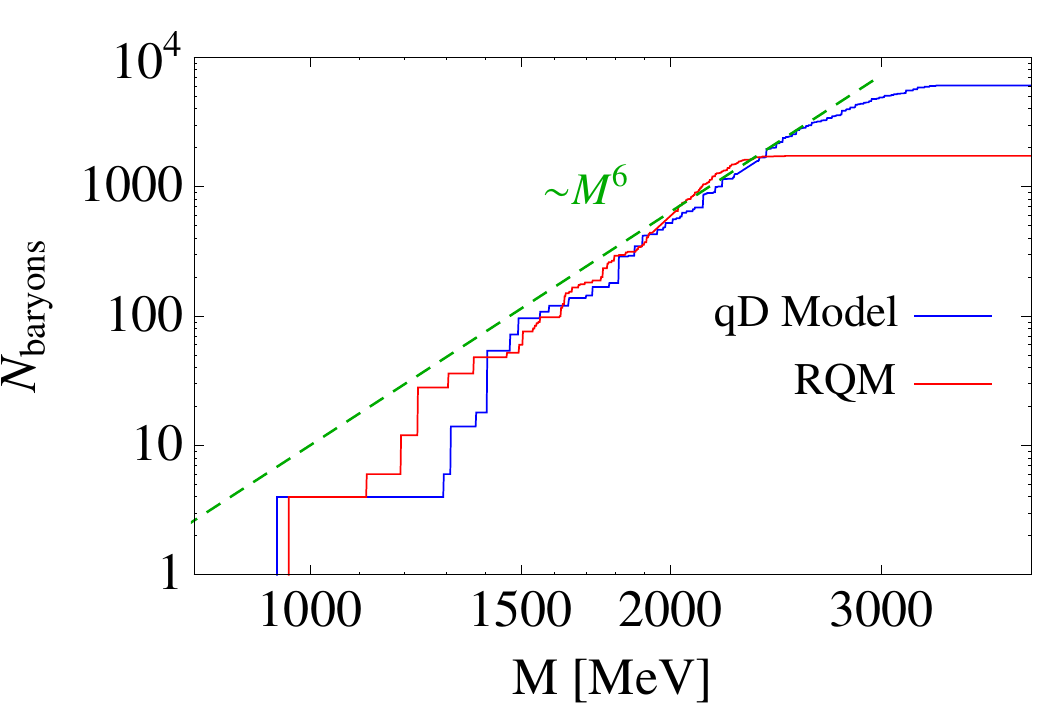}
%\vspace{-0.4cm}
\caption{Cumulative number for the spectrum of baryons as a function
  of the mass in a log-log plot comparing the
  RQM~\cite{Capstick:1986bm}, and the quark-diquark model used in this
  paper. We have used the parameters in Eq.~(\ref{eq:param}). We also
draw the $\sim M^6$ line for illustration. }
\label{fig:spectrumQD}
\end{figure}

After considering a convenient choice of the parameters of the model, for
instance
\begin{equation}
\begin{aligned}
&& m_{D, \textrm{{ns}}} = 0.6 \GeV \,, \quad m_{u,d} = 0.3 \GeV \,, \\ 
&&\hat{m}_s = 0.10 \GeV \,, \quad \mu = -0.459 \GeV \,,
\end{aligned} \label{eq:param}
\end{equation}
and following the procedure mentioned above, we get the spectrum of baryons
that is shown in Fig.~\ref{fig:spectrumQD}. In this figure we compare this
result to the RQM spectrum~\cite{Capstick:1986bm} on a log-log plot where the
$\sim M^6$ growth of the quark-diquark baryonic spectrum can be clearly
identified. It is quite remarkable that below~$M < 2400 \MeV$ the
quark-diquark spectrum is in good agreement with the RQM spectrum. While the
authors of Ref.~\cite{Capstick:1986bm} do not compute baryon masses heavier
than this, with the present quark-diquark model we have obtained further
states up to~$M \approx 3400 \MeV$. Within the HRG picture, these states will
contribute to the EoS of QCD as well as to other thermal observables like the
fluctuations. The choice of parameters of Eq.~(\ref{eq:param}) is motivated by
a comparison with the lattice results for the thermal fluctuations, as we will
see in the next section.

Let us mention that, unless otherwise stated, in the following we will use the
empirical value of the mass for the nucleon, $M_n = 938 \MeV$,  and apply the
quark-diquark model only for the other baryons. The effect of using the
empirical mass for all the baryons in the $1/2^+$ octet is also
discussed. The justification for this is that it is expected that the
quark-diquark picture will be reliable only for excited states. As we will show
in the next subsection, this is confirmed from the analysis of the lattice
data for the baryonic fluctuations.

%%%%%%%%%%%%
%%%%%%%%%%%%%

%%%%%%%%%%%%%
%\begin{figure}[t]
% \includegraphics[width=0.43\textwidth]{plotSpectrumDq.pdf}
%\vspace{-0.4cm}
% \caption{ Baryon spectrum from the quark-diquark model of Eq.~(\ref{eq:H_qD}).}
%\label{fig:spectrumQD}
%\end{figure}
%%%%%%%%%%%%%%

\subsection{Baryonic fluctuations}
\label{subsec:baryonic_fluctuations}

\begin{figure}[t]
\centering
 \includegraphics[width=0.43\textwidth]{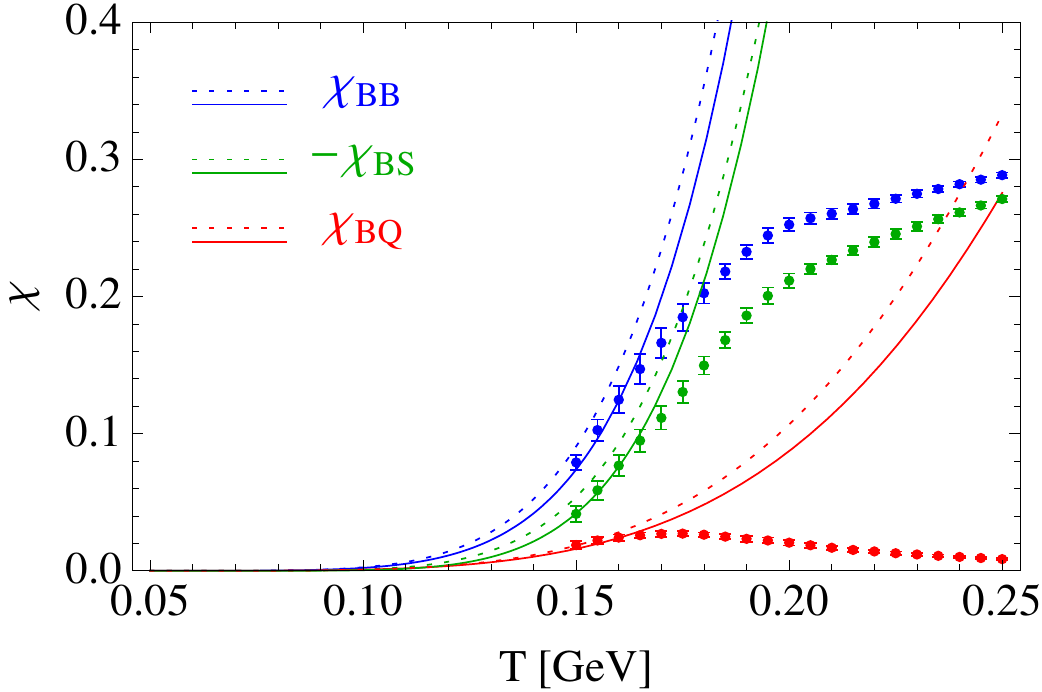}
%\vspace{-0.4cm}
\caption{Baryonic susceptibilities from the quark-diquark model (solid)
   compared to the lattice data of Ref.~\cite{Bazavov:2012jq}. We have used the parameters in Eq.~(\ref{eq:param}). We display also as dotted lines the result from the spectrum of the RQM~\cite{Capstick:1986bm}.}
\label{fig:BaryonicFluctuations}
\end{figure}

From the spectrum of the quark-diquark model, we can obtain the
baryonic fluctuations by using the HRG approach given by
Eq.~(\ref{eq:chi_HRGM}). Our goal is to reproduce the lattice results
for these quantities, at least for the lowest temperature values. A
typical fit of the model prediction with lattice data is shown in
Fig.~\ref{fig:BaryonicFluctuations}.  While ideally one would like to
have temperatures as low as possible so as to determine in a clean
way the low-lying states with the proper quantum numbers, in practice
we find that at the lowest available temperatures, the contribution of
excited states becomes individually small but collectively
important. This is a typical problem in intermediate temperature
analyses, and this is the reason why all possible constraints on the
model, such as the identity of quark-diquark and quark-antiquark
potentials discussed above, are particularly welcome.

%%%%%%%%%%%%
%%%%%%%%%%%%%

In order to perform the best fit to the data, we have chosen to minimize the
function
\begin{equation}
\bar\chi^2 = \bar\chi_{BB}^2 +   \bar\chi_{BQ}^2 +  \bar\chi_{BS}^2 \,, 
\label{eq:barchi1}
\end{equation}
where
\begin{equation}
\bar\chi_{ab}^2 = \sum_{j=1}^{j_{\max}} \frac{\left( \chi_{ab}^{\lat}(T_j) -
  \chi_{ab}^\HRG(T_j)\right)^2}{(\Delta \chi_{ab}^{\lat}(T_j))^2} 
\,.
\label{eq:barchi2}
\end{equation}
Here the $T_j$ are the temperatures used in the lattice calculations.
The lowest temperature of the data is $T_1=125\MeV$ for
Ref.~\cite{Borsanyi:2011sw} and $T_1 = 150\MeV$ for
Ref.~\cite{Bazavov:2012jq}, while $j_{\max}$ is the number of data
points used in the fit for each of the susceptibilities. $\Delta
\chi_{ab}^{\lat}$ are the uncertainties of the lattice results.

Since a hadronic model is not expected to reproduce the QCD crossover
we fit the data corresponding to the lower temperatures in lattice
measurements. Statistical considerations~\cite{RuizArriola:2017kqs}
indicate that those data points should be included for which
\begin{equation}
1 - \sqrt{\frac{2}{\nu}} <  \frac{\bar\chi^2}{\nu}   <  
1 + \sqrt{\frac{2}{\nu}}  
\,, 
\label{eq:chi2_bounds}
\end{equation}
where $\nu$ is the number of degrees of freedom. This condition fixes our
choice of $j_{\max}$. In our case, $\nu = n_{\rm obs} j_{\max} - n_{\param}$,
with $n_{\rm obs}=3$ since three baryonic fluctuations are being fitted, and
$n_{\param}$ is the number of free parameters of the model.

The fits turn out to be very sensitive to the parameter~$\mu$ in
Eq.~(\ref{eq:VqD}); hence it is convenient to always minimize $\bar\chi^2$
with respect to this parameter. Whenever we provide or plot a value for the
function~$\bar\chi^2/\nu$, it should be understood that this
function is already minimized with respect to the parameter~$\mu$. Typical
values of this parameter are $-0.7 \GeV \lesssim \mu \lesssim 0 \GeV$.

\begin{figure}[t]
\centering
 \includegraphics[width=0.43\textwidth]{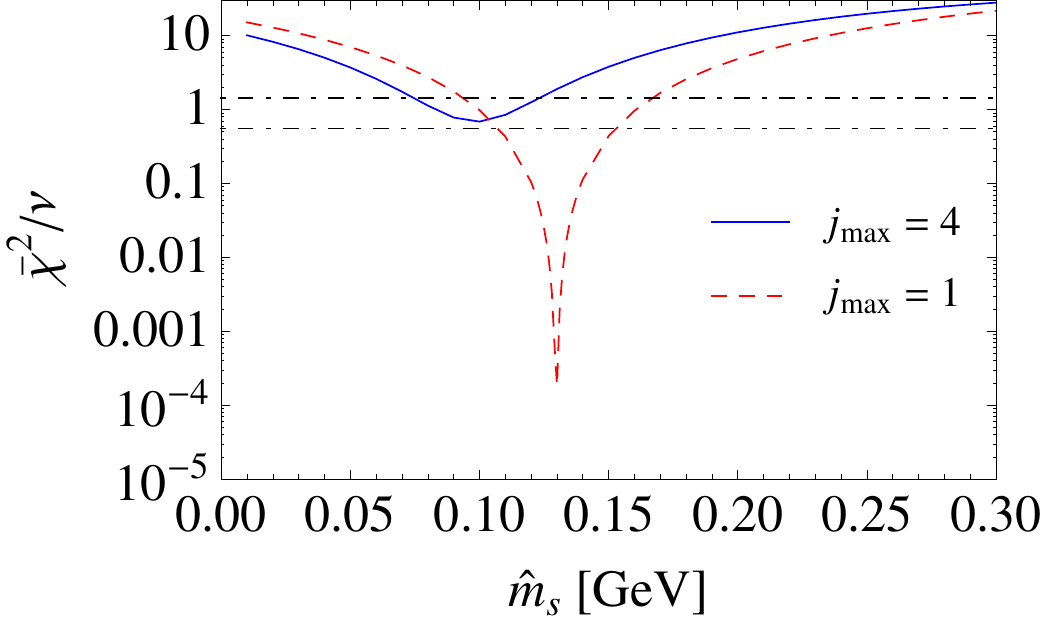}
%\vspace{-0.4cm}
 \caption{$\bar\chi^2/\nu$ as a function of
   $\hat{m}_s$ with $m_{\cons} = 0.3 \GeV$, and $m_{D, \textrm{{ns}}}
   =2m_{\cons}$. The dashed red line corresponds to $j_{\max} = 1$ while the
   solid blue line corresponds to $j_{\max}=4$. The horizontal lines
   correspond to the upper and lower bounds of Eq.~(\ref{eq:chi2_bounds}). We have used the lattice data of Ref.~\cite{Bazavov:2012jq}.%
  }
\label{fig:chi2v1}
\end{figure}

As a first step in the analysis, we show in Fig.~\ref{fig:chi2v1} a plot of
$\bar\chi^2/\nu$ as a function of the current quark mass, while fixing the
constituent mass to $m_{\cons} = 0.3 \GeV$ and $m_{D, \textrm{{ns}}} =
2m_{\cons}$.  When including in the fits only the lowest temperature point of
the data, i.e. $j_{\max} = 1$, one gets an exceedingly good fit with
$\bar\chi^2/\nu = 1.0 \times 10^{-4}$ for $\hat{m}_s = 0.130 \GeV$. This is
rather surprising, as it means that three central values of the data can be
fitted almost exactly with just two parameters: $\hat{m}_s$ and $\mu$. The fit
deteriorates but it is still acceptable when the four lowest temperatures are
used, $j_{\max} = 4$. We find that the model and the lattice data are
compatible with Eq.~(\ref{eq:chi2_bounds}) for temperatures~$T \lesssim 165
\MeV$, either if we analyze the lattice data of Ref.~\cite{Borsanyi:2011sw} or
\cite{Bazavov:2012jq}. The best fit in this case with $\bar\chi^2/\nu = 0.68$ is for $\hat{m}_s = 0.099 \GeV$ and $\mu = -0.459 \GeV$.
%%%%%%%%%%%%
%%%%%%%%%%%%%

\begin{figure*}[htp]
\centering
 \begin{tabular}{c@{\hspace{2.5em}}c}
 \includegraphics[width=0.43\textwidth]{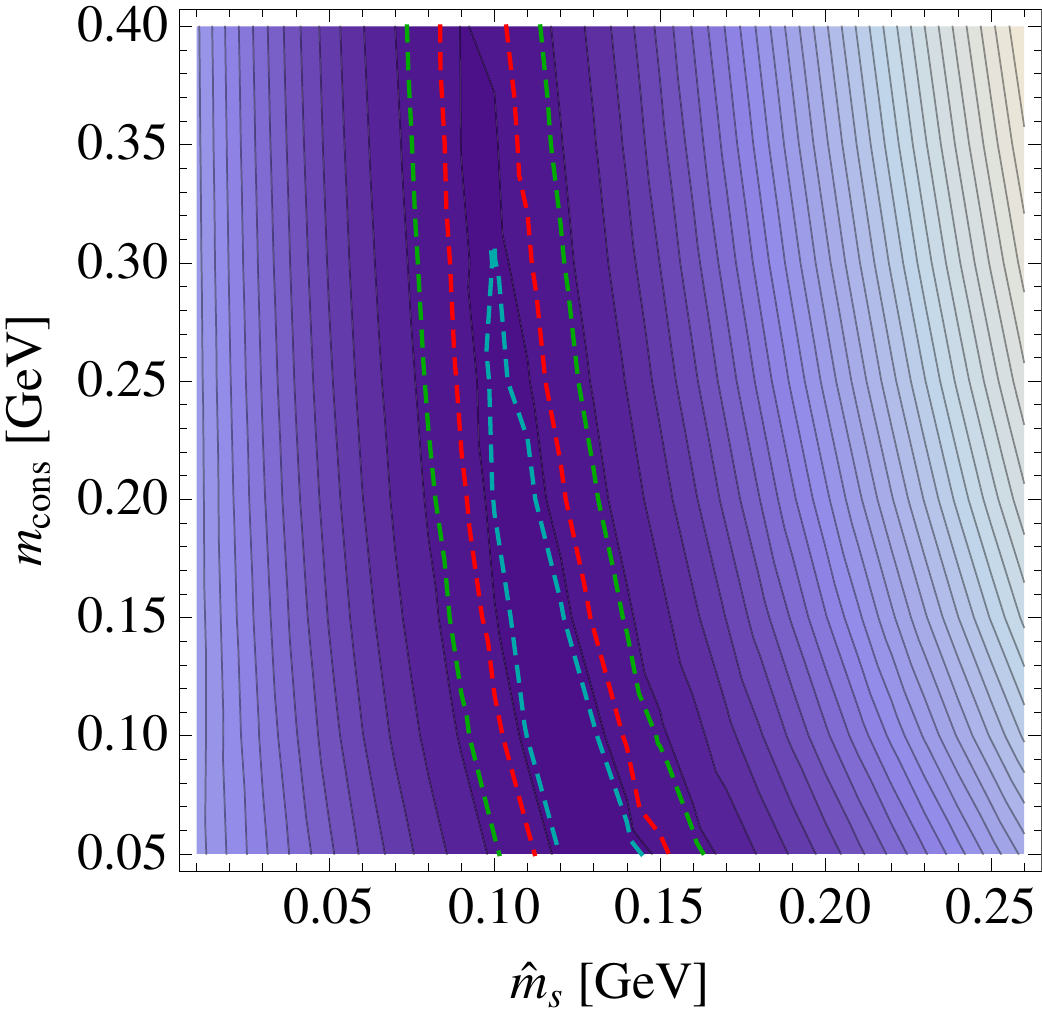} &
 \includegraphics[width=0.43\textwidth]{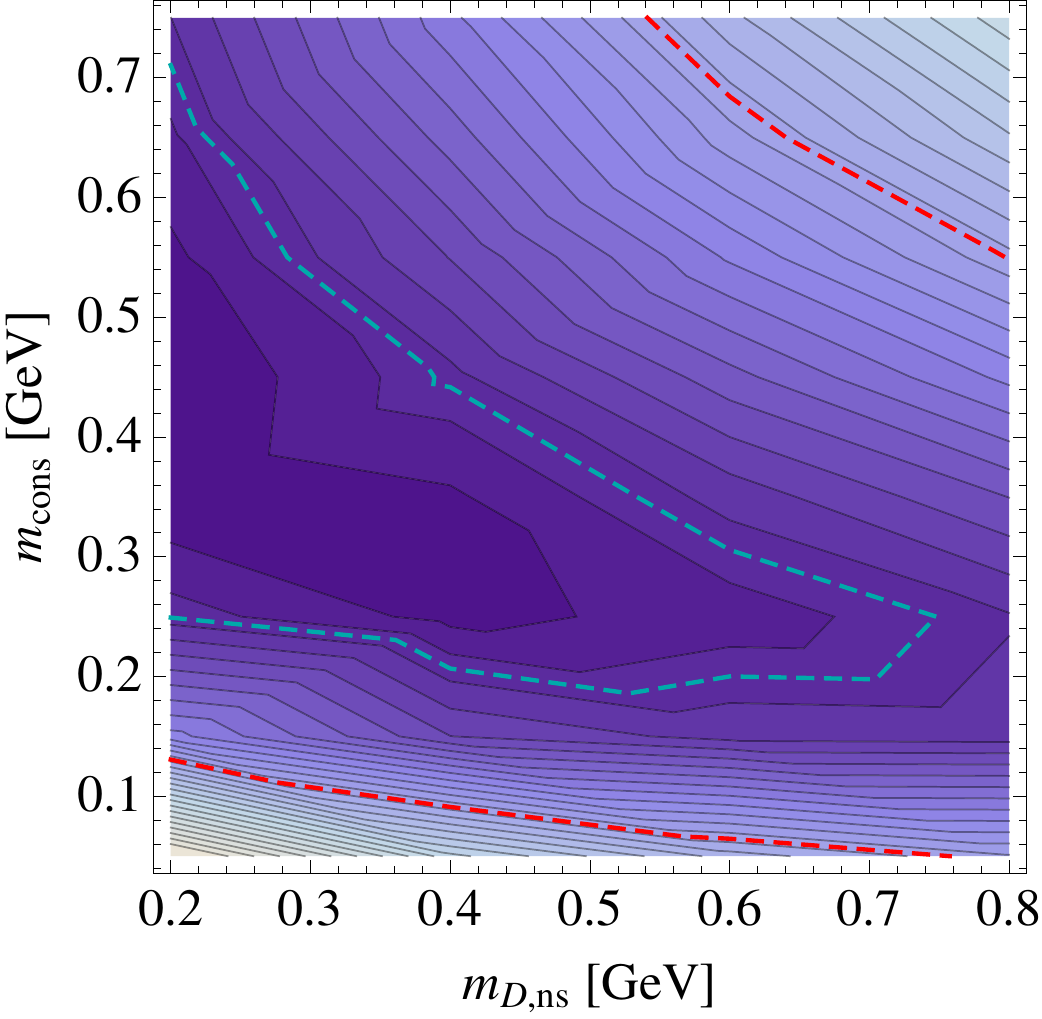}
\end{tabular}
 \caption{$\bar\chi^2/\nu$ from a fit to the lattice data of the
   baryonic fluctuations from Ref.~\cite{Bazavov:2012jq} with
   $j_{\max} = 4$. The dashed lines correspond to $\bar\chi^2/\nu = 0.77$
   (blue), $\bar\chi^2/\nu = 1$ (red) and $1+\sqrt{2/\nu}$
   (green). $\mu$ is determined from minimization.  Left panel: plane
   $(\hat{m}_s, m_{\cons})$ as free parameters with $m_{D,
     \textrm{{ns}}}=2m_{\cons}$. Right panel: plane $(m_{D,
     \textrm{{ns}}}, m_{\cons})$ with $\hat{m}_s = 0.10 \GeV$. }
\label{fig:chi2}
\end{figure*}

Fig.~\ref{fig:chi2} shows plots of $\bar\chi^2/\nu$ in two
versions. The left panel corresponds to the plane $(\hat{m}_s,
m_{\cons})$ with $m_{D, \textrm{{ns}}}=2 m_{\cons}$, while the right
panel corresponds to the plane $(m_{D, \textrm{{ns}}}, m_{\cons})$
with $\hat{m}_s= 0.10\GeV$.

The left panel clearly indicates that the current quark mass takes a value compatible with the PDG, i.e. $80 \MeV \lesssim \hat{m}_s \lesssim 120 \MeV$. In addition, the value of the constituent quark mass cannot be determined with precision, but at least we can ensure that it is in the regime $ 100 \MeV  \lesssim m_{\cons} \lesssim 400 \MeV$. 

In the right panel we have fixed $\hat{m}_s = 0.10\GeV$, and used the
scalar mass of non strange diquarks, $m_{D, \textrm{{ns}}}$, as a free
parameter. The figure shows that the most probable scalar diquark mass
is of the order of $m_{D, \textrm{{ns}}} \simeq 0.4 - 0.5 \GeV$ and the mass
of the constituent quarks $m_{\cons} \simeq 0.3 \GeV$. However, these
values depend of the choice of the current quark mass, so that when increasing the value of $\hat{m}_s$, the best fits happen for lower values of $m_{\cons}$. For instance, for $\hat{m}_s = 0.12\GeV$ the best fit is obtained when $m_{\cons} \lesssim 0.2 \GeV$, while for $\hat{m}_s = 0.09\GeV$ one obtains values $m_{\cons} \simeq 0.5 - 0.6 \GeV$.

%Finally, let us comment on the choice of $m_n$. 

As mentioned above, for our fits we have taken the value of the PDG for the
nucleon mass, $M_n = 938 \MeV$, and used the quark-diquark model values for
the remaining baryonic states. One could further adopt the empirical values
for the masses of the $1/2^+$ baryonic octet, that is, $M_{\Lambda^0} = 1.116
\GeV$, $M_\Sigma = 1.193 \GeV$, and $M_\Xi = 1.318 \GeV$, in view of the fact
that they are stable under strong interactions.  Such states correspond to 16
of the 18 states of the type $[qq]q$ in
Table~\ref{tab:degeneracy}. Specifically $N$, $\Sigma$ and $\Xi$ correspond to
$[nn]n$, $[ns]n$ and $[ns]s$. The four remaining states in $[nn]s$ and $[ns]n$
correspond to the $\Lambda^0$ of the octet plus an ${\rm SU}(3)$ singlet
$\Lambda$-like state~\cite{Santopinto:2014opa}. Since the mass of $[nn]s$ is
slightly lighter than $[ns]n$ we assign the $\Lambda^0$ to this multiplet; the
opposite assignation produces very similar results. The effect of using the
empirical masses for the whole octet is that the best fits presented in
Figs.~\ref{fig:chi2v1} and \ref{fig:chi2} (left) are shifted to larger values
of $\hat{m}_s$, namely $\Delta \hat{m}_s \simeq 24 \MeV$. The quality of the
fits is similar.

\begin{figure}[t]
\centering
 \includegraphics[width=0.43\textwidth]{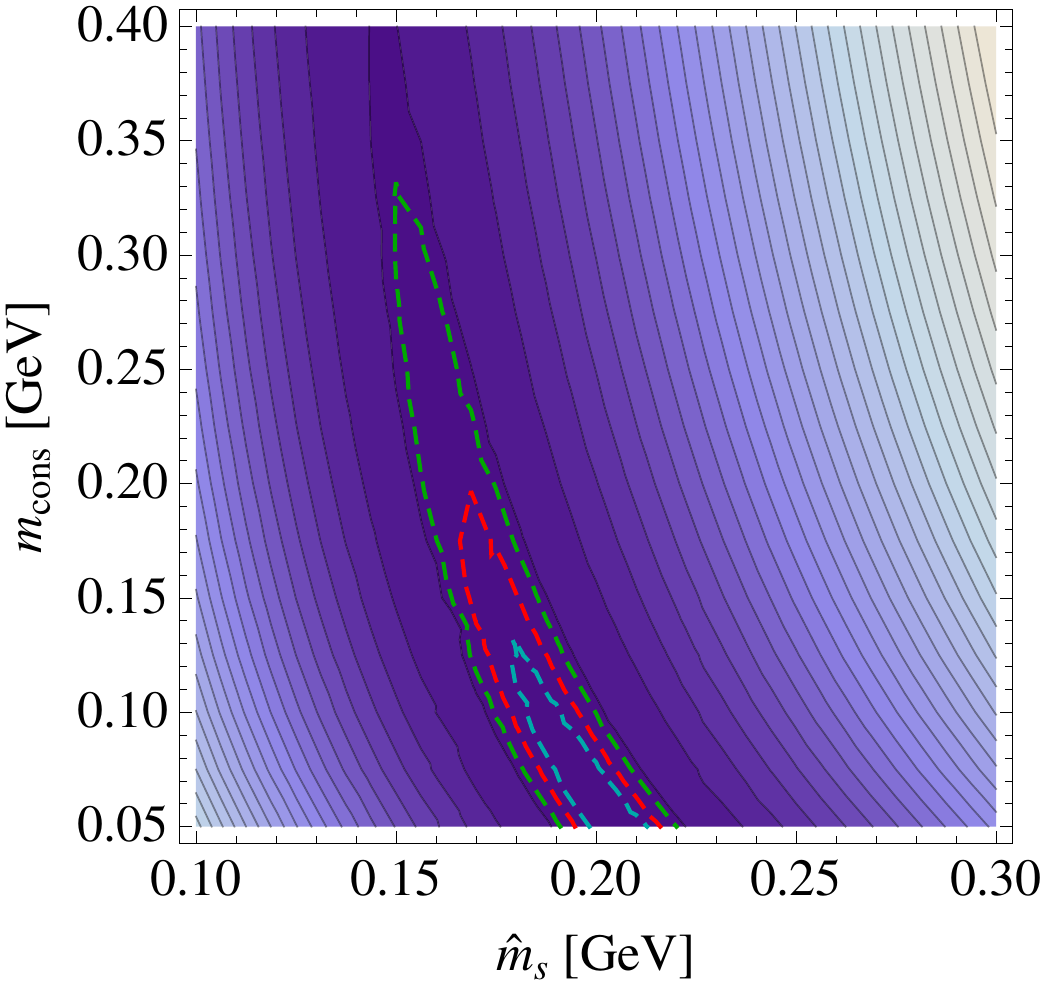}
%\vspace{-0.4cm}
 \caption{ $\bar\chi^2/\nu$ in the plane $(\hat{m}_s, m_{\cons})$ from a fit
   to the lattice data of the baryonic fluctuations from
   Ref.~\cite{Bazavov:2012jq} with $j_{\max} = 4$, and using the value of the
   nucleon mass as obtained from the quark-diquark model.  The dashed lines
   correspond to $\bar\chi^2/\nu = 1.275$ (blue), $\bar\chi^2/\nu = 1.35$
   (red) and $1+\sqrt{2/\nu}$ (green).}
\label{fig:chi2v2}
\end{figure}

The value of the nucleon mass as predicted by our version of the quark-diquark
model is $M_n \simeq 1.16 \GeV$. If this value is used instead of the
empirical one, the fits to the susceptibilities worsen. We show in
Fig.~\ref{fig:chi2v2} a plot of $\bar{\chi}^2/\nu$ in the plane
$(\hat{m}_s,m_{\cons})$ using the nucleon mass as given by the model. One can
see that in this case the best fits would correspond to non physical values of
the current strange-quark mass, $150 \MeV \lesssim \hat{m}_s \lesssim 220
\MeV$.

%The discussion presented in Sec.~\ref{sec:VqD_potential} shows that the
%string tension in a heavy quark-diquark system has the same value as the one
%in a heavy quark-antiquark system. It is a matter of discussion whether this
%is also true for dynamical quarks~\cite{Santopinto:2014opa,Masjuan:2017fzu},
%suggesting in some cases a lower value for the string tension. In order to
%complete the discussion, we have performed these analyses above by
%considering $\sigma = (0.42\GeV)^2/2$. In this case the plots turn out to be
%very similar to the ones presented in
%Figs.~\ref{fig:chi2v1}-\ref{fig:chi2v2}, so that our conclusions are not
%especially sensible to the value of the string tension.

In our model, as in other quark models such as that of Capstick and
Isgur~\cite{Capstick:1986bm}, we have assumed that the string tension
for the light quarks coincides with the one obtained from heavy
quarks, and hence for the qD system we have taken $\sigma= (0.42
\GeV)^2 $.  There are models where this value is reduced by a factor
of 2 when discussing baryon spectroscopy, as good spectra are obtained
for $(u,s,d)$ with $\sigma_{qD}= 2.15
\fm^{-2}$~\cite{Ferretti:2011zz}, for $(u,d)$ $\sigma_{qD}= 1.57
\fm^{-2}$\cite{DeSanctis:2014ria} and for $(c,b)$ with $\sigma_{qD}=
4.5 \fm^{-2}$~\cite{Ebert:2011kk}. We note in passing that this factor
of two re-scaling is also needed in the slope of radial Regge
trajectories~\cite{Arriola:2006sv}. Motivated by these observations we
have repeated our analysis taking $\sigma= (0.42 \GeV)^2 / 2 $, with
minor modifications but slightly worse fit quality.

\begin{figure*}[htb]
\centering
 \begin{tabular}{c@{\hspace{2.5em}}c}
 \includegraphics[width=0.43\textwidth]{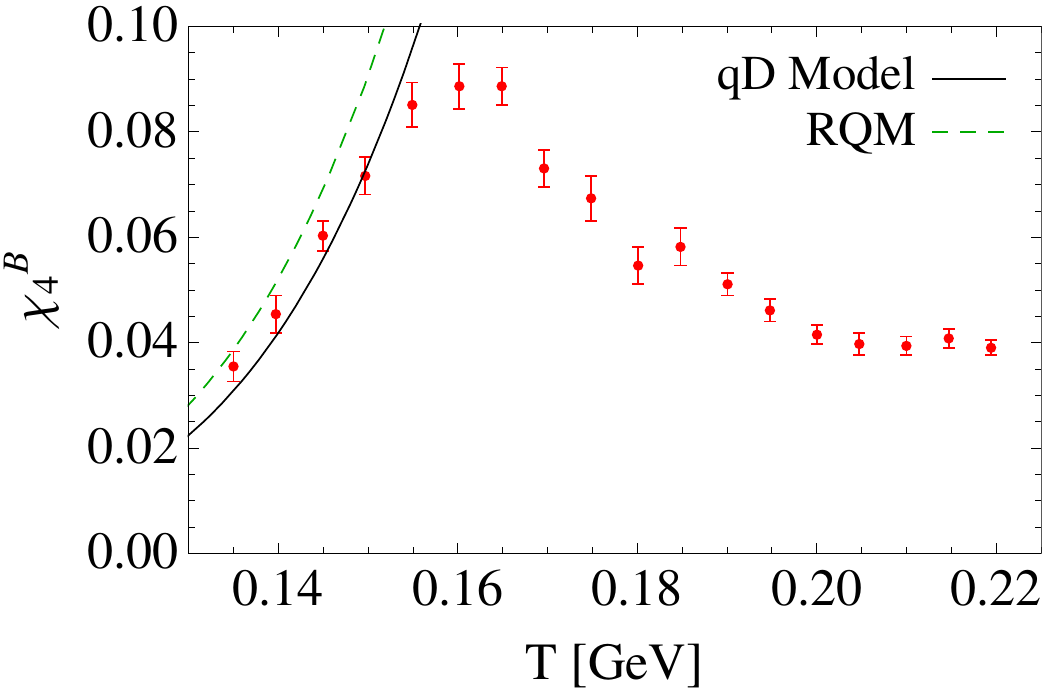} &
 \includegraphics[width=0.43\textwidth]{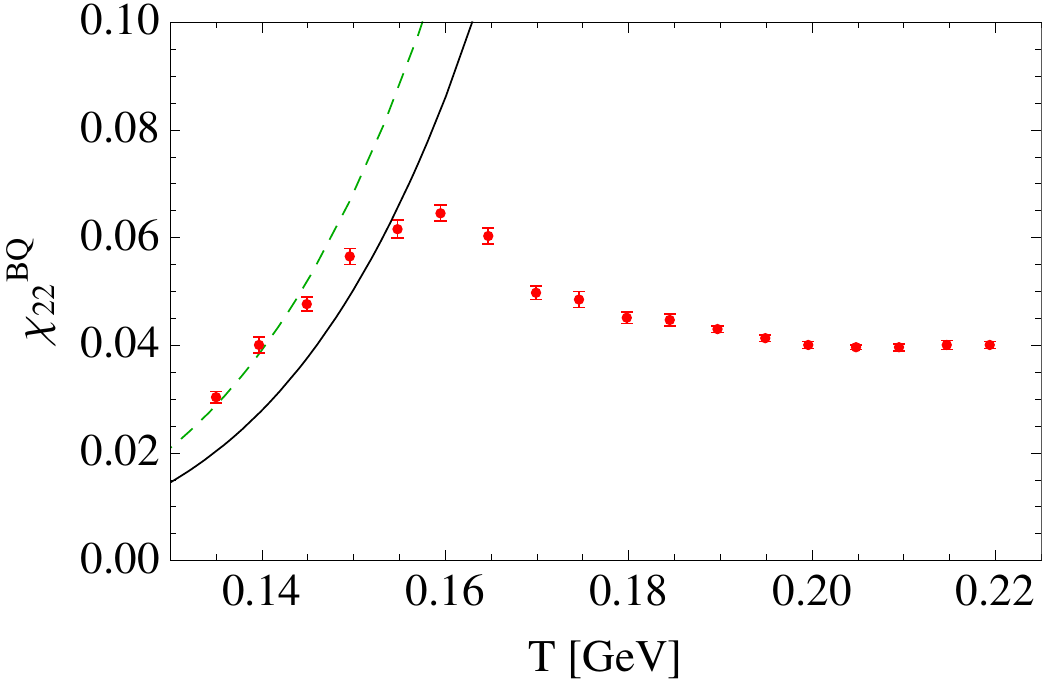} \\
 \includegraphics[width=0.43\textwidth]{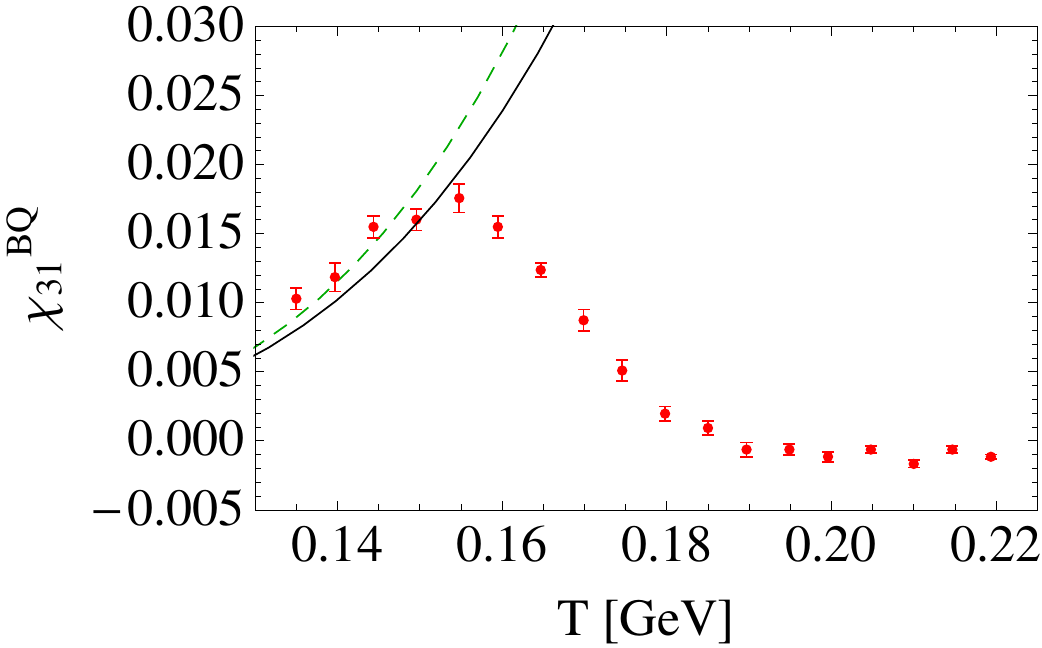} &
 \includegraphics[width=0.43\textwidth]{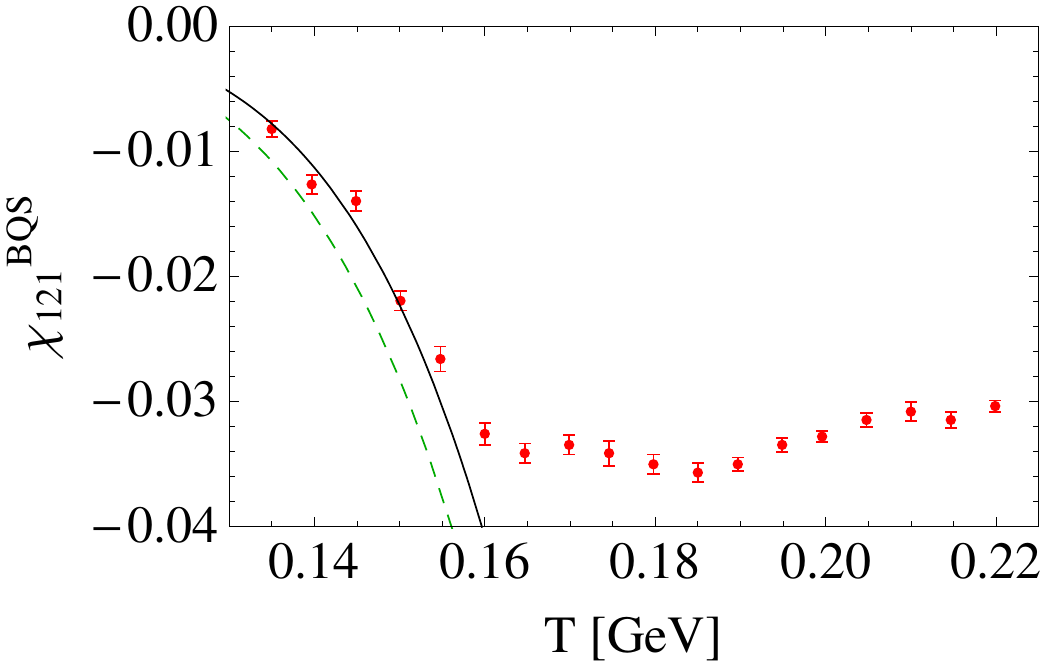}
\end{tabular}
%\vspace{-0.4cm}
 \caption{Baryonic susceptibilities of fourth order, $\chi_4^B$,
   $\chi_{22}^{BQ}$, $\chi_{31}^{BQ}$ and $\chi_{121}^{BQS}$, from the
   quark-diquark model (solid black) compared to the lattice data of
   Ref.~\cite{Borsanyi:2018grb}. We have used the parameters in
   Eq.~(\ref{eq:param}). We display also as dashed green lines the result
   from the spectrum of the RQM~\cite{Capstick:1986bm}.}
\label{fig:chi4}
\end{figure*}

We have studied as well the baryonic fluctuations of fourth order.
Fig.~\ref{fig:chi4} shows the results for $\chi_4^B$,
$\chi_{22}^{BQ}$, $\chi_{31}^{BQ}$ and $\chi_{121}^{BQS}$, computed
from the quark-diquark model and RQM model by using
Eq.~(\ref{eq:chi_HRGMhigher}), and compared with lattice data from
Ref.~\cite{Borsanyi:2018grb}. It can be noted that while the agreement
is reasonable, these lattice data are typically affected by larger
error bars than those of the second order fluctuations studied above,
and the behavior of the data turn out to be noisier, hence no firm
conclusions can be extracted from a fit to these quantities.

Finally, let us stress at this point that the lattice data used in the
present study might have strong correlations. Besides correlations
between susceptibilities at a single temperature, the data at
different temperatures could also be correlated as a consequence of
the reweighting technique and interpolations used in the lattice
simulations. These correlations could modify the results of the fit
and values of $\bar\chi^2$. Unfortunately, to our knowledge numerical
data on such correlations are not available, so such analysis is not
possible at present.

%%%%%%%%%%%%
%%%%%%%%%%%%%

%%%%%%%%%%%%
%%%%%%%%%%%%%

%%%%%%%%%%%%%%%%%%%%%%%%%%%%%%%%%%%%%%%%%%%%%%%%%%%%%%%%%%%%%%%%%%%
%%%%%%%%%%%%%%%%%%%%%%%%%%%%%%%%%%%%%%%%%%%%%%%%%%%%%%%%%%%%%%%%%%%

\section{Conclusions}
\label{sec:Conclusions}

The missing resonance problem, i.e., the apparent overcounting of
excited baryonic states by the quark model compared to the
experimentally found resonances has been a long standing puzzle which
has motivated a wealth of theoretical analysis and experimental work,
mainly grounded in the individual identification of resonance states
in the production process.  This viewpoint demands a good knowledge of
the scattering amplitude and its analytical properties in the complex
energy plane. Indeed, resonances are uniquely characterized as process-independent complex energy poles in unphysical sheets but the
extrapolation from the physical axis, where measurements are actually
made, to the complex plane is subjected to potentially large
uncertainties due to the role played by the background.

The thermodynamic approach to the missing states problem for baryons
has several advantages over the more conventional individual states
analysis, since it addresses the completeness of states problem from
the point of view of quark-hadron duality. It is rather insensitive to
resonance energy profiles as fine details of the level density are
washed by the Boltzmann factor. Lattice QCD has produced thermodynamic
quantities, such as the trace anomaly, where with the currently rather
small uncertainties one is not able to tell the difference between the
current PDG spectrum and a quark model spectrum such as the RQM which
was inferred already a few decades ago. Differences in the comparison
become more visible when susceptibilities involving baryon number,
electric charge and strangeness are considered as different subsets of
states are selected.

For zero density, conserved charges such as $B,Q,S$ have zero
expectation values but fluctuate statistically in a hot vacuum. Those
fluctuations are characterized by susceptibilities that have been
determined numerically in QCD on the lattice by the HotQCD and WB
Collaborations to discriminate among hadronic models with sufficient
accuracy, and thus can be used as a benchmark comparison.

We have argued that the asymptotic three-body phase space for confined
$qqq$ systems $\sim M^{12}$ is much larger than the one actually
determined in RQM $\sim M^{6}$ which resembles instead a two body
system with a linearly growing potential. This strongly suggests a
dominance of quark-diquark dynamics for excited baryons.  Therefore, we
have considered a quark-diquark model with a linearly confining
interaction.

By analyzing the free energy of heavy quark sources characterized by
Polyakov loops, we have been able to disclose under what conditions
the poorly known quark-diquark string tension should coincide with the
much familiar quark-antiquark string tension, in agreement with
previous and recent lattice results. 

Using this {\it a priori} fixed quark-diquark potential we have
determined the remaining model parameters from conserved charges
susceptibilities.  The results are reasonable and fall in the bulk of
previous intensive studies where a detailed description of the
spectrum was pursued.

Finally, let us mention that the study of nonbaryonic
susceptibilities would require a specific model for mesons which, in
principle, would not be related to the quark-diquark dynamics
exploited in this manuscript. Such study is worth pursuing but goes
beyond the scope of the present analysis.

\vspace{1cm}

\begin{acknowledgments}
This work is supported by the Spanish MINECO and European FEDER funds
(Grants No. FIS2014-59386-P and No. FIS2017-85053-C2-1-P), Junta de
Andaluc\'{\i}a (Grant No. FQM-225), and by the Consejer\'{\i}a de
Conocimiento, Investigaci\'on y Universidad of the Junta de
Andaluc\'{\i}a and European Regional Development Fund (ERDF) Grant
No. SOMM17/6105/UGR. The research of E.M. is also supported by the
Ram\'on y Cajal Program of the Spanish MINECO (Grant No. RYC-2016-20678).
\end{acknowledgments}

\appendix

\section{WKB estimates of the susceptibilities}
\label{App:WKB}

In this appendix we compute the susceptibilities within a semiclassical
expansion. Eq.~(\ref{eq:chi_HRGM}) can be expressed as
\begin{equation}
\chi_{ab}(T) =  \sum_{\zeta=\pm} \int_0^\infty dM 
\rho^{\zeta}_{ab}(M) \Phi_\zeta(M/T)
\label{eq:chi_HRGM2}
\end{equation}
where
\begin{equation}
\Phi_\zeta(z) = \frac{z^2}{2\pi^2} \sum_{k=1}^\infty \zeta^{k+1}K_2(kz)
\end{equation}
and
%$\zeta=\pm$ correspond to mesons and baryons respectively, and
\begin{equation}
\rho^\zeta_{ab}(M) = 
 \sum_{i } g_i q_i^a q_i^b \delta(M-M_i) \,,
\end{equation}
and the sum is over mesons or baryons for $\zeta=\pm$, respectively.

For the baryonic susceptibility $\chi_{BB}$, which is the case we are going to
consider in the following, $\rho_{BB}(M)$ is equal to the density of states
$\rho(M)$, as $B= \pm 1$ for (anti)baryons. The density of states can by
computed in a derivative expansion~\cite{Caro:1994ht}, an approximation that
is closely related to a semiclassical expansion in the high mass
spectrum. In this approach it is best to start from the cumulative number
\begin{equation}
N(M) = \Tr(\Theta(M-\hat{H}))
\end{equation}
where $\hat{H}$ is the Hamiltonian and the trace is taken in the
center of mass system subspace. The density is then obtained from
$\rho(M) = dN(M)/dM$.

In the quark-diquark model, the space of states is divided in sectors,
$\lambda$, shown in Table~\ref{tab:degeneracy}. Each sector contains a tower
of multiplets all with degeneracy $g_\lambda$ displayed in the Table. This gives
\begin{equation}
N(M) = \sum_{\lambda} g_\lambda N^{(\lambda)}(M)
\label{eq:sectors}
\end{equation}
where the sum is over sectors and $N^{(\lambda)}(M)$ sums over the tower of
states including just one state in each multiplet. For the sake of clarity, in
what follows we will drop the label $\lambda$. It is understood that the
aggregated expressions are obtained by combining the results of the various
sectors as in \Eq{sectors}.

Within the semiclassical expansion, the cumulative number can be computed as
\begin{equation}
N(M) = \int \frac{d^3x d^3p}{(2\pi)^3} \Theta(M - H)  + \cdots \,, 
\label{eq:ncum2}
\end{equation} 
where $H$ is the (classical) Hamiltonian of the two-body system in the sector
$\lambda$. The zeroth order term has been made explicit while the dots stand for
higher order contributions in the derivative expansion.

The Hamiltonian takes the form
\begin{equation}
H = \sqrt{\vp^2 + m_q^2 } + \sqrt{\vp^2 + m_D^2} + \sigma r 
-\frac{4\alpha_S}{3r}
+ \mu\,, 
\label{eq:Hmassive}
\end{equation}
with parameters corresponding to the sector $\lambda$.  The effect of the
constant additive term $\mu$ on the cumulative number is just a shift $N(M)
\to N(M-\mu)$. So we disregard $\mu$ in our explicit expressions in the
following.

For simplicity, let us consider first massless quarks and diquarks and treat
the Coulomb term perturbatively, that is, $H = H_0 + H_1$ with
\begin{equation}
H_0 = 2p  + \sigma r \,, \qquad  H_1 =  -\frac{4\alpha_S}{3r} \,,
\end{equation}
($p = |\vp|$) and
\begin{eqnarray}
\Theta(M - H) &=& \Theta(M - H_0) - \delta(M - H_0) H_1 \nonumber \\
&&+ \frac{1}{2}\delta^\prime(M - H_0) H_1^2 + \cdots \,. \label{eq:Theta}
\end{eqnarray}
A straightforward computation of the integral in Eq.~(\ref{eq:ncum2}), leads
to the following contribution of the first term in the rhs of
Eq.~(\ref{eq:Theta})
\begin{equation}
N_0(M) = \frac{M^6}{720 \pi \sigma^3}  \,.
\end{equation}
The computation of the contribution of the term~$\propto H_1$ in
Eq.~(\ref{eq:Theta}) is also straightforward using
\begin{equation}
\int_0^\infty dp p^2 \delta(M - 2p -\sigma r) = \frac{1}{8}(M - \sigma r)^2 \,.
\end{equation}
Finally, for the term~$\propto H_1^2$ in Eq.~(\ref{eq:Theta}) one has
\begin{equation}
\int_0^\infty dp p^2 \delta^\prime(M - 2p -\sigma r) 
= \frac{1}{4}(M - \sigma r)  \,,
\end{equation}
using integration by parts. This gives for the semiclassical expansion of the
cumulative number
\begin{equation}
N(M)= \frac{M^6}{720 \pi \sigma^3} + \alpha_S \frac{M^4}{36 \pi
  \sigma^2} + \alpha_S^2 \frac{2M^2}{9\pi \sigma} - \frac{M^2}{9\pi \sigma} +
\cdots \,.
\label{eq:N_semi}
\end{equation}
The last term is obtained with the next-to-leading-order correction in the
semiclassical expansion of Eq.~(\ref{eq:ncum2})~\cite{Arriola:2014bfa}.
Plugging this result into Eq.~(\ref{eq:chi_HRGM2}) with $\zeta =
-1$ (for baryons) and after performing the integration in $M$, one gets
\begin{eqnarray}
\hspace{-1cm} &&\chi_{BB}(T) =  \frac{127 \pi^5}{94500} \left( \frac{T^2}{\sigma} \right)^3 + \frac{31 \pi^3}{5670} \alpha_S \left( \frac{T^2}{\sigma} \right)^2 \nonumber \\
&&\qquad\qquad\qquad + \frac{7\pi}{405} \alpha_S^2 \frac{T^2}{\sigma} - \frac{7\pi}{810} \frac{T^2}{\sigma} + \cdots  \,.    \label{eq:chi_HRGM3}
\end{eqnarray}
This is the contribution just for baryons. A factor of two has to be included to account for the antibaryons.

\begin{figure*}[htp]
\centering
 \begin{tabular}{c@{\hspace{2.5em}}c}
 \includegraphics[width=0.43\textwidth]{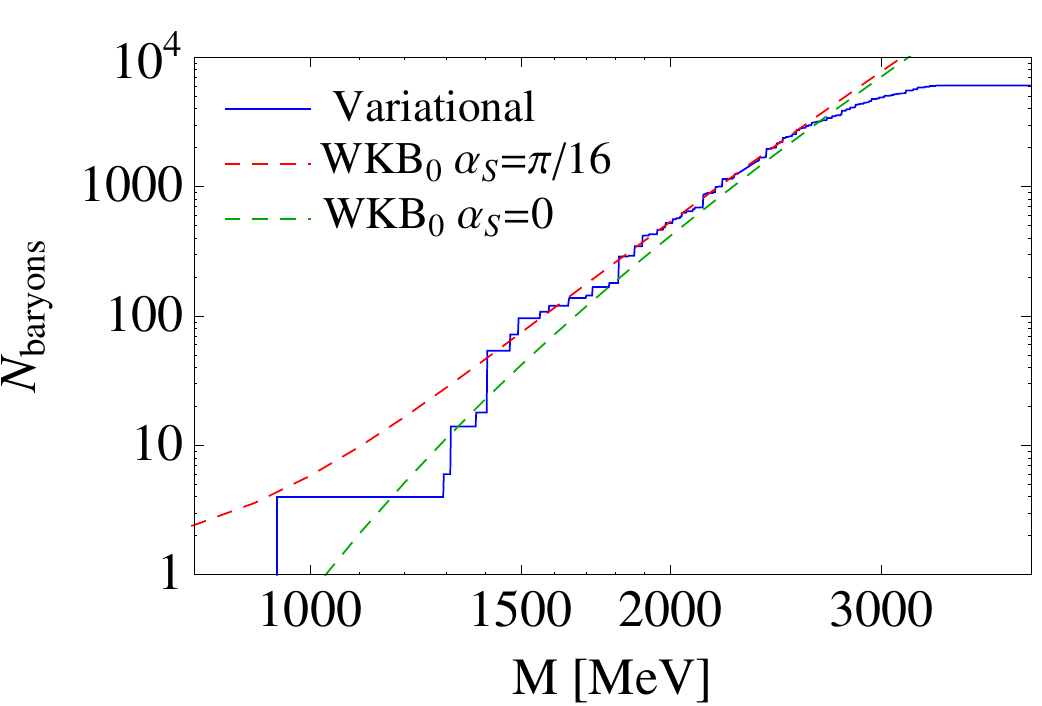} &
 \includegraphics[width=0.43\textwidth]{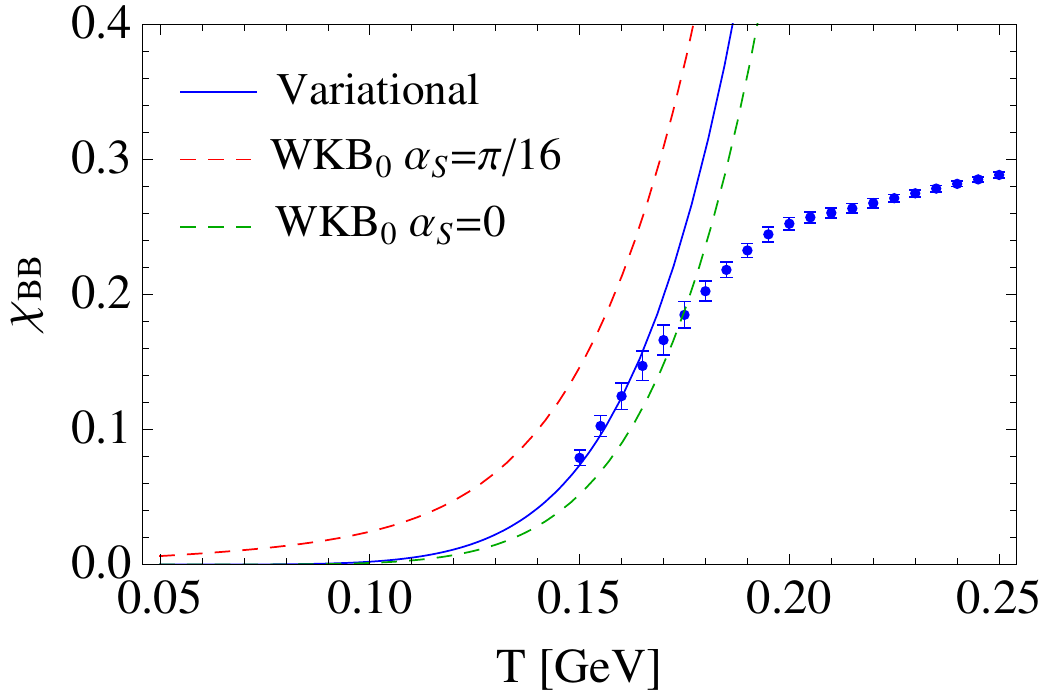}
\end{tabular}
 \caption{Cumulative number for the spectrum of baryons with the quark-diquark
   model (left panel) and baryonic $\chi_{BB}$ susceptibility obtained with
   that spectrum (right panel). We display as solid lines the result from the
   variational procedure of Sec.~\ref{subsec:Spectrum}, and as dashed lines
   the result from the WKB approximation at leading order. The points in the
   right panel are the lattice data of Ref.~\cite{Bazavov:2012jq}. We have
   used the parameters in Eq.~(\ref{eq:param}).}
\label{fig:spectrumQDWKB}
\end{figure*}

The leading contribution to $N(M)$ (which will be denoted WKB$_0$) can be
obtained analytically in the massive case if $\alpha_S$ is set to zero.  This
computation can be done easily by considering a change of variables in the
momentum integrals, $p \to E$, where $E$ is the summation of the two kinetic
terms in this equation. The expression is rather lengthy, thus instead of
presenting the full analytical result, we will show some numerics, which allows
to include the Coulomb term as well.  We display in Fig.~\ref{fig:spectrumQDWKB}
(left) the result of the quark-diquark model spectrum computed with the
variational procedure of Sec.~\ref{subsec:Spectrum}, and the one obtained with
the WKB$_0$ approximation. For the case WKB$_0$ with $\alpha_S = 0$, we have
used the analytical result mentioned above for the Hamiltonian of
Eq.~(\ref{eq:Hmassive}), while for $\alpha_S = \pi/16$ we have computed
numerically the integral in Eq.~(\ref{eq:ncum2}). Note that the agreement
between the variational and the WKB$_0$ approaches is rather good, especially
for the heaviest states, as expected. Finally, we also display in
Fig.~\ref{fig:spectrumQDWKB} (right) the baryonic $\chi_{BB}$ susceptibility
obtained with this spectrum within the different approaches.

%%%%%%%%%%%%
%%%%%%%%%%%%%

For completeness we provide some asymptotic results for finite quark and
diquark mass corrections. For the cumulative number entering in the rhs of
Eq.~(\ref{eq:N_semi}) one obtains
\begin{equation}
\Delta N_0(M) = - \frac{M^4(m_q^2 + m_D^2)}{48\pi \sigma^3}+ \cdots \,,
\label{eq:N0smallm}
\end{equation}
for $m_q^2 + m_D^2 \ll  M^2$. This translates into the following correction for
the susceptibility on the rhs of Eq.~(\ref{eq:chi_HRGM3})
\begin{equation}
\Delta \chi_{BB}(T) = -\frac{31\pi^3}{7560}(m_q^2 + m_D^2) \frac{T^4}{\sigma^3} + \cdots
\end{equation}
for $m_q^2 + m_D^2 \ll T^2$.

More interesting is the behavior of the cumulative number when $M$ is near
and above the (classical) threshold $M_{\rm threshold}=m_q+m_D$:
\begin{equation}
N_0(M) = \frac{64\sqrt{2}}{945 \pi \sigma^3} \left( \frac{m_q m_D}{m_q + m_D} \right)^{3/2} (M-m_q-m_D)^{9/2} +
\cdots \,.
\label{eq:N0largem}
\end{equation}
This region dominates the behavior of the susceptibility at small
temperatures, namely:
\begin{equation}
\chi_{BB}(T) = \frac{(m_q m_D)^{3/2} T^3}{\pi^2 \sigma^3} e^{-(m_q + m_D)/T}  + \cdots \label{eq:chiBB_largem}
\end{equation}
for $T \ll m_q + m_D $.  This temperature region is appropriate for the values
of $m_q$, $m_D$ and $T$ considered in this work.\footnote{For typical values
  of $m_q \simeq 0.3 \GeV$ and $m_D \simeq 0.6 \GeV$, the result of
  Eq.~(\ref{eq:chiBB_largem}) is a factor $\sim 1/4$ of the one computed from
  the full analytical expression in the regime $T \simeq 150 \MeV$.}

\section{Sign of the cross susceptibility for degenerate flavors in QCD}
\label{app:B}

\newcommand{\cD}{{\mathcal D}}
\newcommand{\esp}[1]{\langle #1 \rangle }

In this Appendix we present a proof of the inequality in
(\ref{eq:chiab_neg}). The arguments hold in presence of a lattice regulator.
We assume three flavors, $a=u,d,s$ with $u$ and $d$ degenerated. The partition
function is
\begin{equation}\begin{split}
Z &= \int \cD U \prod_a\cD \bar\psi_a\cD \psi_a \, e^{-S_g(U)
-
\sum_a \bar\psi_a D_a(U) \psi_a}
\\
& = \int \cD U  \, e^{-S_g(U)} \prod_a \det D_a(U)
.
\end{split}\end{equation}
$S_g(U)$ is the Euclidean gluonic action and (for simplicity here we use a
notation of QCD in the continuous formulation),
\begin{equation}
D_a(U) = \gamma_\mu(\partial_\mu
+ i A_\mu) + m_a\,,
\end{equation}
where the gluon field $A_\mu(x)$ is a Hermitian matrix and $m_a$ is the mass
of the flavor $a$. The Dirac matrices are Hermitian.

As is well known \cite{Montvay:1994cy}, the identity
\begin{equation}
\gamma_5 D_a(U)\gamma_5 = D_a(U)^\dagger
\end{equation}
implies that the eigenvalues of the Dirac operator are either real or come in
conjugated pairs, hence $\det D_a(U)$ is real. Because $u$ and $d$ are
degenerated, the weight $\det D_u(U)\det D_d(U)$ is positive. The weight $\det
D_s(U)$ will be assumed to be positive too as required to be able to apply
importance sampling Monte Carlo with dynamical quarks. This allows to define
the real action
\begin{equation}
S_q(U) = -\sum_a \log \det D_a(U)
,
\end{equation}
so that
\begin{equation}
Z = \int \cD U \, e^{-S_g(U) -S_q(U) }
.
\end{equation}

The operator counting the flavor $a$ is
\begin{equation}
Q_a = \int d^3x\, \bar\psi_a(x)\gamma_0\psi_a(x)
\quad a=u,d,s
.
\end{equation}
Because this quantity is conserved, we can use equivalently
\begin{equation}
Q_a = T \int d^4x\, \bar\psi_a(x)\gamma_0\psi_a(x)
\end{equation}
where $T=1/\beta$ is the temperature. Its expectation value can be obtained
using Wick's theorem
\begin{equation}
\esp{Q_a} = 
\frac{T}{Z}
\int \cD U e^{-S_g(U)-S_q(U)} (-1) \Tr(\gamma_0 D_a^{-1}(U))
.
\end{equation}
Of course this expectation value vanishes due to charge conjugation. Namely,
using $D_a(U^c) = C D_a(U)^T C^{-1}$ with $U^c \equiv U^*$ (or $A_\mu^c =
-A_\mu^T$), and $C \gamma_\mu^T C^{-1} = -\gamma_\mu$, it follows that the
measure including the actions are even under $U\to U^c$, whereas 
$\Tr(\gamma_0 D_a^{-1}(U))$ is odd.

For the $ud$ correlation, again applying Wick contractions,
\begin{equation}
%\frac{1}{T^2}
\esp{\Delta Q_u\Delta Q_d}
%\beta^2 \chi_{ud} 
= 
\frac{T^2}{Z}
\int \cD U e^{-S_g(U)-S_q(U)}  \Tr(\gamma_0 D_u^{-1}(U))^2
.
\end{equation}
This quantity is negative definite because $\Tr(\gamma_0 D_a^{-1}(U))$ is
purely imaginary, as follows from
\begin{equation}
(\gamma_0 D_a(U))^\dagger = 
-(\gamma_5 \gamma_0) \gamma_0 D_a(U) (\gamma_5 \gamma_0)^{-1}
.
\label{eq:B10}
\end{equation}

One observation is that the fact that $\Tr(\gamma_0 D_a^{-1}(U))$ is
purely imaginary provides another proof of $\esp{Q_a}=0$ since this quantity
is real, because $Q_a$ is Hermitian in the real time formulation.

Another observation is that $\chi_{ud} \le 0$ will always hold in a Monte Carlo
calculation, since $\Tr(\gamma_0 D_a^{-1}(U))^2 \le 0$ for every configuration,
even if the assumption $\det D_s(U)>0$ were violated. On the other hand, for
$\esp{Q_u^2}$ there are two Wick contractions
\begin{equation}\begin{split}
\langle (\Delta Q_u)^2 \rangle &= \frac{T^2}{Z}
\int \cD U e^{-S_g(U)-S_q(U)}  \Big( \Tr(\gamma_0 D_u^{-1}(U))^2
\\& \quad 
- \Tr( \gamma_0 D_u^{-1}(U) \gamma_0 D_u^{-1}(U)) \Big)
.
\end{split}\end{equation}
\Eq{B10} again implies that the second term is real and the inequality
$\chi_{uu} \ge |\chi_{ud}|$ implies that this second term is not only positive
but at least twice as large (in average) as minus the first one. Since there
is no reason to expect that $\Tr(\gamma_0 D_u^{-1}(U))^2 - \Tr( \gamma_0
D_u^{-1}(U) \gamma_0 D_u^{-1}(U))$ is definite positive for each gauge
configuration, the condition $\chi_{uu} \ge |\chi_{ud}|$ could fail to hold
for nonpositive definite $\det D_s(U)$, thereby providing a test on this.

\newpage

\bibliography{diquarks}

\begin{thebibliography}{52}%
\makeatletter
\providecommand \@ifxundefined [1]{%
 \@ifx{#1\undefined}
}%
\providecommand \@ifnum [1]{%
 \ifnum #1\expandafter \@firstoftwo
 \else \expandafter \@secondoftwo
 \fi
}%
\providecommand \@ifx [1]{%
 \ifx #1\expandafter \@firstoftwo
 \else \expandafter \@secondoftwo
 \fi
}%
\providecommand \natexlab [1]{#1}%
\providecommand \enquote  [1]{``#1''}%
\providecommand \bibnamefont  [1]{#1}%
\providecommand \bibfnamefont [1]{#1}%
\providecommand \citenamefont [1]{#1}%
\providecommand \href@noop [0]{\@secondoftwo}%
\providecommand \href [0]{\begingroup \@sanitize@url \@href}%
\providecommand \@href[1]{\@@startlink{#1}\@@href}%
\providecommand \@@href[1]{\endgroup#1\@@endlink}%
\providecommand \@sanitize@url [0]{\catcode `\\12\catcode `\$12\catcode
  `\&12\catcode `\#12\catcode `\^12\catcode `\_12\catcode `\%12\relax}%
\providecommand \@@startlink[1]{}%
\providecommand \@@endlink[0]{}%
\providecommand \url  [0]{\begingroup\@sanitize@url \@url }%
\providecommand \@url [1]{\endgroup\@href {#1}{\urlprefix }}%
\providecommand \urlprefix  [0]{URL }%
\providecommand \Eprint [0]{\href }%
\providecommand \doibase [0]{http://dx.doi.org/}%
\providecommand \selectlanguage [0]{\@gobble}%
\providecommand \bibinfo  [0]{\@secondoftwo}%
\providecommand \bibfield  [0]{\@secondoftwo}%
\providecommand \translation [1]{[#1]}%
\providecommand \BibitemOpen [0]{}%
\providecommand \bibitemStop [0]{}%
\providecommand \bibitemNoStop [0]{.\EOS\space}%
\providecommand \EOS [0]{\spacefactor3000\relax}%
\providecommand \BibitemShut  [1]{\csname bibitem#1\endcsname}%
\let\auto@bib@innerbib\@empty
%</preamble>
\bibitem [{\citenamefont {Patrignani}\ \emph {et~al.}(2016)\citenamefont
  {Patrignani} \emph {et~al.}}]{Patrignani:2016xqp}%
  \BibitemOpen
  \bibfield  {author} {\bibinfo {author} {\bibfnamefont {C.}~\bibnamefont
  {Patrignani}} \emph {et~al.} (\bibinfo {collaboration} {Particle Data
  Group}),\ }\href {\doibase 10.1088/1674-1137/40/10/100001} {\bibfield
  {journal} {\bibinfo  {journal} {Chin. Phys.}\ }\textbf {\bibinfo {volume}
  {C40}},\ \bibinfo {pages} {100001} (\bibinfo {year} {2016})}\BibitemShut
  {NoStop}%
%%CITATION = CHPHD,C40,100001;%%
\bibitem [{\citenamefont {Bohm}\ and\ \citenamefont
  {Sato}(2005)}]{Bohm:2004zi}%
  \BibitemOpen
  \bibfield  {author} {\bibinfo {author} {\bibfnamefont {A.~R.}\ \bibnamefont
  {Bohm}}\ and\ \bibinfo {author} {\bibfnamefont {Y.}~\bibnamefont {Sato}},\
  }\href {\doibase 10.1103/PhysRevD.71.085018} {\bibfield  {journal} {\bibinfo
  {journal} {Phys. Rev.}\ }\textbf {\bibinfo {volume} {D71}},\ \bibinfo {pages}
  {085018} (\bibinfo {year} {2005})},\ \Eprint
  {http://arxiv.org/abs/hep-ph/0412106} {arXiv:hep-ph/0412106 [hep-ph]}
  \BibitemShut {NoStop}%
%%CITATION = HEP-PH/0412106;%%
\bibitem [{\citenamefont {Hagedorn}(1965)}]{Hagedorn:1965st}%
  \BibitemOpen
  \bibfield  {author} {\bibinfo {author} {\bibfnamefont {R.}~\bibnamefont
  {Hagedorn}},\ }\href@noop {} {\bibfield  {journal} {\bibinfo  {journal}
  {Nuovo Cim. Suppl.}\ }\textbf {\bibinfo {volume} {3}},\ \bibinfo {pages}
  {147} (\bibinfo {year} {1965})}\BibitemShut {NoStop}%
%%CITATION = NUCUA,3,147;%%
\bibitem [{\citenamefont {Hagedorn}(1985)}]{Hagedorn:1984hz}%
  \BibitemOpen
  \bibfield  {author} {\bibinfo {author} {\bibfnamefont {R.}~\bibnamefont
  {Hagedorn}},\ }\bibfield  {booktitle} {\emph {\bibinfo {booktitle} {{Melting
  Hadrons, Boiling Quarks - From Hagedorn Temperature to Ultra-Relativistic
  Heavy-Ion Collisions at CERN: With a Tribute to Rolf Hagedorn}}},\ }\href
  {\doibase %%10.1007/978-3-319-17545-4_25} {\bibfield  {journal} {\bibinfo
  {journal} {Lect. Notes Phys.}\ }\textbf {\bibinfo {volume} {221}},\ \bibinfo
  {pages} {53} (\bibinfo {year} {1985})},\ \bibinfo {note}
  {[,287(2016)]}\BibitemShut {NoStop}%
%%CITATION = LNPHA,221,53;%%
\bibitem [{\citenamefont {Borsanyi}\ \emph {et~al.}(2014)\citenamefont
  {Borsanyi}, \citenamefont {Fodor}, \citenamefont {Hoelbling}, \citenamefont
  {Katz}, \citenamefont {Krieg},\ and\ \citenamefont
  {Szabo}}]{Borsanyi:2013bia}%
  \BibitemOpen
  \bibfield  {author} {\bibinfo {author} {\bibfnamefont {S.}~\bibnamefont
  {Borsanyi}}, \bibinfo {author} {\bibfnamefont {Z.}~\bibnamefont {Fodor}},
  \bibinfo {author} {\bibfnamefont {C.}~\bibnamefont {Hoelbling}}, \bibinfo
  {author} {\bibfnamefont {S.~D.}\ \bibnamefont {Katz}}, \bibinfo {author}
  {\bibfnamefont {S.}~\bibnamefont {Krieg}}, \ and\ \bibinfo {author}
  {\bibfnamefont {K.~K.}\ \bibnamefont {Szabo}},\ }\href {\doibase
  10.1016/j.physletb.2014.01.007} {\bibfield  {journal} {\bibinfo  {journal}
  {Phys. Lett.}\ }\textbf {\bibinfo {volume} {B730}},\ \bibinfo {pages} {99}
  (\bibinfo {year} {2014})},\ \Eprint {http://arxiv.org/abs/1309.5258}
  {arXiv:1309.5258 [hep-lat]} \BibitemShut {NoStop}%
%%CITATION = ARXIV:1309.5258;%%
\bibitem [{\citenamefont {Bazavov}\ \emph {et~al.}(2014)\citenamefont {Bazavov}
  \emph {et~al.}}]{Bazavov:2014pvz}%
  \BibitemOpen
  \bibfield  {author} {\bibinfo {author} {\bibfnamefont {A.}~\bibnamefont
  {Bazavov}} \emph {et~al.} (\bibinfo {collaboration} {HotQCD}),\ }\href
  {\doibase 10.1103/PhysRevD.90.094503} {\bibfield  {journal} {\bibinfo
  {journal} {Phys. Rev.}\ }\textbf {\bibinfo {volume} {D90}},\ \bibinfo {pages}
  {094503} (\bibinfo {year} {2014})},\ \Eprint {http://arxiv.org/abs/1407.6387}
  {arXiv:1407.6387 [hep-lat]} \BibitemShut {NoStop}%
%%CITATION = ARXIV:1407.6387;%%
\bibitem [{\citenamefont {Ruiz~Arriola}\ \emph {et~al.}(2014)\citenamefont
  {Ruiz~Arriola}, \citenamefont {Salcedo},\ and\ \citenamefont
  {Megias}}]{Arriola:2014bfa}%
  \BibitemOpen
  \bibfield  {author} {\bibinfo {author} {\bibfnamefont {E.}~\bibnamefont
  {Ruiz~Arriola}}, \bibinfo {author} {\bibfnamefont {L.~L.}\ \bibnamefont
  {Salcedo}}, \ and\ \bibinfo {author} {\bibfnamefont {E.}~\bibnamefont
  {Megias}},\ }\bibfield  {booktitle} {\emph {\bibinfo {booktitle} {{54th
  Cracow School of Theoretical Physics: QCD meets experiment}}},\ }\href
  {\doibase 10.5506/APhysPolB.45.2407} {\bibfield  {journal} {\bibinfo
  {journal} {Acta Phys. Polon.}\ }\textbf {\bibinfo {volume} {B45}},\ \bibinfo
  {pages} {2407} (\bibinfo {year} {2014})},\ \Eprint
  {http://arxiv.org/abs/1410.3869} {arXiv:1410.3869 [hep-ph]} \BibitemShut
  {NoStop}%
%%CITATION = ARXIV:1410.3869;%%
\bibitem [{\citenamefont {Godfrey}\ and\ \citenamefont
  {Isgur}(1985)}]{Godfrey:1985xj}%
  \BibitemOpen
  \bibfield  {author} {\bibinfo {author} {\bibfnamefont {S.}~\bibnamefont
  {Godfrey}}\ and\ \bibinfo {author} {\bibfnamefont {N.}~\bibnamefont
  {Isgur}},\ }\href {\doibase 10.1103/PhysRevD.32.189} {\bibfield  {journal}
  {\bibinfo  {journal} {Phys. Rev.}\ }\textbf {\bibinfo {volume} {D32}},\
  \bibinfo {pages} {189} (\bibinfo {year} {1985})}\BibitemShut {NoStop}%
%%CITATION = PHRVA,D32,189;%%
\bibitem [{\citenamefont {Capstick}\ and\ \citenamefont
  {Isgur}(1986)}]{Capstick:1986bm}%
  \BibitemOpen
  \bibfield  {author} {\bibinfo {author} {\bibfnamefont {S.}~\bibnamefont
  {Capstick}}\ and\ \bibinfo {author} {\bibfnamefont {N.}~\bibnamefont
  {Isgur}},\ }\bibfield  {booktitle} {\emph {\bibinfo {booktitle}
  {{Proceedings, International Conference on Hadron Spectroscopy: College Park,
  Maryland, April 20-22, 1985}}},\ }\href@noop {} {\bibfield  {journal}
  {\bibinfo  {journal} {Phys. Rev.}\ }\textbf {\bibinfo {volume} {D34}},\
  \bibinfo {pages} {2809} (\bibinfo {year} {1986})}\BibitemShut {NoStop}%
\bibitem [{\citenamefont {Ruiz~Arriola}\ \emph {et~al.}(2012)\citenamefont
  {Ruiz~Arriola}, \citenamefont {Broniowski},\ and\ \citenamefont
  {Masjuan}}]{Arriola:2012vk}%
  \BibitemOpen
  \bibfield  {author} {\bibinfo {author} {\bibfnamefont {E.}~\bibnamefont
  {Ruiz~Arriola}}, \bibinfo {author} {\bibfnamefont {W.}~\bibnamefont
  {Broniowski}}, \ and\ \bibinfo {author} {\bibfnamefont {P.}~\bibnamefont
  {Masjuan}},\ }\bibfield  {booktitle} {\emph {\bibinfo {booktitle}
  {{Proceedings, Conference on Modern approaches to nonperturbative gauge
  theories and their applications (Light Cone 2012): Cracow, Poland, July 8-13,
  2012}}},\ }\href {\doibase 10.5506/APhysPolBSupp.6.95} {\  (\bibinfo {year}
  {2012}),\ 10.5506/APhysPolBSupp.6.95},\ \bibinfo {note} {[Acta Phys. Polon.
  Supp.6,95(2013)]},\ \Eprint {http://arxiv.org/abs/1210.7153} {arXiv:1210.7153
  [hep-ph]} \BibitemShut {NoStop}%
%%CITATION = ARXIV:1210.7153;%%
\bibitem [{\citenamefont {Broniowski}(2016)}]{Broniowski:2016hvt}%
  \BibitemOpen
  \bibfield  {author} {\bibinfo {author} {\bibfnamefont {W.}~\bibnamefont
  {Broniowski}},\ }\bibfield  {booktitle} {\emph {\bibinfo {booktitle}
  {{Proceedings, Mini-Wokshop Bled 2016: Quarks, Hdrons, Matter: Bled,
  Slovenia, July 3-10, 2016}}},\ }\href@noop {} {\bibfield  {journal} {\bibinfo
   {journal} {Bled Workshops Phys.}\ }\textbf {\bibinfo {volume} {17}},\
  \bibinfo {pages} {1} (\bibinfo {year} {2016})},\ \Eprint
  {http://arxiv.org/abs/1610.09676} {arXiv:1610.09676 [nucl-th]} \BibitemShut
  {NoStop}%
%%CITATION = ARXIV:1610.09676;%%
\bibitem [{\citenamefont {Borsanyi}\ \emph {et~al.}(2012)\citenamefont
  {Borsanyi}, \citenamefont {Fodor}, \citenamefont {Katz}, \citenamefont
  {Krieg}, \citenamefont {Ratti},\ and\ \citenamefont
  {Szabo}}]{Borsanyi:2011sw}%
  \BibitemOpen
  \bibfield  {author} {\bibinfo {author} {\bibfnamefont {S.}~\bibnamefont
  {Borsanyi}}, \bibinfo {author} {\bibfnamefont {Z.}~\bibnamefont {Fodor}},
  \bibinfo {author} {\bibfnamefont {S.~D.}\ \bibnamefont {Katz}}, \bibinfo
  {author} {\bibfnamefont {S.}~\bibnamefont {Krieg}}, \bibinfo {author}
  {\bibfnamefont {C.}~\bibnamefont {Ratti}}, \ and\ \bibinfo {author}
  {\bibfnamefont {K.}~\bibnamefont {Szabo}},\ }\href {\doibase
  10.1007/JHEP01(2012)138} {\bibfield  {journal} {\bibinfo  {journal} {JHEP}\
  }\textbf {\bibinfo {volume} {01}},\ \bibinfo {pages} {138} (\bibinfo {year}
  {2012})},\ \Eprint {http://arxiv.org/abs/1112.4416} {arXiv:1112.4416
  [hep-lat]} \BibitemShut {NoStop}%
%%CITATION = ARXIV:1112.4416;%%
\bibitem [{\citenamefont {Bazavov}\ \emph {et~al.}(2012)\citenamefont {Bazavov}
  \emph {et~al.}}]{Bazavov:2012jq}%
  \BibitemOpen
  \bibfield  {author} {\bibinfo {author} {\bibfnamefont {A.}~\bibnamefont
  {Bazavov}} \emph {et~al.} (\bibinfo {collaboration} {HotQCD}),\ }\href
  {\doibase 10.1103/PhysRevD.86.034509} {\bibfield  {journal} {\bibinfo
  {journal} {Phys. Rev.}\ }\textbf {\bibinfo {volume} {D86}},\ \bibinfo {pages}
  {034509} (\bibinfo {year} {2012})},\ \Eprint {http://arxiv.org/abs/1203.0784}
  {arXiv:1203.0784 [hep-lat]} \BibitemShut {NoStop}%
%%CITATION = ARXIV:1203.0784;%%
\bibitem [{\citenamefont {Ruiz~Arriola}\ \emph {et~al.}(2016)\citenamefont
  {Ruiz~Arriola}, \citenamefont {Broniowski}, \citenamefont {Megias},\ and\
  \citenamefont {Salcedo}}]{RuizArriola:2016qpb}%
  \BibitemOpen
  \bibfield  {author} {\bibinfo {author} {\bibfnamefont {E.}~\bibnamefont
  {Ruiz~Arriola}}, \bibinfo {author} {\bibfnamefont {W.}~\bibnamefont
  {Broniowski}}, \bibinfo {author} {\bibfnamefont {E.}~\bibnamefont {Megias}},
  \ and\ \bibinfo {author} {\bibfnamefont {L.~L.}\ \bibnamefont {Salcedo}},\
  }in\ \href {https://inspirehep.net/record/1505434/files/arXiv:1612.07091.pdf}
  {\emph {\bibinfo {booktitle} {{Workshop on Excited Hyperons in QCD
  Thermodynamics at Freeze-Out (YSTAR2016) Mini-Proceedings}}}}\ (\bibinfo
  {year} {2016})\ pp.\ \bibinfo {pages} {128--139},\ \Eprint
  {http://arxiv.org/abs/1612.07091} {arXiv:1612.07091 [hep-ph]} \BibitemShut
  {NoStop}%
%%CITATION = ARXIV:1612.07091;%%
\bibitem [{\citenamefont {Capstick}(1992)}]{Capstick:1992uc}%
  \BibitemOpen
  \bibfield  {author} {\bibinfo {author} {\bibfnamefont {S.}~\bibnamefont
  {Capstick}},\ }\href {\doibase 10.1103/PhysRevD.46.2864} {\bibfield
  {journal} {\bibinfo  {journal} {Phys. Rev.}\ }\textbf {\bibinfo {volume}
  {D46}},\ \bibinfo {pages} {2864} (\bibinfo {year} {1992})}\BibitemShut
  {NoStop}%
%%CITATION = PHRVA,D46,2864;%%
\bibitem [{\citenamefont {Hey}\ and\ \citenamefont {Kelly}(1983)}]{Hey:1982aj}%
  \BibitemOpen
  \bibfield  {author} {\bibinfo {author} {\bibfnamefont {A.~J.~G.}\
  \bibnamefont {Hey}}\ and\ \bibinfo {author} {\bibfnamefont {R.~L.}\
  \bibnamefont {Kelly}},\ }\href {\doibase 10.1016/0370-1573(83)90114-X}
  {\bibfield  {journal} {\bibinfo  {journal} {Phys. Rept.}\ }\textbf {\bibinfo
  {volume} {96}},\ \bibinfo {pages} {71} (\bibinfo {year} {1983})}\BibitemShut
  {NoStop}%
%%CITATION = PRPLC,96,71;%%
\bibitem [{\citenamefont {Capstick}\ and\ \citenamefont
  {Roberts}(2000)}]{Capstick:2000qj}%
  \BibitemOpen
  \bibfield  {author} {\bibinfo {author} {\bibfnamefont {S.}~\bibnamefont
  {Capstick}}\ and\ \bibinfo {author} {\bibfnamefont {W.}~\bibnamefont
  {Roberts}},\ }\href {\doibase 10.1016/S0146-6410(00)00109-5} {\bibfield
  {journal} {\bibinfo  {journal} {Prog. Part. Nucl. Phys.}\ }\textbf {\bibinfo
  {volume} {45}},\ \bibinfo {pages} {S241} (\bibinfo {year} {2000})},\ \Eprint
  {http://arxiv.org/abs/nucl-th/0008028} {arXiv:nucl-th/0008028 [nucl-th]}
  \BibitemShut {NoStop}%
%%CITATION = NUCL-TH/0008028;%%
\bibitem [{\citenamefont {Anselmino}\ \emph {et~al.}(1993)\citenamefont
  {Anselmino}, \citenamefont {Predazzi}, \citenamefont {Ekelin}, \citenamefont
  {Fredriksson},\ and\ \citenamefont {Lichtenberg}}]{Anselmino:1992vg}%
  \BibitemOpen
  \bibfield  {author} {\bibinfo {author} {\bibfnamefont {M.}~\bibnamefont
  {Anselmino}}, \bibinfo {author} {\bibfnamefont {E.}~\bibnamefont {Predazzi}},
  \bibinfo {author} {\bibfnamefont {S.}~\bibnamefont {Ekelin}}, \bibinfo
  {author} {\bibfnamefont {S.}~\bibnamefont {Fredriksson}}, \ and\ \bibinfo
  {author} {\bibfnamefont {D.~B.}\ \bibnamefont {Lichtenberg}},\ }\href
  {\doibase 10.1103/RevModPhys.65.1199} {\bibfield  {journal} {\bibinfo
  {journal} {Rev. Mod. Phys.}\ }\textbf {\bibinfo {volume} {65}},\ \bibinfo
  {pages} {1199} (\bibinfo {year} {1993})}\BibitemShut {NoStop}%
%%CITATION = RMPHA,65,1199;%%
\bibitem [{\citenamefont {Klempt}\ \emph {et~al.}(2017)\citenamefont {Klempt},
  \citenamefont {Sarantsev},\ and\ \citenamefont {Thoma}}]{Klempt:2017lwq}%
  \BibitemOpen
  \bibfield  {author} {\bibinfo {author} {\bibfnamefont {E.}~\bibnamefont
  {Klempt}}, \bibinfo {author} {\bibfnamefont {A.~V.}\ \bibnamefont
  {Sarantsev}}, \ and\ \bibinfo {author} {\bibfnamefont {U.}~\bibnamefont
  {Thoma}},\ }\bibfield  {booktitle} {\emph {\bibinfo {booktitle}
  {{Proceedings, SFB/TRR16 Symposium: Subnuclear Structure of Matter:
  Achievements and Challenges: Bonn, Germany, June 6-9, 2016}}},\ }\href@noop
  {} {\bibfield  {journal} {\bibinfo  {journal} {EPJ Web Conf.}\ }\textbf
  {\bibinfo {volume} {134}},\ \bibinfo {pages} {02002} (\bibinfo {year}
  {2017})}\BibitemShut {NoStop}%
%%CITATION = 00776,134,02002;%%
\bibitem [{\citenamefont {Fleck}\ \emph {et~al.}(1988)\citenamefont {Fleck},
  \citenamefont {Silvestre-Brac},\ and\ \citenamefont
  {Richard}}]{Fleck:1988vm}%
  \BibitemOpen
  \bibfield  {author} {\bibinfo {author} {\bibfnamefont {S.}~\bibnamefont
  {Fleck}}, \bibinfo {author} {\bibfnamefont {B.}~\bibnamefont
  {Silvestre-Brac}}, \ and\ \bibinfo {author} {\bibfnamefont {J.~M.}\
  \bibnamefont {Richard}},\ }\href {\doibase 10.1103/PhysRevD.38.1519}
  {\bibfield  {journal} {\bibinfo  {journal} {Phys. Rev.}\ }\textbf {\bibinfo
  {volume} {D38}},\ \bibinfo {pages} {1519} (\bibinfo {year}
  {1988})}\BibitemShut {NoStop}%
%%CITATION = PHRVA,D38,1519;%%
\bibitem [{\citenamefont {Santopinto}(2005)}]{Santopinto:2004hw}%
  \BibitemOpen
  \bibfield  {author} {\bibinfo {author} {\bibfnamefont {E.}~\bibnamefont
  {Santopinto}},\ }\href {\doibase 10.1103/PhysRevC.72.022201} {\bibfield
  {journal} {\bibinfo  {journal} {Phys. Rev.}\ }\textbf {\bibinfo {volume}
  {C72}},\ \bibinfo {pages} {022201} (\bibinfo {year} {2005})}\BibitemShut
  {NoStop}%
\bibitem [{\citenamefont {Ferretti}\ \emph {et~al.}(2011)\citenamefont
  {Ferretti}, \citenamefont {Vassallo},\ and\ \citenamefont
  {Santopinto}}]{Ferretti:2011zz}%
  \BibitemOpen
  \bibfield  {author} {\bibinfo {author} {\bibfnamefont {J.}~\bibnamefont
  {Ferretti}}, \bibinfo {author} {\bibfnamefont {A.}~\bibnamefont {Vassallo}},
  \ and\ \bibinfo {author} {\bibfnamefont {E.}~\bibnamefont {Santopinto}},\
  }\href {\doibase 10.1103/PhysRevC.83.065204} {\bibfield  {journal} {\bibinfo
  {journal} {Phys. Rev.}\ }\textbf {\bibinfo {volume} {C83}},\ \bibinfo {pages}
  {065204} (\bibinfo {year} {2011})}\BibitemShut {NoStop}%
%%CITATION = PHRVA,C83,065204;%%
\bibitem [{\citenamefont {Gutierrez}\ and\ \citenamefont
  {De~Sanctis}(2014)}]{Gutierrez:2014qpa}%
  \BibitemOpen
  \bibfield  {author} {\bibinfo {author} {\bibfnamefont {C.}~\bibnamefont
  {Gutierrez}}\ and\ \bibinfo {author} {\bibfnamefont {M.}~\bibnamefont
  {De~Sanctis}},\ }\href {\doibase 10.1140/epja/i2014-14169-7} {\bibfield
  {journal} {\bibinfo  {journal} {Eur. Phys. J.}\ }\textbf {\bibinfo {volume}
  {A50}},\ \bibinfo {pages} {169} (\bibinfo {year} {2014})}\BibitemShut
  {NoStop}%
%%CITATION = EPHJA,A50,169;%%
\bibitem [{\citenamefont {Alexandrou}\ \emph {et~al.}(2006)\citenamefont
  {Alexandrou}, \citenamefont {de~Forcrand},\ and\ \citenamefont
  {Lucini}}]{Alexandrou:2006cq}%
  \BibitemOpen
  \bibfield  {author} {\bibinfo {author} {\bibfnamefont {C.}~\bibnamefont
  {Alexandrou}}, \bibinfo {author} {\bibfnamefont {P.}~\bibnamefont
  {de~Forcrand}}, \ and\ \bibinfo {author} {\bibfnamefont {B.}~\bibnamefont
  {Lucini}},\ }\href {\doibase 10.1103/PhysRevLett.97.222002} {\bibfield
  {journal} {\bibinfo  {journal} {Phys. Rev. Lett.}\ }\textbf {\bibinfo
  {volume} {97}},\ \bibinfo {pages} {222002} (\bibinfo {year} {2006})},\
  \Eprint {http://arxiv.org/abs/hep-lat/0609004} {arXiv:hep-lat/0609004
  [hep-lat]} \BibitemShut {NoStop}%
%%CITATION = HEP-LAT/0609004;%%
\bibitem [{\citenamefont {DeGrand}\ \emph {et~al.}(2008)\citenamefont
  {DeGrand}, \citenamefont {Liu},\ and\ \citenamefont
  {Schaefer}}]{DeGrand:2007vu}%
  \BibitemOpen
  \bibfield  {author} {\bibinfo {author} {\bibfnamefont {T.}~\bibnamefont
  {DeGrand}}, \bibinfo {author} {\bibfnamefont {Z.}~\bibnamefont {Liu}}, \ and\
  \bibinfo {author} {\bibfnamefont {S.}~\bibnamefont {Schaefer}},\ }\href
  {\doibase 10.1103/PhysRevD.77.034505} {\bibfield  {journal} {\bibinfo
  {journal} {Phys. Rev.}\ }\textbf {\bibinfo {volume} {D77}},\ \bibinfo {pages}
  {034505} (\bibinfo {year} {2008})},\ \Eprint {http://arxiv.org/abs/0712.0254}
  {arXiv:0712.0254 [hep-ph]} \BibitemShut {NoStop}%
%%CITATION = ARXIV:0712.0254;%%
\bibitem [{\citenamefont {Eichmann}\ \emph {et~al.}(2016)\citenamefont
  {Eichmann}, \citenamefont {Fischer},\ and\ \citenamefont
  {Sanchis-Alepuz}}]{Eichmann:2016hgl}%
  \BibitemOpen
  \bibfield  {author} {\bibinfo {author} {\bibfnamefont {G.}~\bibnamefont
  {Eichmann}}, \bibinfo {author} {\bibfnamefont {C.~S.}\ \bibnamefont
  {Fischer}}, \ and\ \bibinfo {author} {\bibfnamefont {H.}~\bibnamefont
  {Sanchis-Alepuz}},\ }\href {\doibase 10.1103/PhysRevD.94.094033} {\bibfield
  {journal} {\bibinfo  {journal} {Phys. Rev.}\ }\textbf {\bibinfo {volume}
  {D94}},\ \bibinfo {pages} {094033} (\bibinfo {year} {2016})},\ \Eprint
  {http://arxiv.org/abs/1607.05748} {arXiv:1607.05748 [hep-ph]} \BibitemShut
  {NoStop}%
%%CITATION = ARXIV:1607.05748;%%
\bibitem [{\citenamefont {Ebert}\ \emph {et~al.}(2011)\citenamefont {Ebert},
  \citenamefont {Faustov},\ and\ \citenamefont {Galkin}}]{Ebert:2011kk}%
  \BibitemOpen
  \bibfield  {author} {\bibinfo {author} {\bibfnamefont {D.}~\bibnamefont
  {Ebert}}, \bibinfo {author} {\bibfnamefont {R.~N.}\ \bibnamefont {Faustov}},
  \ and\ \bibinfo {author} {\bibfnamefont {V.~O.}\ \bibnamefont {Galkin}},\
  }\href@noop {} {\bibfield  {journal} {\bibinfo  {journal} {Phys. Rev.}\
  }\textbf {\bibinfo {volume} {D84}},\ \bibinfo {pages} {014025} (\bibinfo
  {year} {2011})}\BibitemShut {NoStop}%
\bibitem [{\citenamefont {Masjuan}\ and\ \citenamefont
  {Ruiz~Arriola}(2017)}]{Masjuan:2017fzu}%
  \BibitemOpen
  \bibfield  {author} {\bibinfo {author} {\bibfnamefont {P.}~\bibnamefont
  {Masjuan}}\ and\ \bibinfo {author} {\bibfnamefont {E.}~\bibnamefont
  {Ruiz~Arriola}},\ }\href {\doibase 10.1103/PhysRevD.96.054006} {\bibfield
  {journal} {\bibinfo  {journal} {Phys. Rev.}\ }\textbf {\bibinfo {volume}
  {D96}},\ \bibinfo {pages} {054006} (\bibinfo {year} {2017})},\ \Eprint
  {http://arxiv.org/abs/1707.05650} {arXiv:1707.05650 [hep-ph]} \BibitemShut
  {NoStop}%
%%CITATION = ARXIV:1707.05650;%%
\bibitem [{\citenamefont {Ruiz~Arriola}\ \emph {et~al.}(2017)\citenamefont
  {Ruiz~Arriola}, \citenamefont {Masjuan},\ and\ \citenamefont
  {Broniowski}}]{RuizArriola:2017ggc}%
  \BibitemOpen
  \bibfield  {author} {\bibinfo {author} {\bibfnamefont {E.}~\bibnamefont
  {Ruiz~Arriola}}, \bibinfo {author} {\bibfnamefont {P.}~\bibnamefont
  {Masjuan}}, \ and\ \bibinfo {author} {\bibfnamefont {W.}~\bibnamefont
  {Broniowski}},\ }\bibfield  {booktitle} {\emph {\bibinfo {booktitle}
  {{Proceedings, 9th Workshop "Excited QCD" 2017: Sintra, Portugal, May 7-13,
  2017}}},\ }\href {\doibase 10.5506/APhysPolBSupp.10.1079} {\bibfield
  {journal} {\bibinfo  {journal} {Acta Phys. Polon. Supp.}\ }\textbf {\bibinfo
  {volume} {10}},\ \bibinfo {pages} {1079} (\bibinfo {year} {2017})},\ \Eprint
  {http://arxiv.org/abs/1707.07508} {arXiv:1707.07508 [hep-ph]} \BibitemShut
  {NoStop}%
%%CITATION = ARXIV:1707.07508;%%
\bibitem [{\citenamefont {Tanabashi}\ \emph {et~al.}(2018)\citenamefont
  {Tanabashi} \emph {et~al.}}]{Tanabashi:2018oca}%
  \BibitemOpen
  \bibfield  {author} {\bibinfo {author} {\bibfnamefont {M.}~\bibnamefont
  {Tanabashi}} \emph {et~al.} (\bibinfo {collaboration} {ParticleDataGroup}),\
  }\href {\doibase 10.1103/PhysRevD.98.030001} {\bibfield  {journal} {\bibinfo
  {journal} {Phys. Rev.}\ }\textbf {\bibinfo {volume} {D98}},\ \bibinfo {pages}
  {030001} (\bibinfo {year} {2018})}\BibitemShut {NoStop}%
%%CITATION = PHRVA,D98,030001;%%
\bibitem [{\citenamefont {Broniowski}\ and\ \citenamefont
  {Florkowski}(2000)}]{Broniowski:2000bj}%
  \BibitemOpen
  \bibfield  {author} {\bibinfo {author} {\bibfnamefont {W.}~\bibnamefont
  {Broniowski}}\ and\ \bibinfo {author} {\bibfnamefont {W.}~\bibnamefont
  {Florkowski}},\ }\href {\doibase 10.1016/S0370-2693(00)00992-8} {\bibfield
  {journal} {\bibinfo  {journal} {Phys. Lett.}\ }\textbf {\bibinfo {volume}
  {B490}},\ \bibinfo {pages} {223} (\bibinfo {year} {2000})},\ \Eprint
  {http://arxiv.org/abs/hep-ph/0004104} {arXiv:hep-ph/0004104 [hep-ph]}
  \BibitemShut {NoStop}%
%%CITATION = HEP-PH/0004104;%%
\bibitem [{\citenamefont {Broniowski}\ \emph {et~al.}(2004)\citenamefont
  {Broniowski}, \citenamefont {Florkowski},\ and\ \citenamefont
  {Glozman}}]{Broniowski:2004yh}%
  \BibitemOpen
  \bibfield  {author} {\bibinfo {author} {\bibfnamefont {W.}~\bibnamefont
  {Broniowski}}, \bibinfo {author} {\bibfnamefont {W.}~\bibnamefont
  {Florkowski}}, \ and\ \bibinfo {author} {\bibfnamefont {L.~{\relax Ya}.}\
  \bibnamefont {Glozman}},\ }\href {\doibase 10.1103/PhysRevD.70.117503}
  {\bibfield  {journal} {\bibinfo  {journal} {Phys. Rev.}\ }\textbf {\bibinfo
  {volume} {D70}},\ \bibinfo {pages} {117503} (\bibinfo {year} {2004})},\
  \Eprint {http://arxiv.org/abs/hep-ph/0407290} {arXiv:hep-ph/0407290 [hep-ph]}
  \BibitemShut {NoStop}%
%%CITATION = HEP-PH/0407290;%%
\bibitem [{\citenamefont {Briceno}\ \emph {et~al.}(2016)\citenamefont {Briceno}
  \emph {et~al.}}]{Briceno:2015rlt}%
  \BibitemOpen
  \bibfield  {author} {\bibinfo {author} {\bibfnamefont {R.~A.}\ \bibnamefont
  {Briceno}} \emph {et~al.},\ }\href {\doibase 10.1088/1674-1137/40/4/042001}
  {\bibfield  {journal} {\bibinfo  {journal} {Chin. Phys.}\ }\textbf {\bibinfo
  {volume} {C40}},\ \bibinfo {pages} {042001} (\bibinfo {year} {2016})},\
  \Eprint {http://arxiv.org/abs/1511.06779} {arXiv:1511.06779 [hep-ph]}
  \BibitemShut {NoStop}%
%%CITATION = ARXIV:1511.06779;%%
\bibitem [{\citenamefont {Edwards}\ \emph {et~al.}(2013)\citenamefont
  {Edwards}, \citenamefont {Mathur}, \citenamefont {Richards},\ and\
  \citenamefont {Wallace}}]{Edwards:2012fx}%
  \BibitemOpen
  \bibfield  {author} {\bibinfo {author} {\bibfnamefont {R.~G.}\ \bibnamefont
  {Edwards}}, \bibinfo {author} {\bibfnamefont {N.}~\bibnamefont {Mathur}},
  \bibinfo {author} {\bibfnamefont {D.~G.}\ \bibnamefont {Richards}}, \ and\
  \bibinfo {author} {\bibfnamefont {S.~J.}\ \bibnamefont {Wallace}} (\bibinfo
  {collaboration} {Hadron Spectrum}),\ }\href {\doibase
  10.1103/PhysRevD.87.054506} {\bibfield  {journal} {\bibinfo  {journal} {Phys.
  Rev.}\ }\textbf {\bibinfo {volume} {D87}},\ \bibinfo {pages} {054506}
  (\bibinfo {year} {2013})},\ \Eprint {http://arxiv.org/abs/1212.5236}
  {arXiv:1212.5236 [hep-ph]} \BibitemShut {NoStop}%
%%CITATION = ARXIV:1212.5236;%%
\bibitem [{\citenamefont {Osterwalder}\ and\ \citenamefont
  {Seiler}(1978)}]{Osterwalder:1977pc}%
  \BibitemOpen
  \bibfield  {author} {\bibinfo {author} {\bibfnamefont {K.}~\bibnamefont
  {Osterwalder}}\ and\ \bibinfo {author} {\bibfnamefont {E.}~\bibnamefont
  {Seiler}},\ }\bibfield  {booktitle} {\emph {\bibinfo {booktitle} {{Salerno
  Soliton Wkshp.1977:0201}}},\ }\href {\doibase 10.1016/0003-4916(78)90039-8}
  {\bibfield  {journal} {\bibinfo  {journal} {Annals Phys.}\ }\textbf {\bibinfo
  {volume} {110}},\ \bibinfo {pages} {440} (\bibinfo {year}
  {1978})}\BibitemShut {NoStop}%
%%CITATION = APNYA,110,440;%%
\bibitem [{\citenamefont {Tawfik}(2005)}]{Tawfik:2004sw}%
  \BibitemOpen
  \bibfield  {author} {\bibinfo {author} {\bibfnamefont {A.}~\bibnamefont
  {Tawfik}},\ }\href {\doibase %%10.1103/PhysRevD.71.054502%%} {\bibfield
  {journal} {\bibinfo  {journal} {Phys.Rev.}\ }\textbf {\bibinfo {volume}
  {D71}},\ \bibinfo {pages} {054502} (\bibinfo {year} {2005})}\BibitemShut
  {NoStop}%
%%CITATION = HEP-PH/0412336;%%
\bibitem [{\citenamefont {Caro}\ \emph {et~al.}(1996)\citenamefont {Caro},
  \citenamefont {Ruiz~Arriola},\ and\ \citenamefont {Salcedo}}]{Caro:1994ht}%
  \BibitemOpen
  \bibfield  {author} {\bibinfo {author} {\bibfnamefont {J.}~\bibnamefont
  {Caro}}, \bibinfo {author} {\bibfnamefont {E.}~\bibnamefont {Ruiz~Arriola}},
  \ and\ \bibinfo {author} {\bibfnamefont {L.}~\bibnamefont {Salcedo}},\ }\href
  {\doibase 10.1088/0954-3899/22/7/006} {\bibfield  {journal} {\bibinfo
  {journal} {J.Phys.}\ }\textbf {\bibinfo {volume} {G22}},\ \bibinfo {pages}
  {981} (\bibinfo {year} {1996})},\ \Eprint
  {http://arxiv.org/abs/nucl-th/9410025} {arXiv:nucl-th/9410025 [nucl-th]}
  \BibitemShut {NoStop}%
%%CITATION = NUCL-TH/9410025;%%
\bibitem [{\citenamefont {Bellwied}\ \emph {et~al.}(2015)\citenamefont
  {Bellwied}, \citenamefont {Borsanyi}, \citenamefont {Fodor}, \citenamefont
  {Katz}, \citenamefont {Pasztor}, \citenamefont {Ratti},\ and\ \citenamefont
  {Szabo}}]{Bellwied:2015lba}%
  \BibitemOpen
  \bibfield  {author} {\bibinfo {author} {\bibfnamefont {R.}~\bibnamefont
  {Bellwied}}, \bibinfo {author} {\bibfnamefont {S.}~\bibnamefont {Borsanyi}},
  \bibinfo {author} {\bibfnamefont {Z.}~\bibnamefont {Fodor}}, \bibinfo
  {author} {\bibfnamefont {S.~D.}\ \bibnamefont {Katz}}, \bibinfo {author}
  {\bibfnamefont {A.}~\bibnamefont {Pasztor}}, \bibinfo {author} {\bibfnamefont
  {C.}~\bibnamefont {Ratti}}, \ and\ \bibinfo {author} {\bibfnamefont {K.~K.}\
  \bibnamefont {Szabo}},\ }\href {\doibase 10.1103/PhysRevD.92.114505}
  {\bibfield  {journal} {\bibinfo  {journal} {Phys. Rev.}\ }\textbf {\bibinfo
  {volume} {D92}},\ \bibinfo {pages} {114505} (\bibinfo {year} {2015})},\
  \Eprint {http://arxiv.org/abs/1507.04627} {arXiv:1507.04627 [hep-lat]}
  \BibitemShut {NoStop}%
%%CITATION = ARXIV:1507.04627;%%
\bibitem [{\citenamefont {Asakawa}\ and\ \citenamefont
  {Kitazawa}(2016)}]{Asakawa:2015ybt}%
  \BibitemOpen
  \bibfield  {author} {\bibinfo {author} {\bibfnamefont {M.}~\bibnamefont
  {Asakawa}}\ and\ \bibinfo {author} {\bibfnamefont {M.}~\bibnamefont
  {Kitazawa}},\ }\href {\doibase 10.1016/j.ppnp.2016.04.002} {\bibfield
  {journal} {\bibinfo  {journal} {Prog. Part. Nucl. Phys.}\ }\textbf {\bibinfo
  {volume} {90}},\ \bibinfo {pages} {299} (\bibinfo {year} {2016})},\ \Eprint
  {http://arxiv.org/abs/1512.05038} {arXiv:1512.05038 [nucl-th]} \BibitemShut
  {NoStop}%
%%CITATION = ARXIV:1512.05038;%%
\bibitem [{\citenamefont {Bakry}\ \emph {et~al.}(2015)\citenamefont {Bakry},
  \citenamefont {Chen},\ and\ \citenamefont {Zhang}}]{Bakry:2014gea}%
  \BibitemOpen
  \bibfield  {author} {\bibinfo {author} {\bibfnamefont {A.~S.}\ \bibnamefont
  {Bakry}}, \bibinfo {author} {\bibfnamefont {X.}~\bibnamefont {Chen}}, \ and\
  \bibinfo {author} {\bibfnamefont {P.-M.}\ \bibnamefont {Zhang}},\ }\href
  {\doibase 10.1103/PhysRevD.91.114506} {\bibfield  {journal} {\bibinfo
  {journal} {Phys. Rev.}\ }\textbf {\bibinfo {volume} {D91}},\ \bibinfo {pages}
  {114506} (\bibinfo {year} {2015})},\ \Eprint {http://arxiv.org/abs/1412.3568}
  {arXiv:1412.3568 [hep-lat]} \BibitemShut {NoStop}%
%%CITATION = ARXIV:1412.3568;%%
\bibitem [{\citenamefont {Bissey}\ \emph {et~al.}(2009)\citenamefont {Bissey},
  \citenamefont {Signal},\ and\ \citenamefont {Leinweber}}]{Bissey:2009gw}%
  \BibitemOpen
  \bibfield  {author} {\bibinfo {author} {\bibfnamefont {F.}~\bibnamefont
  {Bissey}}, \bibinfo {author} {\bibfnamefont {A.~I.}\ \bibnamefont {Signal}},
  \ and\ \bibinfo {author} {\bibfnamefont {D.~B.}\ \bibnamefont {Leinweber}},\
  }\href {\doibase 10.1103/PhysRevD.80.114506} {\bibfield  {journal} {\bibinfo
  {journal} {Phys. Rev.}\ }\textbf {\bibinfo {volume} {D80}},\ \bibinfo {pages}
  {114506} (\bibinfo {year} {2009})},\ \Eprint {http://arxiv.org/abs/0910.0958}
  {arXiv:0910.0958 [hep-lat]} \BibitemShut {NoStop}%
%%CITATION = ARXIV:0910.0958;%%
\bibitem [{\citenamefont {Bakry}\ \emph {et~al.}(2018)\citenamefont {Bakry},
  \citenamefont {Chen}, \citenamefont {Deliyergiyev}, \citenamefont {Galal},
  \citenamefont {Xu},\ and\ \citenamefont {Zhang}}]{Bakry:2017jna}%
  \BibitemOpen
  \bibfield  {author} {\bibinfo {author} {\bibfnamefont {A.}~\bibnamefont
  {Bakry}}, \bibinfo {author} {\bibfnamefont {X.}~\bibnamefont {Chen}},
  \bibinfo {author} {\bibfnamefont {M.}~\bibnamefont {Deliyergiyev}}, \bibinfo
  {author} {\bibfnamefont {A.}~\bibnamefont {Galal}}, \bibinfo {author}
  {\bibfnamefont {S.}~\bibnamefont {Xu}}, \ and\ \bibinfo {author}
  {\bibfnamefont {P.~M.}\ \bibnamefont {Zhang}},\ }\bibfield  {booktitle}
  {\emph {\bibinfo {booktitle} {{Proceedings, 17th International Conference on
  Hadron Spectroscopy and Structure (Hadron 2017): Salamanca, Spain, September
  25-29, 2017}}},\ }\href {\doibase 10.22323/1.310.0211} {\bibfield  {journal}
  {\bibinfo  {journal} {PoS}\ }\textbf {\bibinfo {volume} {Hadron2017}},\
  \bibinfo {pages} {211} (\bibinfo {year} {2018})},\ \Eprint
  {http://arxiv.org/abs/1712.03109} {arXiv:1712.03109 [hep-lat]} \BibitemShut
  {NoStop}%
%%CITATION = ARXIV:1712.03109;%%
\bibitem [{\citenamefont {Koma}\ and\ \citenamefont
  {Koma}(2017)}]{Koma:2017hcm}%
  \BibitemOpen
  \bibfield  {author} {\bibinfo {author} {\bibfnamefont {Y.}~\bibnamefont
  {Koma}}\ and\ \bibinfo {author} {\bibfnamefont {M.}~\bibnamefont {Koma}},\
  }\href {\doibase 10.1103/PhysRevD.95.094513} {\bibfield  {journal} {\bibinfo
  {journal} {Phys. Rev.}\ }\textbf {\bibinfo {volume} {D95}},\ \bibinfo {pages}
  {094513} (\bibinfo {year} {2017})},\ \Eprint
  {http://arxiv.org/abs/1703.06247} {arXiv:1703.06247 [hep-lat]} \BibitemShut
  {NoStop}%
%%CITATION = ARXIV:1703.06247;%%
\bibitem [{\citenamefont {Megias}\ \emph {et~al.}(2014)\citenamefont {Megias},
  \citenamefont {Ruiz~Arriola},\ and\ \citenamefont
  {Salcedo}}]{Megias:2013xaa}%
  \BibitemOpen
  \bibfield  {author} {\bibinfo {author} {\bibfnamefont {E.}~\bibnamefont
  {Megias}}, \bibinfo {author} {\bibfnamefont {E.}~\bibnamefont
  {Ruiz~Arriola}}, \ and\ \bibinfo {author} {\bibfnamefont {L.~L.}\
  \bibnamefont {Salcedo}},\ }\href@noop {} {\bibfield  {journal} {\bibinfo
  {journal} {Phys. Rev.}\ }\textbf {\bibinfo {volume} {D89}},\ \bibinfo {pages}
  {076006} (\bibinfo {year} {2014})}\BibitemShut {NoStop}%
\bibitem [{\citenamefont {Santopinto}\ and\ \citenamefont
  {Ferretti}(2015)}]{Santopinto:2014opa}%
  \BibitemOpen
  \bibfield  {author} {\bibinfo {author} {\bibfnamefont {E.}~\bibnamefont
  {Santopinto}}\ and\ \bibinfo {author} {\bibfnamefont {J.}~\bibnamefont
  {Ferretti}},\ }\href {\doibase 10.1103/PhysRevC.92.025202} {\bibfield
  {journal} {\bibinfo  {journal} {Phys. Rev.}\ }\textbf {\bibinfo {volume}
  {C92}},\ \bibinfo {pages} {025202} (\bibinfo {year} {2015})},\ \Eprint
  {http://arxiv.org/abs/1412.7571} {arXiv:1412.7571 [nucl-th]} \BibitemShut
  {NoStop}%
%%CITATION = ARXIV:1412.7571;%%
\bibitem [{\citenamefont {Luscher}(1981)}]{Luscher:1980ac}%
  \BibitemOpen
  \bibfield  {author} {\bibinfo {author} {\bibfnamefont {M.}~\bibnamefont
  {Luscher}},\ }\href {\doibase 10.1016/0550-3213(81)90423-5} {\bibfield
  {journal} {\bibinfo  {journal} {Nucl. Phys.}\ }\textbf {\bibinfo {volume}
  {B180}},\ \bibinfo {pages} {317} (\bibinfo {year} {1981})}\BibitemShut
  {NoStop}%
%%CITATION = NUPHA,B180,317;%%
\bibitem [{\citenamefont {Jaffe}(2005)}]{Jaffe:2004ph}%
  \BibitemOpen
  \bibfield  {author} {\bibinfo {author} {\bibfnamefont {R.~L.}\ \bibnamefont
  {Jaffe}},\ }\bibfield  {booktitle} {\emph {\bibinfo {booktitle}
  {{Proceedings, 6th International Conference on Hyperons, charm and beauty
  hadrons (BEACH 2004): Chicago, USA, June 27-July 3, 2004}}},\ }\href
  {\doibase 10.1016/j.physrep.2004.11.005} {\bibfield  {journal} {\bibinfo
  {journal} {Phys. Rept.}\ }\textbf {\bibinfo {volume} {409}},\ \bibinfo
  {pages} {1} (\bibinfo {year} {2005})},\ \bibinfo {note} {[,191(2004)]},\
  \Eprint {http://arxiv.org/abs/hep-ph/0409065} {arXiv:hep-ph/0409065 [hep-ph]}
  \BibitemShut {NoStop}%
%%CITATION = HEP-PH/0409065;%%
\bibitem [{\citenamefont {Ruiz~Arriola}\ \emph {et~al.}(2018)\citenamefont
  {Ruiz~Arriola}, \citenamefont {Amaro},\ and\ \citenamefont
  {Navarro~Pérez}}]{RuizArriola:2017kqs}%
  \BibitemOpen
  \bibfield  {author} {\bibinfo {author} {\bibfnamefont {E.}~\bibnamefont
  {Ruiz~Arriola}}, \bibinfo {author} {\bibfnamefont {J.~E.}\ \bibnamefont
  {Amaro}}, \ and\ \bibinfo {author} {\bibfnamefont {R.}~\bibnamefont
  {Navarro~Pérez}},\ }\bibfield  {booktitle} {\emph {\bibinfo {booktitle}
  {{Proceedings, 17th International Conference on Hadron Spectroscopy and
  Structure (Hadron 2017): Salamanca, Spain, September 25-29, 2017}}},\ }\href
  {\doibase 10.22323/1.310.0134} {\bibfield  {journal} {\bibinfo  {journal}
  {PoS}\ }\textbf {\bibinfo {volume} {Hadron2017}},\ \bibinfo {pages} {134}
  (\bibinfo {year} {2018})},\ \Eprint {http://arxiv.org/abs/1711.11338}
  {arXiv:1711.11338 [nucl-th]} \BibitemShut {NoStop}%
%%CITATION = ARXIV:1711.11338;%%
\bibitem [{\citenamefont {De~Sanctis}\ \emph {et~al.}(2016)\citenamefont
  {De~Sanctis}, \citenamefont {Ferretti}, \citenamefont {Santopinto},\ and\
  \citenamefont {Vassallo}}]{DeSanctis:2014ria}%
  \BibitemOpen
  \bibfield  {author} {\bibinfo {author} {\bibfnamefont {M.}~\bibnamefont
  {De~Sanctis}}, \bibinfo {author} {\bibfnamefont {J.}~\bibnamefont
  {Ferretti}}, \bibinfo {author} {\bibfnamefont {E.}~\bibnamefont
  {Santopinto}}, \ and\ \bibinfo {author} {\bibfnamefont {A.}~\bibnamefont
  {Vassallo}},\ }\href@noop {} {\bibfield  {journal} {\bibinfo  {journal} {Eur.
  Phys. J.}\ }\textbf {\bibinfo {volume} {A52}},\ \bibinfo {pages} {121}
  (\bibinfo {year} {2016})}\BibitemShut {NoStop}%
\bibitem [{\citenamefont {Ruiz~Arriola}\ and\ \citenamefont
  {Broniowski}(2007)}]{Arriola:2006sv}%
  \BibitemOpen
  \bibfield  {author} {\bibinfo {author} {\bibfnamefont {E.}~\bibnamefont
  {Ruiz~Arriola}}\ and\ \bibinfo {author} {\bibfnamefont {W.}~\bibnamefont
  {Broniowski}},\ }\bibfield  {booktitle} {\emph {\bibinfo {booktitle} {{Quarks
  and nuclear physics. Proceedings, 4th International Conference, QNP 2006,
  Madrid, Spain, June 5-10, 2006}}},\ }\href {\doibase
  10.1140/epja/i2006-10184-7} {\bibfield  {journal} {\bibinfo  {journal} {Eur.
  Phys. J.}\ }\textbf {\bibinfo {volume} {A31}},\ \bibinfo {pages} {739}
  (\bibinfo {year} {2007})},\ \Eprint {http://arxiv.org/abs/hep-ph/0609266}
  {arXiv:hep-ph/0609266 [hep-ph]} \BibitemShut {NoStop}%
%%CITATION = HEP-PH/0609266;%%
\bibitem [{\citenamefont {Borsanyi}\ \emph {et~al.}(2018)\citenamefont
  {Borsanyi}, \citenamefont {Fodor}, \citenamefont {Guenther}, \citenamefont
  {Katz}, \citenamefont {Szabo}, \citenamefont {Pasztor}, \citenamefont
  {Portillo},\ and\ \citenamefont {Ratti}}]{Borsanyi:2018grb}%
  \BibitemOpen
  \bibfield  {author} {\bibinfo {author} {\bibfnamefont {S.}~\bibnamefont
  {Borsanyi}}, \bibinfo {author} {\bibfnamefont {Z.}~\bibnamefont {Fodor}},
  \bibinfo {author} {\bibfnamefont {J.~N.}\ \bibnamefont {Guenther}}, \bibinfo
  {author} {\bibfnamefont {S.~K.}\ \bibnamefont {Katz}}, \bibinfo {author}
  {\bibfnamefont {K.~K.}\ \bibnamefont {Szabo}}, \bibinfo {author}
  {\bibfnamefont {A.}~\bibnamefont {Pasztor}}, \bibinfo {author} {\bibfnamefont
  {I.}~\bibnamefont {Portillo}}, \ and\ \bibinfo {author} {\bibfnamefont
  {C.}~\bibnamefont {Ratti}},\ }\href {\doibase 10.1007/JHEP10(2018)205}
  {\bibfield  {journal} {\bibinfo  {journal} {JHEP}\ }\textbf {\bibinfo
  {volume} {10}},\ \bibinfo {pages} {205} (\bibinfo {year} {2018})},\ \Eprint
  {http://arxiv.org/abs/1805.04445} {arXiv:1805.04445 [hep-lat]} \BibitemShut
  {NoStop}%
%%CITATION = ARXIV:1805.04445;%%
\bibitem [{\citenamefont {Montvay}\ and\ \citenamefont
  {Munster}(1997)}]{Montvay:1994cy}%
  \BibitemOpen
  \bibfield  {author} {\bibinfo {author} {\bibfnamefont {I.}~\bibnamefont
  {Montvay}}\ and\ \bibinfo {author} {\bibfnamefont {G.}~\bibnamefont
  {Munster}},\ }\href {\doibase 10.1017/CBO9780511470783} {\emph {\bibinfo
  {title} {{Quantum fields on a lattice}}}},\ Cambridge Monographs on
  Mathematical Physics\ (\bibinfo  {publisher} {Cambridge University Press},\
  \bibinfo {year} {1997})\BibitemShut {NoStop}%
%%CITATION = INSPIRE-378182;%%
\end{thebibliography}%

\end{document}